\documentclass[10pt,twocolumn,twoside]{IEEEtran}

\usepackage{etex}

\usepackage{times,enumerate} 
\usepackage[usenames,dvipsnames,svgnames,table]{xcolor}

\usepackage{subfloat}

\usepackage{graphicx}
\usepackage{setspace}
\usepackage{bbm}
\usepackage{mathdots,mathrsfs}
\usepackage{amssymb,latexsym,amsfonts,amsmath,cite,comment,relsize,enumitem}
\usepackage{stmaryrd}
\usepackage[dvips]{psfrag}
\usepackage{amsthm}

\usepackage{adjustbox}

\usepackage{setspace}
\usepackage{amsmath}
\usepackage{url}
\usepackage{soul}
\usepackage{blindtext}
\usepackage{graphicx}
\usepackage{lipsum}

\usepackage{adjustbox}
\usepackage{tikz}
\usepackage{pgfplots}
\pgfplotsset{every tick label/.append style={font=\small}}

\usepackage{relsize}
\usetikzlibrary{patterns}

\usepackage{epsfig} 
\usepackage{subfig}
\usepackage{algorithm,algorithmic}
\algsetup{linenosize=\tiny}

\usepackage{setspace}
\let\Algorithm\algorithm
\renewcommand\algorithm[1][]{\Algorithm[#1]\setstretch{0.7}}

\usepackage{mathtools}

\newtheorem{theorem}{Theorem}
\newtheorem*{theorem*}{Theorem}
\newtheorem{lemma}{Lemma}

\newtheorem{proposition}{Proposition}
\newtheorem{corollary}{Corollary}

\theoremstyle{remark}
\newtheorem*{continuancex}{Example \continuanceref  ~(Continued)}  
\newenvironment{continuance}[1]
  {\newcommand\continuanceref{\ref{#1}}\continuancex}
  {\endcontinuancex}

\theoremstyle{definition}
\newtheorem{definition}{Definition}

\theoremstyle{remark}
\newtheorem{example}{Example}
\newtheorem{remark}{Remark}

\newcommand{\BC}{\color{blue}}
\newcommand{\BR}{\color{brown}}

\tikzstyle{vertex}=[circle, draw, inner sep=0pt, minimum size=6pt]
\usepackage[sans]{dsfont}
\usepackage[font={small,it}]{caption}

\newcommand{\R}{\mathbb{R}}
\newcommand{\suchthat}{\;\ifnum\currentgrouptype=16 \middle\fi|\;}

\title{
Explicit Characterization of Performance of a Class of  Networked Linear Control Systems %Part II: Sparsification and Design   %with Guaranteed $\mathcal{H}_2$ Norm Bounds
}

\author{Hossein K. Mousavi and Nader Motee$^{1}$% <-this % stops a space 
\thanks{$^{1}$The Authors are with the Department of Mechanical Engineering and Mechanics, Lehigh University,
        Bethlehem, PA 18015, USA
        {\tt\small \{mousavi, motee\}@lehigh.edu}}% 
}

\begin{document}

\maketitle
\thispagestyle{empty}
\pagestyle{empty}

%%%%%%%%%%%%%%%%%%%%%%%%%%%%%%%%%%%%%%%%%%%%%%%%%%%%%%%%%%%%%%%%%%%%%%%%%%%%%%%%
\begin{abstract} We show that the steady-state variance as a performance measure for a class of networked linear control systems is expressible as the summation of a rational function over the Laplacian eigenvalues of the network graph. Moreover, we characterize the role of connectivity thresholds for the feedback (and observer) gain design of these networks. We use our framework to derive bounds and scaling laws for the performance of the dynamical network. Our approach generalizes and unifies the previous results on the performance measure of these networks for the case of arbitrary nodal dynamics. We bring extensions of our methodology for the case of decentralized observer-based output feedback as well as a class of composite networks. Numerous examples support our theoretical contributions. 

%We study the decentralized control of a group of identical control systems, where each system is disturbed by an additive  disturbance process. We consider a consensus-based control architecture where each subsystem may acquire the feedback on an output variable of their neighbors over the undirected graph of the network. We utilize  the variance of the network output in the steady-state as a performance measure and show that it can be written as the sum of a so-called performance function over the nonzero Laplacian eigenvalues. Moreover, we demonstrate that the notion of minimum connectivity threshold is useful for design of the feedback gains for these networks. Additionally, we investigate the fundamental limitation on the best attainable value of the performance measure. This framework allows us to analyze the performance of a wide range of noisy dynamical networks, as well as a class of networks of networks.  Furthermore, these performance functions let us  derive graph-theoretic bounds and approximations to the performance measure over general and specific topologies, which help us find scaling laws for this measure and deduce design tradeoffs. Numerous examples have been included to support and clarify our theoretical contributions. 
\end{abstract}

%\begin{keywords}
% Nonlinear Dynamical Systems, Network Theory, Operator Theory
%\end{keywords}

%%%%%%%%%%%%%%%%%%%%%%%%%%%%%%%%%%%%%%%%%%%%%%%%%%%%%%%%%%%%%%%%%%%%%%%%%%%%%%%%
\section{Introduction}  \label{sec:int}

Developing tools to reduce design complexity has been in the center of recent research in networked control systems \cite{siami2018network, siami2017growing, jadbabaie2018scaling,fardad2014design,andreasson2017performance}. In several important applications, network design problem reduces to finding an optimal communication (graph) topology among a network of identical subsystems that are coupled to each other through some common mission-related control objectives. Examples include formation control in a cooperative team of robots, the platoon of vehicles in automated highways, space-time rendezvous in a team of robots, and networks of synchronous oscillators in power networks. The design problem usually involves the optimization of a measure of performance or robustness while respecting various constraints. Due to their combinatorial nature, most network design problems become intractable as network size increases and suffer from high computational complexities. Possibility of characterizing performance and robustness measures in closed and explicit forms will significantly facilitate the design process by allowing the network designer to identify relevant functional properties of the measures and their behaviors with respect to the interconnection topology. In this paper, we present explicit expressions for the $\mathcal{H}_2$-norm, as a performance and robustness measure,  of a class of interconnected network of linear control systems.

The authors of \cite{bamieh2012coherence} consider coherency of a platoon of vehicles by evaluating the $\mathcal{H}_2$-norm of second-order consensus algorithms and propose several scaling laws for various scenarios of coordination.  In \cite{jovanovic2005ill}, ill-posedness of a certain class of platoons is investigated and shown that stabilizability deteriorates as the size of the platoon increases.  The string stability of a class of formation problems with limited communication range is studied in \cite{middleton2010string}, where a fundamental limit on the disturbance rejection quality of the network in the frequency domain is derived. The stability and robustness of 
large platoon of vehicles with double-integrator dynamics are considered in \cite{hao2013stability}, where it is shown that how scaling of a robustness measure (in terms of the platoon size) to external disturbances improves from geometric to polynomial growth when vehicles are allowed to communicate with their two immediate neighbors. In \cite{zheng2016platooning}, robustness analysis and distributed $\mathcal{H}_\infty$ controller design of platoon of vehicles with third-order models and undirected communication topologies are considered. In  \cite{siami2016fundamental}, several graph theoretic bounds on the $\mathcal{H}_2$-based performance of linear consensus networks with first- and second-order dynamics are characterized and it is shown how the performance measure scales with the network size and depends on structural properties of the communication topology.  %\cite{xiao2007} \cite{lin2012optimal} \cite{patterson2010leader} \cite{siami2013robustness} \cite{fardad2014design} \cite{siami2016tractable}\cite{moghaddam2015interior} 
In \cite{li2011h}, the authors consider distributed  $\mathcal{H}_2$ and $\mathcal{H}_\infty$ controller design for a multi-agent system whose subsystems   have general linear time-invariant dynamics. Using a consensus-like algorithm and notion of the grounded graph (e.g., see \cite{pirani2015smallest}) to model coupling of agents to leaders, it is shown under what conditions such controllers exist and how they can be suboptimally designed.

In this paper, we consider a network of identical subsystems that are connected over an undirected graph and subject to external disturbance and measurement noise.   We propose a methodology to express the steady-state variance of the output of  a class of interconnected linear time-invariant networks as a rational function of their Laplacian eigenvalues. Our method extends the existing results in the literature for first- and second-order linear consensus network models (cf. \cite{siami2017growing} and reference in there).    We illustrate that the notion of minimum connectivity threshold is useful for the design of the feedback gains for these networks. It turns out that stabilizability of the nodal dynamics (and detectability in case of observer-based output-feedback)  guarantee the existence of such designs. Using these developments, it is shown that fundamental limits may emerge for networks whose subsystems are non-minimum phase.  We find graph-theoretic bounds for the performance of the network, which paves the way to find scaling laws for the performance measure. Moreover, a tradeoff between the graph sparsity and performance measure is revealed. Additionally, for networks over path or cycle graphs, we find the asymptotic trend of the performance measure. We bring two extensions of the analyses for the cases of observer-based output feedback as well as a class of composite networks.  We have included several parametric and numerical examples to support our theoretical contributions\footnote{The proofs are included in the appendices A to P of the paper.}. Our approach is advantageous for the design of these dynamical networks. Our spectral expressions  can facilitate solving of underlying optimal control problems: instead of dealing directly with optimization problems with high-dimensional matrices, our method leverages the structure of the control system and decouples the roles of typically low-dimensional feedback gains and the eigenspectrum of the communication graph. This is the outgrowth of the preliminary results that were presented in the conference version \cite{mousavi2017spectral}.

{\BR 
%In this paper, we consider a similar linear consensus algorithm over an undirected graph for a network of identical control systems, where the subsystems are able to get noisy feedback on an output variable from their neighbors. A random disturbance process also exists on each subsystem. 
 %The main distinction  of our work from \cite{li2011h} is focus on derivation and further applications of spectral expressions for the $\mathcal{H}_2$ performance measure in terms of eigenvalues of the graph Laplacian. 
 %  We show that the performance measure  can be expressed as the sum of a so-called performance function over the Laplacian eigenvalues of the network graph. We illustrate that notion of minimum connectivity threshold is useful for  design of the feedback gains for these networks, and that  stabilizability   (and detectability) of the nodal dynamics guarantee the existence of this design.  These are the essential tools that let us find  bounds on the best achievable performance of the dynamical network. We also uncover the capability of this  framework for analysis of a class of networks of networks.  We find graph-theoretic   bounds for the performance of the network, that paves the way to find scaling laws for the performance measure. Moreover, they reveal a tradeoff between the graph sparsity and performance measure. Additionally, for networks over  path or cycle graphs, we find  the asymptotic trend of the performance measure. We have included several parametric and numerical examples to support our theoretical contributions.
}

\section{Notations and Preliminaries}

% A linear time-invariant system $\dot x=\mathbf{A}x$ is called asymptotically stable if the eigenvalues of $\mathbf{A}$ have negative  real parts. 
%Such a system is called stable if  for every $t>0$, $\| \mathrm{e}^{ \mathbf{A} t} \|_2<B$ for some $B>0$. 
%A stable system that is not asymptotically stable is called marginally stable.  
%We use set of Reals $\mathbb{R}$,  Integers $\mathbb{Z}$ and Natural numbers $\mathbb{N}$ in this work. 
The subscripts $_+$ and $_{++}$ subscripts denote the nonnegative and positive subsets of a set, respectively (e.g. $\R_+$). The  operator$\mathrm{Tr}(.)$ represents matrix trace. The partial ordering on the cone of positive-semidefinite matrices is denoted via  $\succ$ and similar operators.   The standard basis for $\R^N$ is denoted by the set of vectors $\{e_1,\dots,e_N\}$. 
 The vector and matrix of ones are denoted by ${1}_N\in \R^N$ and ${\mathbf{J}_N}\in \R^{N\times N}$, respectively. Also, $\mathbf{I}_N$ and ${\mathbf{M}_N}=\mathbf{I}_N-\mathbf{J}_N/N$ are identity and centering matrices, respectively.  The vectorization  is denoted by $\mathrm{vec}(\mathbf S)$. The Kronecker product is denoted by $\mathbf{A}\otimes {\mathbf{B}}$. %If the products $AC$ and ${\mathbf{B}}D$ are well-defined, then 
%$
%(A\otimes {\mathbf{B}})(C\otimes D) =(AC)\otimes ({\mathbf{B}}D) .
%$  Moreover,  
%\begin{gather*}%\label{eq:vec_kron}
%$
%\mathrm{vec}(A{\mathbf{B}}C)=(C^T\otimes A)\mathrm{vec}({\mathbf{B}}).
%\end{gather*}
%$
The matrix transpose and conjugate transpose are denoted by $(.)^T$ and $(.)^*$ superscripts, respectively. 
 %$S(n,\R)$ denotes the group of symmetric $n\times n$ matrices.  
 %The vectorization of a matrix $S$ is represented by $\mathrm{vec}(S)$. $A\otimes {\mathbf{B}}$ denotes the Kronecker product of $A$ and ${\mathbf{B}}$. These two operators are linked via the identity  
%\begin{gather}\label{eq:vec_kron}
%\mathrm{vec}(A{\mathbf{B}}C)=(C^T\otimes A)\mathrm{vec}({\mathbf{B}}).
%\end{gather}
 %The vector and matrix of ones are denoted by $\mathbb{1}_n\in \R^n$ and ${\mathbf{J}_n}\in \R^{n\times n}$, respectively. Also, $\mathbf{I}_n$ and $\mathbf{M}_n=\mathbf{I}_n-\mathbf{J}_n/n$ are identity and centering matrices, respectively. %The spectral radius of $A$ is defined as $\rho(A):=\max_i |\lambda_i|.$
% Let $S_n^{++}$ denote the set of positive-definite and symmetric $n\times n$ matrices. Similar to \cite{xiao2007}, we define set $\phi$, which is a subset of symmetric and doubly-stochastic matrices, as 
%\begin{gather*}
%\phi:=\{W\in S_n^{++}\suchthat \rho(W)\leq 1, ~W\mathds{1}_n=\mathds{1}_n\}.
%\end{gather*}
%Symmetric matrices are diagonalizable.
A weighted undirected graph over $N$ nodes is a collection $\mathcal{G}={(\mathcal{V},\mathcal{E},k)}$ with the following components: the set of nodes $\mathcal{V}=\{1,2,\dots,N\}$, the set of edges $\mathcal{E}\subset\big\{\{i,j\}\suchthat i,j\in \mathcal{V}\}$, and the weight function $k:~\mathcal{E}\rightarrow $  $\R_{+}$. % We call graph $\mathcal{G}_1$ with the same nodes a subgraph of $\mathcal{G}$ and denote it as $\mathcal{G}_1 \subset \mathcal{G}$ if $\mathcal{E}_1 \subset \mathcal{E}$. 
%For an undirected graph over $N$ nodes with $M$ links, assuming any numbering for the links of the graph, the loopy incidence matrix $\beta \in \R^{M \times N}$ of the graph is
%$$
%\beta=\begin{pmatrix} \mathbf{e}_1 &\mathbf{e}_2 & \hdots &\mathbf{e}_M\end{pmatrix}^T,
%$$ where if link $k\in \{1,\dots,M\}$ connects nodes $i$ and $j$, then 
%\begin{align}
%\mathbf{e}_k=\left \{ \arraycolsep=1pt\def\arraystretch{1.5} \begin{array}{ll}
%        e_i-e_j & ~~\mbox{if } i\neq j \\
%        e_i & ~~\mbox{if } i= j
%    \end{array}\right. ,
%\end{align}
%which we call it the vector corresponding to link $a_{ij}$. 
We define $k_{ij}:=k(\{i,j\})=k_{ji}$  and
% stack the weights of all edges in a vector $E:=\mathrm{vec}(w_{ij})$. The degree of node $i\in \mathcal{V}$ is 
% $d_i:=\sum_{e=\{i,j\}\in \mathcal{E}}^n w(e).$
%which induces the degree matrix $D_{\mathcal{G}}=\mathbf{diag}(d_i)$.
% The adjacency matrix of a graph $A_{\mathcal{G}}$ is defined with $(A_{\mathcal{G}})_{ij}=w\left(\{i,j\}\right)$ if $\{i,j\}\in \mathcal{V}$, and $(A_{\mathcal{G}})_{ij}=0$ otherwise.
form the (symmetric) graph Laplacian  ${\mathbf{L}}=\left [l_{ij}\right ]\in \R^{N\times N}$ with entries %that are given by 
\begin{align}
l_{ij}=\left\{
    \begin{array}{ll}
        \displaystyle \sum_{\{i,j\}\in \mathcal{E}} k_{ij} & \mbox{if } i=j \\
        ~-k_{ij} & \mbox{if } i\neq j
    \end{array}
\right. .
\end{align}
%In cases that we allow self-loops (or multi-graph \cite{li2011h}). 
%For both loopless and loopy graph Laplacian, if we put the weight of the links $\mathbf{a}_k=a_{ij}$ in the $k^{\text{th}}$ diagonal of the weight matrix $W \in \R^{M \times M}$ (with zero off-diagonal entries), we may alternatively evaluate $\mathbf{L}$ as 
%$$
%L=\beta^T W \beta=\sum_{k=1}^M \mathbf{a}_k \mathbf{e}_k \mathbf{e}_k^T,
%$$ 
%which in the case of an unweighted graph, $W=\mathbf{I}_M$, thus
%$$
%L=\beta^T \beta=\sum_{k=1}^M \mathbf{e}_k \mathbf{e}_k^T.
%$$ 
The set of neighbors 
of a node is 
%denoted by $\mathcal{N}_i$ that is
$
 \mathcal{N}_i:=\left \{j\in \mathcal{V}~\big|~  \{i,j\}\in \mathcal{E}\right \}
$ for $i \in \mathcal{V}$. 
%The spectrum of the graph Laplacian plays a crucial role in many applications.
The eigenvalues of  ${\mathbf{L}}$ are denoted by $\lambda_1\leq \dots\leq \lambda_N$, which 
are real and nonnegative for a weighted undirected graph.  For a connected graph, $\lambda_1=0$ with eigenvector $1_N$, and  $\lambda_2>0$. The Laplacian eigendecomposition is $\mathbf L=\mathbf{U}\mathbf{\Lambda} \mathbf{U}^T,$ where $\mathbf{U}$ is its orthonormal matrix of eigenvectors and $\mathbf{\Lambda}=\mathrm{diag}(\lambda_1,\dots,\lambda_N).$ 
%A loopy Laplacian can be decomposed into $L=\mathbf{L}_0+\mathbf{L}_d$, where $\mathbf{L}_0$ is the loopless Laplacian of a graph with all links excluding the self-loops, and $\mathbf{L}_d$ is the diagonal matrix with self-loops on the diagonal. %For a loopy graph, we denote the weight sum of the loops and largest self loop through  
%$$
%a_L:=\sum_{i\in \mathcal{V}} a_{ii},~~a_{\max}:=\max_{j=1,\dots,N} a_{jj}.
%$$
For positive  sequences $\{p_n\}_{n \in \mathbb{Z}_+}$ and $\{q_n\}_{n \in \mathbb{Z}_+}$, we adapt 
$
q_n=O(p_n) $ if    ${{q_n}/{p_n}} \leq C \text{ for some } C>0$. Moreover,  
	$q_n=O(p_n) \Leftrightarrow p_n=\Omega (q_n)$. Additionally, $ 
	q_n=O(p_n),~ q_n=\Omega (p_n) \Leftrightarrow q_n=\Theta(p_n)$.  % p_n=\Theta(q_n) 
	Finally, we consider  $
 q_n \sim p_n \Leftrightarrow  \displaystyle\lim_{n \rightarrow \infty } {q_n}/{p_n} =1. 
$
%\begin{align*}
%&q_n=O(p_n) \Leftrightarrow  {{q_n}/{p_n}} \leq m \text{ for some } m>0   \\
%&\begin{adjustbox}{max width=240pt}$
%q_n=O(p_n) \Leftrightarrow p_n=\Omega (q_n),$
%\end{adjustbox} \\
%&
%\begin{adjustbox}{max width=240pt}$
%q_n=O(p_n),~ q_n=\Omega (p_n) \Leftrightarrow q_n=\Theta(p_n) \Leftrightarrow p_n=\Theta(q_n) 
%$
%\end{adjustbox}  \\
%& q_n \sim p_n \Leftrightarrow \displaystyle \lim_{n \rightarrow \infty } {q_n}/{p_n} =1
%\end{align*}

%\noindent {\it Probability Theory:}
%Consider a probability space $(\Omega,\mathcal{F},\mathbb{P})$ where every random variable defined in $\Omega$ is $\mathcal{F}$-measurable. For a discrete random variable $\mathcal{X}(\omega):\Omega\rightarrow \Gamma$, a probability mass function $p:\Gamma \rightarrow \R_{+}$ implies that  for all $X\in \Gamma$, 
%$$\mathrm{Pr}(\mathcal{X}(\omega)=X)=p(X),$$
%where $\mathrm{Pr}(\text{A})$ is the probability of an event $A$ (see \cite{durrett2013probability}). Also, $\mathbb{E}\{.\}$ represents the expected value operator . Let $\mathcal{W}$ be a discrete random variable $\mathcal{W}(\omega):\Omega \rightarrow \phi$
%with a finite range $\phi_m:=\{\mathcal{W}_1,\mathcal{W}_2,\dots,\mathcal{W}_m\}=\mathcal{W}(\Omega)$, for $m \geq 1$
%and a probability mass function $\pi:\phi\rightarrow \R_{+}$. For the input noise, we define the random variable
%$\mathcal{K}:\Omega \rightarrow \R^n,$ 
%whose components are  Gaussian and i.i.d. with zero mean and variance $\sigma^2=1$.

 \section{Problem Statement}\label{sec:problem}
%In this section, after introducing the underlying model for the dynamical networks that we study in this paper, we define the performance and input measures. Then we state the objectives of this paper with regards to these metrics. 

%\subsection{Model and Measure Description}

We consider an interconnected network of $N$ subsystems where the dynamics of the $i$'th subsystem is governed by  %\cite{li2011h}
 \begin{align}\label{eq:subsystem}
S_i: \left \{ \begin{array} {l}
 \dot x_i(t)\,=\, \mathbf{A}\,x_i(t)+{\mathbf{B}}\, u_i(t)+{\mathbf{E}}\,  \xi_i(t)%\\
\\  y_i(t)\, =\, {\mathbf{H}}\, x_i(t)+ \sigma\, \eta_i(t) \\
 z_i(t)\, =\, \mathbf{C}\, x_i(t) ,
\end{array}   \right . ,
%&y_i={\mathbf{C}} x_i, \notag
\end{align}
%where the transfer-function from the input to the output is
%\begin{align}
%\mathcal{F}(s)=C(s\mathbf{I}_n-A)^{-1}{\mathbf{B}}.
%\end{align}
for $i=1,\dots,N$, in which $x_i(t)\in \R^n$ is the state vector of the subsystem, $u_i(t)\in \R^p$ is the control input, $\xi_i(t) \in \R^{m_1}$ is the exogenous {disturbance} input, $\eta_i(t) \in \R^{m_3}$ is the measurement  noise,  $y_i(t)\in \R^{q}$ is the  measurable output, and  $z_i(t) \in \R^{m_2}$ is the performance  output. Parameter $\sigma \geq 0$ dictates the magnitude of the  measurement noise. 
 %We form a network of $N$ subsystem with identical dynamics (\ref{eq:subsystem}). Then, the network 
% We pick $N$ identical instances of (\ref{eq:subsystem}) and this gives us a heterogeneous multi-agent system, which 
%We form a network that consists of $N$ identical subsystems. Then, the network state vector  is given by  %; that are %$and the consensus output
%$y_c$ as follows. 
The state of the entire network is 
\begin{align*}
x(t):=\left [x_1(t)^T,x_2(t)^T,\dots,x_N(t)^T\right ]^T \in \R^{Nn}. 
\end{align*} 
The vectors representing the network input, disturbance, feedback noise, feedback output, and controlled output  are similarly defined and denoted by $u$, $\xi$, $\eta$, $y$, and $z$, respectively. 

The control objective for the network is to achieve synchronization (or consensus), i.e.,  $x_i(t)-x_j(t)\rightarrow 0$ as $t\rightarrow \infty$ for all $i,j \in \{1,\dots,N\}$. To realize this goal, we employ the following feedback control law 
\begin{align}\label{eq:laplacian_control_c}
u_i(t)=-\sum_{j\in \mathcal{N}_i}\mathbf{K}_{ij} \big(y_i(t)-y_j(t) \big)
\end{align}
for each subsystem $i \in \{1,\dots,N\}$. The subsystems are allowed to exchange  their relative output measurements information over an undirected communication graph $\mathcal{G}$. It is assumed that the structure of the feedback gain matrices $\mathbf{K}_{ij} \in \R^{p \times q}$ are restricted to  $\mathbf{K}_{ij}=k_{ij} \mathbf{K} $ where $k_{ij}$'s are nonnegative scalars {(i.e., the weights of graph $\mathcal{G}$)} and $\mathbf{K} $ is the common factor among all feedback gain matrices. 

When stabilizing feedback control law \eqref{eq:laplacian_control_c} exists and there is no disturbance and noise, one can show that $x_i(t)-x_j(t)\rightarrow 0$ as $t\rightarrow \infty$ holds for the closed-loop network. However, in the presence of disturbance or noise, the state variables will fluctuate around the consensus state. To quantify these fluctuations, we look at the deviations from the average  of the output states subsystems, which are  given by  
\begin{align}\label{eq:11}
\nu_i(t): =z_i(t)-\dfrac{1}{N}\sum_{j=1}^N z_i(t)  
\end{align}
for every $i\in \mathcal{V}$. We can represent \eqref{eq:11} in vector form as
\begin{align} \label{eq:outputdefine}
\nu(t)=({\mathbf{M}_N} \otimes \mathbf{I}_{m_1})\,z(t)=( {\mathbf{M}_N} \otimes \mathbf{C}) \, x(t),
\end{align}
where ${\mathbf{M}_N}$ is the centering matrix {of size $N$}.
%Let us denote the transfer matrix from $[\xi^T,\eta^T]^T$ to this performance output $\nu$ by ${\mathbf{G}}(s) \in \R^{N(m_2+m_3) \times Nm_1}$.
The network \eqref{eq:subsystem} and \eqref{eq:laplacian_control_c} asymptotically reaches consensus if and only if $\nu(t)$ asymptotically goes to zero.  Since   \eqref{eq:laplacian_control_c} can be rewritten as 
\begin{align}
u(t)=-({\mathbf{L}\, \otimes\,} {\mathbf{K}}{\mathbf{H}})\, x(t),
\end{align}  the controller synthesis breaks into two components:  designing a feedback gain $\mathbf{K}$ and designing a weighted undirected graph with Laplacian  $\mathbf{L}$.  
%Fig. \ref{fig:bd}  illustrates the essence of this control 
%architecture. 
It is assumed that measurement noise and noise input are both Gaussian, uncorrelated, and with independent components with unit variance. In order to   measure the aggregate fluctuations in the network,  we adopt the steady-state variance of the deviation from the average  as a  measure of performance for the design, which is defined by 
\begin{align}
 \rho(\mathbf{L},\mathbf{K}):= \, \lim_{t\rightarrow \infty}\mathbb{E}\left \{ \|\nu(t) \|_2^2 \right \}. \label{perf-meas}
 \end{align}

The \textit{research problems} are to characterize  performance measure \eqref{perf-meas} in terms of Laplacian eigenvalues of the underlying communication graph of the network, illustrate  role of feedback (and observer) gains in stability and emergence of fundamental limits on performance and design tradeoffs, and derive scaling laws {for the performance} as  the network   grows. 
 
 \section{Stability and Performance Measure Characterization} \label{sec:stab_perf}
 We look at the stability criteria for these dynamical networks.  Moreover, we derive and characterize spectral expressions for the performance measure. For brevity, we remove the time argument from the variables.
 
%  First, we need to pose an assumption. 
%
%\begin{assumption}\label{eq:ofs} The system ${S}_i$ with triplet $(\mathbf{A},{\mathbf{B}},{\mathbf{H}})$ is output-feedback stabilizable.   
%\end{assumption} 

%This assumption is standard since it is also required for an output feedback design of a single subsystem in the first place.
 %Because we use only relative-feedback, the subsystems must be stable on their own, as indicted in the following assumption. 

%\begin{assumption} \label{eq:assumption} The linear system $\dot x=\mathbf{A}x$ is stable.  
%\end{assumption} 

Once we apply feedback control protocol (\ref{eq:laplacian_control_c}), the   closed-loop dynamics of the
network  are given by  
 \begin{align}\label{eq:CL}
& \dot x=
(\mathbf{I}_N \otimes \mathbf{A} -
 {\mathbf{L}\,\otimes\,} {\mathbf{B}}  {\mathbf{K}}{\mathbf{H}})\,x+
 (\mathbf{I}_n \otimes {\mathbf{E}} )\,\xi
 -(\mathbf{L} \, \otimes \sigma \mathbf{I}_{m_3}  ) \eta.%\\
%& y=(\mathbf{I}_N \otimes \mathbf{C})x. \notag
 \end{align}
We define the auxiliary  variables $r$, $\chi$, and $\gamma$ to be %By means of a suitable change of variable 
 \begin{align}\label{eq:ch_var}
 r:=(\mathbf{U}^T\otimes \mathbf{I}_n)\,x,~\chi:=(\mathbf{U}^T\otimes \mathbf{I}_{m_1})\, \xi,~\gamma:=(\mathbf{U}^T\otimes \mathbf{I}_{m_3}) \eta. %~y=(\mathbf{U} \otimes \mathbf{I}_{m_2})\hat y,
 \end{align}
 Then, the following dynamical decoupling is realized (see \cite{li2011h} for the case of state-feedback {without the measurement noise}).
 %decouples the dynamics into
%$N$ subsystems of the same dimension of dimension $n$. 
 \begin{proposition} \label{prop:change}
By the change of variables (\ref{eq:ch_var}), the resulting closed-loop network dynamics given by (\ref{eq:CL})
are decoupled into $N$  systems
\begin{align}\label{eq:decoupled}
\Sigma_i:~ \left . \begin{array}{ll}
\dot r_i\,=\,(\mathbf{A}-\lambda_i\,{\mathbf{B}} {\mathbf{K}}{\mathbf{H}})\, r_i+{\mathbf{E}} \,\chi_i -\lambda_i {\mathbf{B}}  \mathbf{K} \sigma \gamma_i , \\
 %{\hat y}_i=C\hat x_i,
\end{array} \right.
\end{align}
for each $i=1,2,\dots,N$. In the absence of disturbance and noise, the network reaches consensus if and only if systems $\Sigma_2,\dots,\Sigma_N$ are asymptotically stable. 
\end{proposition}

 %% some part to save space 
% Equivalently, one may write (\ref{eq:tf_trans}) as
 %\begin{align}\label{eq:tf_trans_2}
%{\mathbf{H}}=(V\otimes \mathbf{I}_{m_1})\mathrm{diag}(\hat {\mathbf{H}}_1,\hat {\mathbf{H}}_2,\dots ,\hat {\mathbf{H}}_N)(V^T\otimes \mathbf{I}_{m_2}).
%\end{align}

We leverage this decoupling to arrive at spectral expressions for the   performance measure of the network. %The following result is implicitly used throughout \cite{li2011h}. 

\begin{theorem} \label{thm:first_main} Suppose that in \eqref{eq:decoupled}   systems $\Sigma_2,\dots,\Sigma_N$ are asymptotically stable. Then,  the performance measure  can be expressed as %for the networks over loopy graphs 
\begin{align}\label{eq:spectral_form_gen}
\rho({\mathbf{L},\mathbf{K}})=\sum_{i=2}^N \phi(\lambda_i,\mathbf{K}),
\end{align}
with  the performance function $\phi(\lambda,\mathbf{K})$ given by 
\begin{align}\label{eq:fidef}
\phi(\lambda,\mathbf{K}):=\mathrm{Tr}\left (\mathbf{C}{\mathbf{P}}(\lambda,\mathbf{K})\mathbf{C}^T\right ),\end{align}
 which is a rational function of  $\lambda$ and entries of $\mathbf{K}$. The map  ${\mathbf{P}}(\lambda,\mathbf{K})$ is the unique positive-definite solution to an algebraic Lyapunov equation given by  
\begin{align}\label{eq:sub_lyap}
& (\mathbf{A}-\lambda {\mathbf{B}}\mathbf{K}\mathbf{H}) {\mathbf{P}}(\lambda,\mathbf{K})+{\mathbf{P}}(\lambda,\mathbf{K})(\mathbf{A}-\lambda {\mathbf{B} {\mathbf{K}} {\mathbf{H}} })^T\\ &~~~~~~~~~~~~~~~~~~~~~~~~~~~~+{\mathbf{E}} {\mathbf{E}} ^T+\lambda^2 \sigma^2 {\mathbf{B}}{\mathbf{K}}  ({\mathbf{B}}  {\mathbf{K}})^T=\mathbf 0, \notag
\end{align}
  for all values of $\lambda$ that make $\mathbf{A}-\lambda {\mathbf{B}}\mathbf{K}\mathbf{H}$ a Hurwitz matrix. 
\end{theorem}

The dimension 
of the dynamics of each subsystem is often small and has nothing to do with the number of subsystems.  %, in contrast to the number of agents that may be large in a network.
Therefore,    evaluation of performance function $\phi(\lambda,\mathbf{K})$ %,% which %only requires matrix operations, could be carried out by standard symbolic software. In fact, we
can be done via symbolically solving  Lyapunov equation (\ref{eq:sub_lyap}) after converting it to a linear system by vectorization (see the proof of Theorem \ref{thm:first_main}).  

{Due to linearity of the Lyapunov equation,} one inspects that the performance function $\phi$ can be decomposed into two components  {according to } 
$$
\phi(\lambda,\mathbf{K})=\phi_\xi(\lambda,\mathbf{K})+\sigma^2 \phi_\eta(\lambda,\mathbf{K}),
$$
in which spectral functions $\phi_\xi$ and $\phi_\eta$ only reflect  the effect of disturbance and {measurement} noise, respectively. 

\begin{remark} In this paper, we occasionally skip   argument $\mathbf{K}$ in $\phi(\lambda,\mathbf{K})$ and denote it as $\phi(\lambda)$. In those cases, we solely consider the dependence of the functions on the eigenvalues of the graph Laplacian (i.e., for a fixed feedback gain $\mathbf{K}$). 
\end{remark}

\begin{remark} A part of the result of Theorem \ref{thm:first_main} is hidden in the analysis provided in \cite{li2011h}, in the case of state-feedback. However, the authors did not explicitly derive the spectral expressions for the performance. %measure. 
\end{remark}

We extend the previous analysis to the output-feedback and   synthesize a decentralized observer. We show that the separation principal in the linear filtering using Luenberger observers is naturally carried into this design as well. Our  procedure consists of four steps: 

\noindent{\it (i)} We augment the dynamics of subsystem $i$ by an observer variable $\hat x_i \in \R^n$, whose dynamics are governed by% that follows the equation 
\begin{align}
\dot {\hat   x}_i =\mathbf{A} \hat x_i +{\mathbf{B}} u_i+ \hat u_i, 
\end{align}
where $\hat u_i\in \R^n$ is an  auxiliary control input for the observer. We will set the value of this input in a decentralized manner in the last step. 

\noindent{\it (ii)}  As it is usual in the observer design, we use $\hat x_i$ to compute% the control input as  
\begin{align} \label{eq:u_i_observer}
u_i=-\mathbf{K} \hat x_i. 
\end{align}
\noindent{\it (iii)}  In addition to the relative output feedback on $\mathbf H x_i$, the subsystems should {share} the value of   $\mathbf H \hat x_i$ with their neighbors. Once we consider these three steps, the augmented dynamics of subsystem $i$ are  
\begin{align}\label{eq:subsystem_dynamic_observable}
\begin{adjustbox}{max width=225pt}$
\hat S_i: \left \{ \begin{array} {l}
\begin{bmatrix}\dot x_i \\ \dot {\hat x}_i \end{bmatrix} =\begin{bmatrix}\mathbf{A}  & -{\mathbf{BK}} \\ \mathbf 0 & \mathbf{A}-{\mathbf{BK}} \end{bmatrix}
\begin{bmatrix} x_i \\ \hat x_i \end{bmatrix}+\begin{bmatrix} \mathbf 0 \\ \mathbf{I}_n \end{bmatrix}  \hat u_i 
+\begin{bmatrix} {\mathbf{E}} \\ \mathbf 0 \end{bmatrix}  \xi_i 
\vspace{1mm} \\ 
\hat y_i= \begin{bmatrix}\mathbf{H} & \mathbf{0} \\ \mathbf{0} & \mathbf{H} \end{bmatrix} \begin{bmatrix}x_i \\ \hat x_i \end{bmatrix}+\begin{bmatrix}{\mathbf{I}_{m_3}} \\ \mathbf 0 \end{bmatrix}  \eta_i  \vspace{1mm} \\ 
z_i=\begin{bmatrix} \mathbf{C}& \mathbf{0} \end{bmatrix}\begin{bmatrix}x_i \\ \hat x_i \end{bmatrix}  \\
\end{array}   \right . ,
$
\end{adjustbox}
%&y_i={\mathbf{C}} x_i. \notag
\end{align}
Variable $\hat y_i$ has the same role as $y_i$ in \eqref{eq:subsystem}; i.e., the augmented subsystems will use the relative-feedback on this variable. \\
\noindent{\it (iv)}  We use the following theorem and  design the  gain of control law (\ref{eq:laplacian_control_c}) when applied on  subsystems $\hat S_i$ in (\ref{eq:subsystem_dynamic_observable}), which in this case will be an observer gain. 

\begin{theorem} \label{lem:obs} Suppose that we apply control law (\ref{eq:laplacian_control_c}) on   augmented subsystems 
	$\hat S_i$  in (\ref{eq:subsystem_dynamic_observable}) by setting% and set
	\begin{align}
	\hat u_i=-\hat {\mathbf{F}} \sum_{j \in \mathcal{N}_i} a_{ij} (\hat y_i-\hat y_j) ,
	\end{align}
	where the observer gain is set to be 
	\begin{align}
	\hat {\mathbf{F}}= \big [-\mathbf{F},~~ \mathbf{F}\,\,\,  \big ] \in \R^{n \times (2q)}.
	\end{align}
	Moreover, assume that  $\mathbf{F} \in  \R^{n \times q}$ is chosen such that   $\mathbf{A}-\lambda_i \mathbf{F} \mathbf{H}$   is Hurwitz for $i=2,\dots,N$.  Then, the estimation and regulation are separated: if we apply control input $u_i$ given in \eqref{eq:u_i_observer} for any $\mathbf{K}$ that makes $\mathbf{A}-{\mathbf{BK}}$ a Hurwitz matrix, then the network with this observer-based relative output-feedback reaches the consensus in the absence of disturbance and noise. 
\end{theorem}

For this design, we denote the performance function by $\phi(\lambda, \mathbf K,\mathbf F)$. This function can be found   similar to the case of simple state-feedback, except  that we need the augmented matrices of  $\hat S_i$ given in (\ref{eq:subsystem_dynamic_observable}) for solving \eqref{eq:sub_lyap} and  evaluation of this function.   

The separation principal together with the duality between the estimation and regulation let us prove similar results for the quality of estimation using this decentralized observer. First, we define the error of estimation as
\begin{align}
e(t):=\hat x(t)-x(t). 
\end{align}
Because we are only employing the relative feedback, we may only control the deviations of the error components from their average. These deviations  are reflected by the variable  
\begin{align}
\delta(t):=(\mathbf M_N \otimes \mathbf I_n)\, e(t). 
\end{align}
Next, we define the estimation measure for network as 
\begin{align}
\mu(\mathbf{L},\mathbf{F}):=\lim_{t \rightarrow \infty} \mathbb{E} \left\{ \|\delta(t)\|_2^2\right\}.
\end{align}
The dual of system $\Sigma_i$ in \eqref{eq:decoupled} is 
\begin{align}\label{eq:decoupled_obs}
\Upsilon_i:~~ \left . \begin{array}{ll}
\dot r_i\,=\,(\mathbf{A}-\lambda_i\,{\mathbf{F}} \mathbf {H})\, r_i+{\mathbf{E}} \,\chi_i -\lambda_i \sigma \mathbf{F}  \gamma_i  \\
%{\hat y}_i=C\hat x_i,
\end{array} \right. ,
\end{align}
which lets us deduce the next result (compare to Theorem \ref{thm:first_main}). 

\begin{theorem} \label{thm:estimation} Suppose that in \eqref{eq:decoupled_obs}   systems $\Upsilon_2,\dots,\Upsilon_N$ are asymptotically stable. Then, we can express the estimation measure as % the estimation measure as %for the networks over loopy graphs 
	\begin{align}\label{eq:spectral_form_gen_obs}
	\mu({\mathbf{L},\mathbf{F}})=\sum_{i=2}^N \psi(\lambda_i,\mathbf{K}),
	\end{align}
	with  the estimation function $\psi(\lambda,\mathbf{K})$ given by 
	\begin{align}
	\psi(\lambda,\mathbf{K}):=\mathrm{Tr}\left ({\mathbf{Q}}(\lambda,\mathbf{K})\right ),\end{align}
	which is a rational function of  $\lambda$ and entries of $\mathbf{F}$. The map  ${\mathbf{Q}}(\lambda,\mathbf{K})$ is the unique positive-definite solution to an algebraic Lyapunov equation given by  
	\begin{align}\label{eq:sub_lyap_obs}
	& (\mathbf{A}-\lambda {\mathbf{F}}\mathbf{H}) {\mathbf{Q}}(\lambda,\mathbf{K})+{\mathbf{Q}}(\lambda,\mathbf{K})(\mathbf{A}-\lambda {\mathbf{F}}\mathbf{H} )^T\\ &~~~~~~~~~~~~~~~~~~~~~~~~~~~~~~+{\mathbf{E}} {\mathbf{E}} ^T+\lambda^2 \sigma^2 {\mathbf{F}}^T {\mathbf{F}}=0, \notag
	\end{align}
	  for all values of $\lambda$ that make $\mathbf{A}-\lambda {\mathbf{F}}\mathbf{H}$ a Hurwitz matrix. 
\end{theorem}

\begin{remark}
	In \cite{li2011consensus}, the authors propose the following observer-based approach. They define their observer variable $v_i$ to follow the dynamics 
	\begin{align}
	\dot v_i=\mathbf Fv_i +\mathbf G y_i +\mathbf T\mathbf B u_i,
	\end{align}
	and define their control law as 
	\begin{align}
	u_i=\mathbf K\mathbf Q_1 \sum_{i=1}^{N}a_{ij}(y_i-y_j)+\mathbf K\mathbf Q_2 \sum_{i=1}^{N}a_{ij}(v_i-v_j).
	\end{align}
	where matrix $\mathbf F$ has no eigenvalue in common with $\mathbf A$, the pair $(\mathbf F,\mathbf G)$ is stabilizable, and $\mathbf T$ is the unique solution to Sylvester equation $\mathbf T\mathbf A-\mathbf F\mathbf T=\mathbf G\mathbf C$. Then, they design $\mathbf K,\mathbf Q_1$ and $\mathbf Q_2$ such that a design with minimum connectivity threshold is achieved. One can see that our design is different and  simpler as we only need a feedback gain $\mathbf K$ and an observer gain $\mathbf F$. Moreover, our approach is built upon the separation principle  between the regulation and estimation, which is also the case in the classical Luenberger (or LQG) observer design. 
\end{remark}

\section{ Design of Control Law Gains}  \label{sec:K}

We investigate the problem of finding {feedback}  gains  and focus on gains inducing a minimum connectivity threshold. This property makes the design process with respect to the  graph more tractable. 
After that, we discuss related performance limitations. %   are discussed.  

\subsection{ Minimum Connectivity Threshold} 
%We may design $\mathbf{K}$ such 
%that by increasing the network connectivity beyond some limit, the network remains stable. To this end,
We define the minimum connectivity threshold $\tilde \lambda(\mathbf{K}) \in [0,\infty]$ for a feedback gain $\mathbf{K}$ to be  
\begin{align}\label{eq:tildelambda}
\tilde \lambda(\mathbf{K}):=\inf_{\lambda>0}\big \{\lambda: (\mathbf{A}-c\, {\mathbf{B}} {\mathbf{K}}{\mathbf{H}}) \mbox{ is Hurwitz for } c >\lambda \big \}.
\end{align}
Similar notions have been reported (e.g. \cite{li2011h}), while our goal is characterization of conditions for finding gains with $\tilde \lambda(\mathbf{K})<\infty$\footnote{$\tilde \lambda(\mathbf{K})=\infty$ corresponds to finding the infimum of the empty set in \eqref{eq:tildelambda}.}. The following definition is for this purpose. 

\begin{definition} \label{eq:def_unbounded}
The feedback gain $\mathbf{K}$ is said to have an unbounded stability region if $\tilde \lambda(\mathbf{K})\in [0,\infty)$. 
\end{definition}

If  $\mathbf{K}$ has an unbounded stability region, then the network is {robust} to all increases in the connectivity: if the network is output-stable for a given graph $\mathcal{G}_1$ with Laplacian $\mathbf{L}_1$, then for every graph $\mathcal{G}_2$ with Laplacian $\mathbf{L}_2$ and $\mathcal{G}_1 \subset \mathcal{G}_2$, the network is still output-stable. The reason is that   $\lambda_i(\mathbf{L}_1) \leq \lambda_i(\mathbf{L}_2)$ for $i=2,\dots,N$ (this has been emphasized in \cite{li2011h} as well). Moreover, this makes the stability analysis with respect to the graph  more {tractable}, since ensuring $\lambda_2(\mathbf{L}) > \tilde \lambda (\mathbf{K})$ guarantees the output-stability of network. {Before bringing methods to find such feedback gains, let us look at a consequence of choosing them.}
% feedback gain that has an unbounded  stability region.% gives the following useful smoothness property to $\phi(\lambda)$.%, that would be useful for 
%consequent analysis. 

\begin{theorem}\label{eq:lip_const}
For a network designed with a feedback gain $\mathbf{K}$ that in endowed by  a connectivity threshold $\tilde \lambda(\mathbf{K})<\infty$, %derived from Proposition \ref{pr:\mathbf{K}_selection}. Then,
 the performance function $\phi(\lambda)$ is analytic on   interval $(\tilde \lambda(\mathbf{K}),\infty)$.  %Moreover, if $\tilde \lambda(\mathbf{K})>0$, then $\Sigma_2$ defined by \eqref{eq:decoupled} is marginally stable for $\lambda_2=\tilde \lambda(\mathbf{K})$.
\end{theorem}

The  openness of the interval of interest in Theorem \ref{eq:lip_const} suggests  that if $\tilde \lambda (\mathbf{K})>0$, we need to maintain a minimum distance  from this value. This will  make sure that the stability margin is large enough.  

\subsection{State-Feedback  Minimum Connectivity Design }

 Let us consider the state-feedback (i.e., $\mathbf{H}=\mathbf{I}_n$ in \eqref{eq:subsystem}).  It turns out that the stabilizability is the necessary and sufficient condition for existence a gain $\mathbf{K}$ that induces a bounded threshold $\tilde \lambda(\mathbf{K})$.  %is the necessary and sufficient condition for the existence of a $\mathbf{K}$ with an unbounded stability region. 
 
 \begin{theorem} \label{lem:kdesign} If $(\mathbf{A},{\mathbf{B}} )$ is stabilizable, then for every value of $c>0$, the choice of feedback gain given by 
\begin{align} \label{eq:K_choice}
\mathbf{K}=\dfrac{1}{2} {\mathbf{B}}  ^T \mathbf{Q}^{-1},
\end{align}
satisfies $\tilde \lambda (\mathbf{K}) \in [0,c]$, where $\mathbf{Q} \succ 0$ is a  solution to the following feasible linear matrix inequality. 
  \begin{align}\label{eq:K_LMI}
&\mathbf{A}\mathbf{Q}+\mathbf{Q}\mathbf{A}^T-c\, {\mathbf{B}} {\mathbf{B}} ^T \prec 0.
\end{align}
Conversely, if there exists a  gain $\mathbf{K}$ with $\tilde \lambda(\mathbf K)<\infty$, then $(\mathbf{A},{\mathbf{B}} )$ is stabilizable. 
  \end{theorem}

%In simple cases, we may look at \emph{parametric} computation of the minimum connectivity and required criteria for choice of $\mathbf{K}$. On the other hand, 

The linear matrix  inequality (LMI)  (\ref{eq:K_LMI}) is a {computational} tool to find {a} gain $\mathbf{K}$ for a given network and graph with a minimum connectivity threshold at most equal to $c$ (see Example \ref{ex:aircraft}). 
 The solvability of   LMI (\ref{eq:K_LMI}) is  called the {quadratic stabilizability} of $(\mathbf{A},{\mathbf{B}} )$ by means of  a linear state-feedback   (see Section 7.2 of \cite{boyd1994linear}).

 \begin{remark} This result is inspired by Theorem 11 in \cite{li2011h}, while our main contribution is in pointing out the role of stabilizability in existence of feedback gains with minimum connectivity thresholds. 
 \end{remark}
 \begin{remark} The optimal choice of $\mathbf{Q}$ is not the concern in Theorem \ref{lem:kdesign}. Instead, we focus on the existence of designs for $\mathbf{K}$ with a minimum connectivity design. In fact, various performance criteria could potentially get addressed.  
For instance, suppose that for some $d>0$, we replace  LMI (\ref{eq:K_LMI}) with
  \begin{align}\label{eq:K_LMI_2}
&\mathbf{A}\mathbf{Q}+\mathbf{Q}\mathbf{A}^T-c {\mathbf{B}} {\mathbf{B}} ^T +2d \mathbf{Q}\prec \mathbf{0}.
\end{align}
Then, for $\mathbf{K}$ computed from \eqref{eq:K_choice} using any solution  to this inequality $\mathbf{Q}\succ \mathbf 0$, not only $\tilde{\lambda}(\mathbf{K})\leq c$, but also  for each eigenvalue $\lambda>\tilde \lambda(\mathbf{K})$, the poles of $\mathbf{A}-\lambda {\mathbf{BK}}$  have real parts less than $-d$ (see \cite{chilali1999robust}).
As another example, authors of \cite{li2011h} brought a version of the matrix inequality which ensures that each decoupled subsystem $ \Sigma_i$ has $\mathcal{H}_2$-norm less than a desired value, which they state that  could be conservative in practice.  Criteria such as   robustness or non-fragility could be potentially added by building on top of  \eqref{eq:K_LMI} as well (e.g. see \cite{famularo2000robust}). 
  \end{remark}
 
\subsection{Observer-Based  Minimum Connectivity Design for Output-Feedback} \label{subsection:observer_design}

 The duality between the derived conditions on  $\mathbf{A}-\lambda_i \mathbf{F} \mathbf{H}$ in Theorem \ref{lem:obs} and  on $\mathbf{A}-\lambda_i {\mathbf{B}}{\mathbf{K}}$ in Theorem \ref{lem:kdesign} lets us conclude the following result that resembles the result of Theorem \ref{lem:kdesign}. 

 \begin{theorem}   \label{lem:kdesign_observable}
  Suppose that $(\mathbf{A},\mathbf{H})$ is detectable. Then, for every $c>0$,  the following observer gain for the settings of Theorem \ref{lem:obs}, has an unbounded stability region with $\tilde \lambda \left (\mathbf{F}\right ) \in \left [0,c\right ]$.
  \begin{align}
\hat {\mathbf{F}}=\left [-\dfrac{1}{2} \mathbf{Q}^{-1}{\mathbf{H}}^T,~ \dfrac{1}{2} \mathbf{Q}^{-1}{\mathbf{H}}^T \right ] \in \R^{n \times (2q)},
\end{align}
  where $\mathbf Q\succ 0$ is a solution to the following feasible LMI.
   \begin{align}\label{eq:K_LMI_obs}
&\mathbf{A}^T\mathbf{Q}+\mathbf{Q}\mathbf{A}-c\, \mathbf{H}^T\mathbf{H}  \prec \mathbf{0}.
\end{align}
Conversely, if under the settings of Theorem 4 an observer gain $\mathbf F$ has a bounded $\tilde \lambda(\mathbf F)$, then $(\mathbf{A},\mathbf{H})$ is detectable. 
 \end{theorem}
 
 The LMI (\ref{eq:K_LMI_obs}) is the quadratic stabilizability condition for the dual pair $(\mathbf{A}^T,{\mathbf{H}}^T)$\footnote{It is stabilizable since pair $(\mathbf{A},\mathbf{H})$ is detectable}. % and value of $\mathbf{F}$ when solving for the performance function from the corresponding Lyapunov equation.

\subsection{Asymptotic Performance and Estimation Bounds }
 
 An important design question is if the performance function $\phi(\lambda,\mathbf{K})$ can be made arbitrarily small, which is related to the notion of {almost disturbance decoupling} \cite{weiland1989almost}:  attenuating the effect of the disturbance in a performance metric as much as desired. %First,
 We study the case of relative state-feedback below. %, if the subsystems are right-invertible and minimum-phase.
  
  \begin{theorem} \label{thm:right_mp} Suppose that $(\mathbf{A},\mathbf{B})$ is stabilizable and  $(\mathbf{A},\mathbf{C})$ is detectable and that $\sigma=0$. %Moreover, under the relative state-feedback, %, the following statements hold. % we can make the performance function arbitrary small according to 
%\noindent {\it (i)} there exist a feedback gain $K_0$
%\begin{align}\label{eq:k_zero}
%\mathbf{K}_0:=\lim_{\epsilon \rightarrow 0} {\mathbf{K}_\epsilon}/{\|\mathbf{K}_\epsilon\|_{2}},
%\end{align} 
% with $\{\mathbf{K}_\epsilon\}_{\epsilon \geq 0}$ a family of controllers with elements 
%\begin{align}
%\mathbf{K}_\epsilon:=\epsilon^{-2} {\mathbf{B}}^T {\mathbf{P}}_\epsilon \text{~~ for } \epsilon>0,
%\end{align} and 
% \noindent {\it (ii)} the feedback gain $\mathbf K_0$ has an unbounded stability region. 
% \begin{align}
% \lim_{\lambda \rightarrow \infty}\phi(\lambda,\mathbf{K}_0)=0, %\min_{K \text{ has } \tilde \lambda (\mathbf{K}) } \lim_{\lambda \rightarrow \infty} \phi(\lambda,\mathbf{K})= 0.
% \end{align}
% \noindent {\it (iii)} 
 For all pairs of $\lambda>0$ and $\mathbf{K}$ for which $\mathbf{A}-\lambda {\mathbf{BK}}$ is Hurwitz,  the performance function resulting from  the relative state-feedback
 is bounded from below according to 
 \begin{align} \label{eq:phi_asymptotic}
 \phi(\lambda,\mathbf{K}) > \mathrm{Tr}\Big ({\mathbf{E}} ^T \mathbf{P}_0 {\mathbf{E}} \Big ),
 \end{align} 
 for a positive semi-definite matrix $\mathbf{P}_0$ given by
 \begin{align}
 \mathbf{P}_0:=\lim_{\epsilon \rightarrow 0} \mathbf{P}_\epsilon
 \end{align}
  where ${\mathbf{P}}_\epsilon$ is the unique positive semi-definite solution to the parametric algebraic Riccati equation
 \begin{align}
 \mathbf{A}^T {\mathbf{P}}_\epsilon +{\mathbf{P}}_\epsilon \mathbf{A}+\mathbf{C}^T \mathbf{C}-\epsilon^{-2}{\mathbf{P}}_\epsilon \mathbf{B} \mathbf{B}^T {\mathbf{P}}_ \epsilon=\mathbf 0.   
 \end{align}
% 
%\noindent {\it (iv)} the choice of feedback gain $\mathbf{K}_0$ given  in \eqref{eq:k_zero} results in 
% \begin{align}
% \lim_{\lambda \rightarrow \infty }{\phi(\lambda,\mathbf{K}_0)} = \mathrm{Tr}\Big ({\mathbf{E}} ^T \mathbf{P}_0 {\mathbf{E}} \Big ).
% \end{align}
 Matrix $\mathbf P_0$ is zero if and only if  transfer matrix $\mathbf C(s\mathbf I_n-\mathbf A)^{-1} \mathbf B$ is right-invertible and minimum-phase.  
  \end{theorem} 

%We mention few remarks about these results:

%\noindent {\it (i)} The formula for $\mathbf{K}_0$ given in  (\ref{eq:k_zero}) provides us with a {direction} in the space of feedback gains, along which the performance function $\phi(\lambda,\mathbf{K}_0)$ asymptotically obtains its minimal value as $\lambda \rightarrow \infty$. 

% \noindent {\it (ii)} 
 
 {For instance, if the transfer matrix is $\mathbf C(s\mathbf I_n-\mathbf A)^{-1} \mathbf B$ non minimum-phase and} the columns of ${\mathbf{E}}$ are not in the null space of $\mathbf{P}_0$, then the bound in (\ref{eq:phi_asymptotic}) is strictly positive. The dual of this result for estimation quality is given below, whose proof is identical to Theorem \ref{thm:right_mp}. % and has been omitted.  % Moreover, this choice of {direction} for $\mathbf{K}$ results in the {minimal} asymptotic value of the performance function as $\lambda \rightarrow \infty$. 

\begin{theorem} \label{thm:cheapestimation} Suppose that $(\mathbf{A},\mathbf{E})$ is stabilizable and  $(\mathbf{A},\mathbf{H})$ is detectable.  
%	 the following arguments hold.	
%	\noindent {\it (i)} there exist an observer gain
%\begin{align}\label{eq:f_zero}
%\mathbf{F}_0:=\lim_{\epsilon \rightarrow 0} {\mathbf{F}_\sigma}/{\|\mathbf{F}_\sigma\|_{2}},
%\end{align} 
%
%\noindent (ii) $\mathbf F_0$ has an unbounded stability region;
%\noindent (iii)
 If for some gain $\mathbf{F}$, $\mathbf A-\lambda_i \mathbf{F} \mathbf H$ are Hurwitz for $i=2,\dots,N$, then %for uncorrelated Gaussian disturbance and noise processes with covariances equal to identity matrix%, the steady-state variance of the estimation error using the observer of Theorem \ref{lem:kdesign_observable} satisfies 
$$
\psi(\lambda,\mathbf F) > \mathrm{Tr}(\mathbf{S}_0),
$$
for a positive semi-definite matrix $\mathbf{S}_0$ given by
\begin{align}
\mathbf{S}_0:=\lim_{\sigma \rightarrow 0} \mathbf{S}_\sigma,
\end{align}
% with $\{\mathbf{F}_\sigma\}_{\sigma \geq 0}$  given by 
%\begin{align}
%\mathbf{F}_\sigma:=\sigma^{-2}   {\mathbf{S}}_\sigma \mathbf{H}^T \text{~~ for } \sigma>0,
%\end{align} 
%and
 where ${\mathbf{S}}_\sigma$ is the unique positive semi-definite solution to the parametric algebraic Riccati equation
\begin{align}
\mathbf{A} {\mathbf{S}}_\sigma +{\mathbf{S}}_\sigma \mathbf{A}^T+\mathbf{E} \mathbf{E}^T-\sigma^{-2}{\mathbf{S}}_\sigma \mathbf{H}^T \mathbf{H} {\mathbf{S}}_ \sigma=\mathbf 0;   
\end{align}
%\noindent (iii) for choice of observer gain $\mathbf F_0$, it holds that 
%\begin{align*}
%\lim_{ \lambda \rightarrow \infty} \psi(\lambda,\mathbf{F}_0)=   \mathrm{Tr}(\mathbf{S}_0).
%\end{align*}
Matrix $\mathbf{S}_0$ is zero if and only if transfer matrix  $\mathbf H(s\mathbf I_n-\mathbf A)^{-1}\mathbf E$ is right-invertible and minimum-phase. 
\end{theorem} 
  
\subsection{ Parametric Evaluation of $\tilde \lambda (\mathbf{K})$} 
%For the multi-agent systems that we study in this paper,
In both relative state or output feedback designs, if  $n$ is not large (e.g. $n \sim $ 1 to 4), we may design $\mathbf{K}$ with an unbounded 
stability region using Routh-Hurwitz criteria and {explicitly} evaluate $\tilde \lambda(\mathbf{K})$. In fact, the characteristic equation of the matrix $\mathbf{A}-\lambda {\mathbf{B}}{\mathbf{K}}\mathbf{H}$ for the decoupled systems for eigenvalue $\lambda$ is
\begin{align}
p_{\lambda}(s)=p(s;\lambda,\mathbf{K})=\mathrm{det} \left (  s\,\mathbf{I}_n -(\mathbf{A}-\lambda {\mathbf{B}}{\mathbf{K}} {\mathbf{H}}) \right ).
\end{align}
They must be Hurwitz polynomials for $\lambda=\lambda_2,\dots,\lambda_N$. As we enforce the Routh-Hurwitz criteria, we find a set of essentially nonlinear inequalities involving $\lambda$ and elements of $\mathbf{K}$, such that the minimum connectivity threshold is realizable and evaluable based on values of $\mathbf{K}$   (see the next section for examples).

\section{Examples of Performance Analysis}\label{sec:example}

In this section, we bring different classes of subsystems and characterize their performance within this framework. We bring additional details of the examples in Appendix Q. %We also include an analysis of a class of network of networks. 

%\subsection{Examples without Feedback Noise}

 First, we consider two single-input single-output controllable subsystems under the relative state-feedback, where the disturbance and control input drive the  dynamics from the same channel {(without the measurement noise)}. 
\begin{table}
\centering
\resizebox{0.35 \textwidth}{!}{%
\begin{tabular}{|c|c|c|}
\hline
   Realization & $\phi(\lambda,\mathbf{K})$  \\ \hline \hline %& $\mu(\lambda,\mathbf{K})$
  &  \\[-1em]
 $\begin{array}{c} \mathfrak{s}_1: \\ \mathbf{A}=-a, \\ {\mathbf{B}}={\mathbf{E}} =1,~{\mathbf{C}}=1\end{array}$              & $\dfrac{1}{2(k\lambda+a)}$           \\ %     & $\dfrac{k^2\lambda^2}{2(k\lambda+a)}$                \\ 
 &  \\[-1em]
 &  \\[-1em] \hline
   &  \\[-1em] 
$\begin{array}{c} \mathfrak{s}_2: \\ \mathbf{A}=\begin{bmatrix}
0 & 1 \\
-a_2 & -a_1
\end{bmatrix}\\ {\mathbf{B}}={\mathbf{E}} =\begin{bmatrix} 0 & 1 \end{bmatrix}^T\\ {\mathbf{C}}=[b_1~ b_0] \end{array}$              & $\dfrac{b_0^2k_1\lambda + a_2b_0^2 + b_1^2}{2( k_2 \lambda+a_1)(k_1 \lambda+a_2)}$  %   \\            %& $\dfrac{\lambda^2(k_1^2 + \lambda k_1 k_2^2 + a_2 k_2^2)}{2(k_2 \lambda+a_1)( k_1 \lambda+a_2)}$               \\ 
%  &  \\[-1em] \hline 
%   &  \\[-1em]
%   &  \\[-1em]
% $\begin{array} {c} \mathfrak{s}_3: \\ \mathbf{A}=\left [ \begin{array} {ccc} 0 & 1 & 0 \\ 0 & 0 & 1 \\ -a_3 & -a_2 & -a_1 \end{array} \right ]  \\ \\ {\mathbf{B}}={\mathbf{E}} =\begin{bmatrix}0 & 0 & 1\end{bmatrix}^T\\ {\mathbf{C}}=\begin{bmatrix} b_2~ b_1~b_0 \end{bmatrix}    \end{array}   $       & 
%$\begin{matrix}
%\dfrac{b_2^2(k_3 \lambda+a_1)}{2 ( k_1 \lambda+a_3) ((k_2 \lambda+a_2)(k_3 \lambda+a_1)- ( k_1 \lambda+a_3))}  \\
%+\dfrac{ b_1^2-2b_0 b_2}{ 2((k_2 \lambda+a_2)(k_3 \lambda+a_1)- ( k_1 \lambda+a_3))} \\
%+\dfrac{b_0^2( k_2 \lambda+a_2)}{2((k_2 \lambda+a_2)(k_3 \lambda+a_1)- ( k_1 \lambda+a_3))}
%\end{matrix}   $   
\\    %   &    
%
%$\begin{matrix}
%\dfrac{k_1^2\lambda^2(k_3 \lambda+a_1)}{2 ( k_1 \lambda+a_3) ((k_2 \lambda+a_2)(k_3 \lambda+a_1)- ( k_1 \lambda+a_3))} \\
%+ \dfrac{ \lambda^2(k_2^2-2k_2 k_3)}{ 2((k_2 \lambda+a_2)(k_3 \lambda+a_1)- ( k_1 \lambda+a_3))} \\
%+ \dfrac{\lambda ^2 k_3^2(k_2 \lambda+a_2)}{2((k_2 \lambda+a_2)(k_3 \lambda+a_1)- ( k_1 \lambda+a_3))}
%\end{matrix}   $          \\ 
  & \\[-1em]
   & \\ \hline
\end{tabular}}
\caption{The  subsystems investigated in  Example \ref{ex:three} together with the  performance functions in the case of relative state-feedback with $\sigma=0$. We assume that $a,a_1,a_2 \geq 0$.  }
\label{table:someformulas}
\end{table}

\begin{example} \label{ex:three} Consider the subsystems given in Table \ref{table:someformulas}, where we have also reported the corresponding performance functions.    For the nodal dynamics $\mathfrak{s}_1$ and $\mathfrak{s}_2$, supposing that $\mathbf{K}=k>0$ and $\mathbf{K}=[k_1,k_2]\succ 0$, respectively, in both cases $\tilde \lambda (\mathbf{K})=0$. Moreover, for $\lambda > \tilde \lambda (\mathbf{K})$,   performance function $\phi(\lambda)$ is strictly convex and strictly decreasing. If all $a_i$'s are zero and $\mathbf{C}=e_1^T$, these subsystems are called single and double-integrators,  respectively. As a numerical example, let us consider  double-integrators with $k_1=k_2=1$. Then,   using the second row of Table \ref{table:someformulas} we get  
	\begin{align} \label{eq:secondorder_state}
	\phi(\lambda)=\dfrac{1}{2 \lambda^2}.
	\end{align}
	This is a well-known result (e.g. see \cite{patterson2010leader}).

%	For $\mathfrak{s}_3$, we choose $\mathbf{K}=[k_1,k_2,k_3]\succ 0$ and consider the quadratic equation 
%	\begin{align}
%	k_2 k_3 \lambda^2 + (a_1 k_2 +a_2 k_3 -k_1) \lambda +(a_1 a_2- a_3)=0.
%	\end{align}
%	If this equation has positive roots, then $\tilde \lambda (\mathbf{K})$ will be equal to its largest positive root. Otherwise, $\tilde \lambda (\mathbf{K})=0$. For $\mathfrak{s}_3$, if $b_1^2-2b_1  b_0 \geq 0$ this would also hold (see the appendix for the details). 

%	For $\mathfrak{s}_3$, we choose $\mathbf{K}=[k_1,k_2,k_3]\succ 0$ and consider the quadratic equation 
%	\begin{align}
%	k_2 k_3 \lambda^2 + (a_1 k_2 +a_2 k_3 -k_1) \lambda +(a_1 a_2- a_3)=0.
%	\end{align}
%	If this equation has positive roots, then $\tilde \lambda (\mathbf{K})$ will be equal to its largest positive root. Otherwise, $\tilde \lambda (\mathbf{K})=0$.
\end{example}
	
\begin{example} \label{ex:dint_observer} Consider double-integrators with relative feedback only on  positions (${\mathbf{H}}=[1,0]$) using the decentralized observer of  Theorem \ref{lem:kdesign_observable}. We let   $\mathbf{K}=[k_1,k_2] \succ 0$ and set the observer gain to be $\mathbf{F}=[f_1,f_2]^T$.  Theorem \ref{lem:obs} requires the stability analysis for matrix $\mathbf{A}-\lambda \mathbf{F} {\mathbf{H}}$,
which is Hurwitz if and only if $f_1,f_2>0$. Then, we get $\tilde \lambda (\mathbf{F})=0$. We can show that 
\begin{align} \label{eq:seconorderrelativefeedback}
\phi(\lambda,\mathbf{K},\mathbf{F})=\dfrac{c_1 \lambda^4+c_2\lambda^3+c_3\lambda^2+c_4\lambda+c_5}{c_6 \lambda^4+c_7\lambda^3+c_8\lambda^2},
\end{align}
where $c_1$ to $c_8$ are polynomials of $k_1,k_2,f_1,$ and $f_2$.  %as given in the appendix. 
Using the observer with $k_1=k_2=f_1=f_2=1$, \eqref{eq:seconorderrelativefeedback}  becomes  
\begin{align} \label{eq:secondorder_output}
\phi(\lambda)=\dfrac{9\lambda^4 + 11\lambda^3 + 9\lambda^2 + 4\lambda + 1}{6\lambda^4 + 2\lambda^2}.
\end{align}
One observes that for weak connectivity regimes (i.e., $\lambda$ near zero), $\phi(\lambda)$ in \eqref{eq:secondorder_output} is close to the function in \eqref{eq:secondorder_state}, while as $\lambda$ increases, the performance function corresponding to relative state-feedback vanishes, while the function from observer design does not.  
\end{example} 

\begin{example}\label{ex:triple}
	We consider a triple-integrator   with dynamics 
	\begin{align}
	\dddot x_i=u_i+\xi_i.
	\end{align} 
	Let us choose the state to be $[x_i,\dot x_i,\ddot x_i]^T$ with element-wise positive   gain $\mathbf K=[k_1,k_2,k_3] \succ 0$. We can show that 
	\begin{align}
	\phi(\lambda,\mathbf{K})=\dfrac{k_3 }{2 (k_1 k_2 k_3 \lambda^2- k_1^2 \lambda)},~ \tilde \lambda (\mathbf{K})=\dfrac{k_1}{k_2 k_3}. 
	\end{align}
\end{example}

 The next two examples also have performance functions that under conditions become strictly decreasing and convex. 
 
 \begin{example} \label{ex:harmonic} The dynamics of a harmonic oscillator of mass $m$ are governed by 
\begin{align}
\ddot x_i=-2\zeta \omega_0 \dot x_i-\omega_0^2 x_i+\dfrac{u_i}{m}+\dfrac{\xi_i}{m},
\end{align}
 where $\zeta$ is the damping ratio and $\omega_0$ is the undamped angular frequency (see   \cite{gitterman2005noisy}). We consider ${\mathbf{C}}=[1,0]$ and  compute $\phi(\lambda)$ with the relative state-feedback on   $[x,\dot x]^T$ with $\mathbf{K}=[k_1,k_2]\succ 0$. Using arguments similar to Example \ref{ex:three}, if we define $ \alpha_1:={m \omega_0^2}/{k_1 }$ and $\alpha_2:={2 m \omega_0 \zeta}/{k_2} $ we get 
\begin{align}
\phi(\lambda,\mathbf{K})=\dfrac{1}{2k_1 k_2\left (\lambda+\alpha_1 \right ) \left (\lambda +\alpha_2 \right )}.
\end{align}
%Now, let us define $\phi_0:={1}/(2k_1k_2\lambda^2)$, that is the result for $\omega_0=\zeta=0$. Then we can nondimensionalize the performance function using the nondimensional parameters 
%$$
%z_1:=\dfrac{m \omega_0^2}{k_1 \lambda },~z_2:=\dfrac{2 m \omega_0 \zeta}{k_2 \lambda },
%$$
%that results in the nondimensional performance index 
%$$
%\phi_{nd}:=\dfrac{\phi}{\phi_0}=\frac{1}{(1+z_1)(1+z_2)}.
%$$ 
%In fact, $\phi_{nd}$ is the \emph{ratio} of the performance function of a network of second order harmonic oscillators to the performance function of agents with zero damping and elasticity. In Fig. \ref{fig:harmonicplot}, the values of $\phi_{nd}$ have been shown for $[z_1,z_2]\in [0,1.5]^2$.
%
%where we have defined the parameters 
%$$
%}. 
%$$
Again, for element-wise positive feedback gains, $\phi(\lambda)$ is strictly convex and strictly decreasing for $\lambda>\tilde \lambda(\mathbf{K})=0$. %We will revisit this example later too. % next section.  
%Now
%\begin{align*}
%\dfrac{\lambda^2 k_1^2}{2k_1 k_2 ( \lambda+\alpha_1)(\lambda+\alpha_2)}+
%\dfrac{\lambda^2 k_2^2 }{2mk_1 k_2 (\lambda+\alpha_2)}
%\end{align*}
\end{example} 
 
 \begin{example} [Platoon of Vehicles] \label{ex:pltn} We consider a network of vehicles, in which the position of $i$'th vehicle is denoted by $p_i \in \R$. It has the third-order dynamics 
\begin{align}
\tau \dddot p_i+\ddot p_i= u_i+\xi_i,
\end{align} 
where the input $u_i \in \R$ is the desired acceleration and $\xi_i \in \R$ is the disturbance. The time-constant $\tau > 0$ characterizes how fast the vehicles  responds to the acceleration command. The state vector is chosen as $[p_i,\dot p_i,\ddot p_i]^T$, where they denote the (errors in)  the position, velocity, and acceleration of the vehicles in the platoon, respectively (see \cite{zheng2016platooning} for more details).
 The state-space matrices are given in the appendix.
  % this case is 
Using relative state-feedback, by application of the Routh-Hurwitz criteria we find that  if $\mathbf{K}=[k_1,k_2,k_3]$ satisfies $k_1,k_2>0, k_3\geq 0,$ we get
\begin{align}\label{eq:bi}
\tilde \lambda(\mathbf{K})=\left\{
    \begin{array}{ll}
        0 & \mbox{if } k_3=0 \\
        \max\left \{0,\mathlarger{\frac{\tau k_1-k_2}{k_2k_3}}\right\} & \mbox{if } k_3>0
    \end{array}
\right. .
\end{align}
We can show that if  $\sigma=0$, we get  
\begin{align}
\phi(\lambda,\mathbf{K})=\dfrac{1}{2k_1k_2}\dfrac{k_3\lambda + 1}{k_3\lambda^3 + (k_2 - k_1\tau)\lambda^2/k_2}.
\end{align}
 If $k_3=0$, the design corresponds to a relative output-feedback on only positions and velocities, with  a performance function 
$$
\phi(\lambda)=\dfrac { 1}{2k_1({k_2 - k_1\tau})\lambda^2},
$$
which is  strictly convex and strictly decreasing for $\lambda>\tilde \lambda(\mathbf{K})=0$. If $k_3>0$, we have the relative state-feedback and for $\lambda >\tilde \lambda(\mathbf{K})$ the same argument holds (see the appendix).% We investigate its performance scaling over a path graph in Section \ref{sec:bounds}. 
%$\dfrac{1}{2\tau}\dfrac{k_2k_3^2\lambda^2 + (\tau k_2^2 - k_1 k_3 \tau)\lambda + k_1 \tau}{k_2 k_3\lambda  + k_2 - k_1\tau}$

%Still to add details

 \end{example}

%    \begin{figure}
%    \centering
%\includegraphics[width=5.5cm]{images/harmonicplot}
%    \caption{The values of nondimensional performance function $\phi_{nd}$ for a harmonic oscillator in terms of nondimensional parameters $z_1$ and $z_2$ (see Example \ref{ex:harmonic}). }
%    \label{fig:harmonicplot}
%\end{figure}

%We focus on the performance limit of a non minimum-phase system. 
 %to 
%a right-hand plane zero in the subsystem dynamics.% of the subsystems. 

\begin{example} \label{ex:exampleone} Consider a network with nodal matrices 
$$
\mathbf{A}=\begin{bmatrix}
0 & 1 \\
0 & 0 
\end{bmatrix},~{\mathbf{B}}=\begin{bmatrix}
0  \\
1 
\end{bmatrix},~{\mathbf{C}}=\begin{bmatrix}
-\zeta  & 1
\end{bmatrix},
$$ 
for $\zeta>0$ and $\sigma=0$. We observer that the subsystems have a non minimum-phase input-output transfer function  
$$\mathbf C(s\,\mathbf I_2-\mathbf A)^{-1}\mathbf B=(s-\zeta)/{s^2},$$
where $\zeta>0$ is the location of the right-hand plane zero. Let us consider the relative state-feedback. 
We can show that in this case, we have  $
\mathbf{P}_0=\mathrm{diag} \left ( 2\zeta ,0 \right )
$. 
%To derive the zero dynamics, we use the transformation  
%$$
%T=\begin{bmatrix}
% -\zeta  & 1 \\
%  1 & 0
%\end{bmatrix}
%$$
%and get  a realization with variables $[y,z]^T=Tx,$ satisfying  
%\begin{align*}
%&\dot y=-\zeta y-\zeta^2 z+u,\\
%&\dot z=y+\zeta z.
%\end{align*}
%Thus,  $A_0=\zeta$, $B_0=1$, and $\mathbf{P}_0=2\zeta.$ We let $E =[\alpha,\beta]^T$ and get $
%D_0=T_zE =
%% \begin{bmatrix}1 & 0\end{bmatrix} [\alpha,\beta]^T=
%\alpha.
%$
%According to  Lemma \ref{lem:cheap},  we have 
%$$
%\|{\mathbf{G}}(s)\|_{2}^2 > 2 \zeta \alpha^2. 
%$$
For a  disturbance matrix 
$
{\mathbf{E}} =\left [\alpha,\beta\right ]^T
$, Theorem \ref{thm:right_mp} gives us the bound 
\begin{align}\label{eq:nmpbound1}
\phi(\lambda,\mathbf{K}) > \mathrm{Tr}\left ({\mathbf{E}} ^T \mathbf{P}_0 {\mathbf{E}} \right )=2\zeta \alpha^2.
\end{align}
Alternatively, if we use the relative state-feedback, we can show that the corresponding performance function ${\phi}(\lambda,\mathbf{K})$ is % given by 
\begin{align*}& \begin{adjustbox}{max width=250pt} 
$\dfrac{\alpha^2(k_1+k_2\zeta)^2\lambda^2+(2k_1\alpha^2\zeta^2+2k_2\alpha\beta \zeta^2+k_1\beta^2)\lambda+\beta^2 \zeta^2}{2k_1k_2\lambda^2},$
\end{adjustbox}
\end{align*}
which is strictly convex and decreasing for $\lambda >\tilde \lambda (\mathbf{K})=0$. Now, for any gain $\mathbf{K}$ with an unbounded stability region %To compute the left hand side of (\ref{eq:limit_2}), first note that 
\begin{align}
\lim_{\lambda \rightarrow \infty }{\phi}(\lambda,\mathbf{K})=\frac{\alpha^2(k_1+k_2\zeta)^2}{2k_1k_2}=\alpha^2\dfrac{(1+r\zeta)^2}{2r},
\end{align}
where 
$r:={k_2}/{k_1}$. By differentiation with respect to $r$, we find that the right side attains 
%$
%{\phi}(\lambda,\mathbf{K})\rightarrow {\alpha^2(1+r\zeta)^2}/(2r), 
%$
its minimum at $r={1}/{\zeta}$. Thus%, (\ref{eq:limit_2}) gives the same bound %, we have
\begin{align}
{\phi}(\lambda,\mathbf{K})>  \lim_{\lambda \rightarrow \infty} \phi(\lambda,\mathbf{K})|_{k_2/k_1=1/\zeta}=2\zeta \alpha^2 , 
\end{align}
which is the same bound as \eqref{eq:nmpbound1}. The bound on the performance function scales with the magnitude of the right-hand plane zero at $\zeta$.
% Note that we can alternatively find the direction of the optimal $\mathbf{K}$ (i.e., $r=1/\zeta$) using the singular Riccati equation argument in Theorem \ref{thm:cheap}.
 One inspects  that if the disturbance enters the subsystem   from the same channel as the control input, we do not face a fundamental limitation on the performance, because in this case it does not touch the zero dynamics of the subsystems (see \cite{schwartz1996sub} for a similar observation in the case of $\mathcal{H}_\infty$-norm).  
\end{example}

%\subsection{Examples with Feedback Noise}

\begin{example} \label{ex:fnoise}
	In this example,  first, we consider two different designs for a network of double-integrator agents {with measurement noise}. Recall that the magnitude of feedback noises is controlled by parameter $\sigma>0$.
	
	\noindent {\it (i)} the relative state-feedback without the filtering (i.e., without the decentralized observer): in this case, using $\mathbf{K}=[k_1,k_2]$, we can show that
	\begin{align}
	\phi(\lambda,\mathbf K)=\dfrac{{1}}{k_1 k_2 \lambda^2}+\sigma^2 \dfrac{k_1^2+k_2^2}{2k_1 k_2},
	\end{align}
	in which the first term can be recovered from Table \ref{table:someformulas} and the second term  appears due to {the measurement} noises. {\BC} 
	
	%(ii) the relative state-feedback with the decentralized observer; and 
	\noindent {\it (ii)} the relative output-feedback on positions with the decentralized observer: in this case
	\begin{align}
	& \phi(\lambda)=\phi_\xi(\lambda,\mathbf K)+\sigma^2 \dfrac{c_9 \lambda^2 +c_{10} \lambda+c_{11}}{c_{12} \lambda^2 +c_{13} \lambda+c_{14}},%\\& %\sigma^2% \dfrac{(f_1^3k_1^2 + 3f_1^2 f_2 k_1 k_2 + f_1 f_2^2k_1 + f_1f_2^2k_2^2)l^2 + (f_1^2k_1^2k_2 + 3f_1f_2 k_1 k_2^2 + f_2^2k_2^3)\lambda + f_2k_1^2k_2}{(2k_1f_1^3k_2 + 2f_1^2 f_2 k_2^2 + 2 f_1 f_2^2 k_2)l^2 + (2 k_1 f_1^2k_2^2 + 2 f_2 f_1 k_2^3 - 4 f_2 k_1 f_1k_2)l + 2f_1 k_1^2 k_2}
	\end{align}
	in which $\phi_\xi$ is the performance function read from \eqref{eq:seconorderrelativefeedback} and $c_9$ to $c_{14}$ are polynomials of $k_i$'s and $f_i$'s. For instance, in the case of $k_1=k_2=f_1=f_2=1$, this function becomes 
	\begin{align}
	\phi(\lambda)=\dfrac{9\lambda^4 + 11\lambda^3 + 9\lambda^2 + 4\lambda + 1}{2(3 \lambda^2 + 1)}+\sigma^2\dfrac{6 \lambda^2 + 5 \lambda + 1}{2(3 \lambda^2 + 1)}. 
	\end{align}
	Next, we find estimation function $\psi(\lambda,\mathbf F)$. We can show that 
	\begin{align}
\psi(\lambda,\mathbf F)=\dfrac{1}{2 f_1 f_2 \lambda^2}+\dfrac{\sigma^2}{2}\left ( f_1 \lambda + \dfrac{f_2}{ f_1} \right ). 
	\end{align}
The first term is  due to the disturbances, while the second term originates from the feedback noises. The transfer matrix $\mathbf H(s\mathbf I-\mathbf A)^{-1}\mathbf E$ is right-invertible and minimum-phase. Hence, as Theorem \ref{thm:cheapestimation} suggests, as $\sigma$ becomes smaller, we can make the estimation arbitrarily  precise  by increasing the magnitude of observer gains $f_1$ and $f_2$.  
\end{example}

\section{ Analysis of Network of Networks } \label{sec:netnet}

We introduce and analyze a class of networks of networks that are built by a repeated application of control law (\ref{eq:laplacian_control_c}).  For simplicity of the developments, we neglect the feedback noises (i.e., set $\sigma=0$). One can show that the same approach works in the presence of those noises as well.

\subsection{Construction Procedure for Composite Networks} 

 \begin{figure}
    \centering
\includegraphics[width=8.4cm]{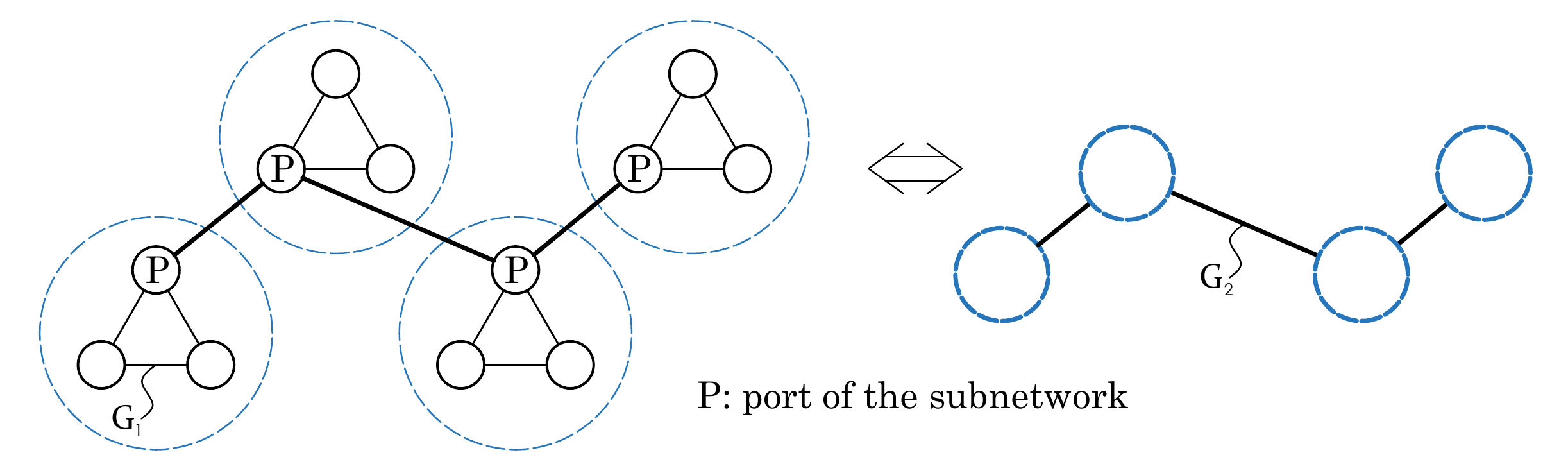}
    \caption{An illustration  of the proposed model for a network of networks, where the subnetworks over graph $\mathcal{G}_1$ are interconnected   via their port nodes (designated with  letter $\mathbf{P}$) over  graph $\mathcal{G}_2$. } 
    \label{fig:netnet}
\end{figure}

First, we build identical networks using  control law (\ref{eq:laplacian_control_c}) over  graph $\mathcal{G}_1$. We denote the number of nodes of $\mathcal{G}_1$ by $m$ and the order of the state-space realization for each subsystem by $n$.  Moreover, we denote the feedback gain used to build each network by $\mathbf{K}_1 \in \R^{p \times q}$. Let us denote the  state of   subsystem $j$ in module or subnetwork $i$ by $x_j^{(i)}\in \R^n$. Similarly, we denote the rest of corresponding variables. The  Laplacian matrix corresponding to $\mathcal{G}_1$ is also denoted by $\mathbf{L}_1$. For the subsequent analysis, let us define 
\begin{align} \label{eq:replacements}
& \tilde {\mathbf{A}} :=\mathbf{I}_m \otimes \mathbf{A}-\mathbf{L}_1 \otimes \mathbf{B}  \mathbf{K}_1\mathbf{H} ,~\tilde {\mathbf{B}}  :=e_m \otimes \mathbf{B} , \\
& \tilde {\mathbf{E}}  :=\mathbf{I}_m \otimes {\mathbf{E}} ,~~\tilde {\mathbf{C}}:=\mathbf{I}_m\otimes \mathbf{C}, ~\tilde {\mathbf{H}}:=e_m^T \otimes \mathbf{H}.  \notag 
\end{align}
According to (\ref{eq:CL}), the dynamics of the subnetwork $i$ are% given by 
\begin{align} \label{eq:before_composite}
\dot x^{(i)}=\tilde {\mathbf{A}}\, x^{(i)}+\tilde {\mathbf{E}}\, \xi^{(i)},%+(e_m \otimes \mathbf{B} ) \tilde u^{(i)} , 
\end{align}
where %$x^{(i)} \in \R^{mn}$  is
$
 x^{(i)}:=\begin{bmatrix} (x_1^{(i)})^T \, \dots \, (x_m^{(i)})^T\end{bmatrix}^T \in \R^{mn}, 
$
 is the state    vector of  module $i$ and disturbance vector $\xi^{(i)} \in \R^{mm_1}$ is defined similarly. Without loss of generality, we designate the last node in  graph $\mathcal{G}_1$ as the  {port of the module}\footnote{~If we wish to choose another node, we can simply relabel the nodes.}, which corresponds to a subsystem that we can add a term to its control input. This converts (\ref{eq:before_composite}) to new {open-loop} dynamics 
\begin{align} 
\dot x^{(i)}= \tilde {\mathbf{A}} \, x^{(i)} +\tilde {\mathbf{B}}\,  u^{(i)} +  \tilde {\mathbf{E}}\,  \xi^{(i)},  \label{eq:open_loop_composite}
\end{align} 
%
%\begin{align}\notag
%\dot x^{(i)}=& (\mathbf{I}_m \otimes A -\mathbf{L}_1\otimes \mathbf{B}  \mathbf{K}_1{\mathbf{H}})x^{(i)} \\
%& +  (\mathbf{I}_m \otimes {\mathbf{E}} ) \xi^{(i)}+(e_m \otimes \mathbf{B} ) u^{(i)} ,  \label{eq:open_loop_composite}
%\end{align} 
wherein $ u^{(i)} \in \R^m$ is the tunable control input to the module. Moreover, we assume that two modules can become interconnected only through their port nodes. Then, the only variable  that  module $i$ can use for relative feedback is the output variable for the port node, which  is  denoted by 
\begin{align}\label{eq:y_composite}
y^{(i)}=\tilde {\mathbf H} x^{(i)}=\mathbf{H} x_m^{(i)}.
\end{align}
We collect $N$ instances of these networks with dynamics \eqref{eq:open_loop_composite} and feedback variables (\ref{eq:y_composite})  to construct a composite network. Therefore, the subsystems   equivalent to $S_i$ in  (\ref{eq:subsystem}) for this network design are
 \begin{align}\label{eq:subsystem_nn}
S^{(i)}: \left \{ \begin{array} {l}
 \dot x^{(i)}=\tilde {\mathbf{A}}x^{(i)}+\tilde {\mathbf{B}} u^{(i)}+\tilde {\mathbf{E}}  \xi^{(i)}%\\
\\ y^{(i)}=\tilde {\mathbf{H}} x^{(i)} \\
 z^{(i)}=\tilde {\mathbf{C}} x^{(i)} 
\end{array}   \right .,
\end{align}
with the structured matrices defined by \eqref{eq:replacements}.
%for the reasons that we just described. 
% Let say we have $N$ modules and modules $i$ and $j$ are connected to each other over  $\mathcal{G}_2$ with the edge set $\mathcal{E}_2$. 
Now, we build a modular network by application of  control law (\ref{eq:laplacian_control_c}) with $N$ modules (or subnetworks) connected over a higher level graph $\mathcal{G}_2$ with  feedback  $\mathbf{K}_2 \in \R^{p \times q}$.  \footnote{~Feedback gains $\mathbf{K}_1$ and $\mathbf{K}_2$ are matrices of the same dimension because we have chosen one node as the port of a subnetwork.} If $\{i,j\} \in \mathcal{E}_2$ has a weight denoted by $b_{ij}$, then the application of  control law \eqref{eq:laplacian_control_c} will be  
\begin{align}\label{eq:netnet_u}
 u^{(i)}=-\mathbf{K}_2  \sum_{\{i,j\} \in \mathcal{E}_2} b_{ij}  \left ( y_m^{(i)}-y_m^{(j)} \right ).
\end{align}
We have $N$ modules and each one consists of $m$ subsystems. Therefore, the consensus output of the network should be %defined as 
\begin{align}
\nu_{\mathrm{nn}}(t):=(\mathbf{M}_{Nm} \otimes \mathbf{C}) x(t). 
\end{align}
Then, we set its steady-state variance as the performance measure  %then, ${\mathbf{G}}_\mathrm{nn}(s)$ denotes the transfer matrix  from the disturbance to the output, inducing the performance measure % denoted by
\begin{align}
\rho_{\mathrm{nn}}(\mathbf{L}_2,\mathbf{K}_2):=\, \lim_{t\rightarrow \infty}\mathbb{E}\left \{ \|\nu_{\mathrm{nn}} (t)\|_2^2 \right \}, %\left \|{\mathbf{G}}_\mathrm{nn}(s)\right \|_{\mathcal{H}_2}^2,
\end{align}
where $\mathbf{L}_2$ is the graph Laplacian of $\mathcal{G}_2$. 
%Let us denote the state, output for feedback, and input vector of node number $l$ in module number $i$ by $x_l^{(i)}$, $z_l^{(i)}$, and $u_l^{(i)}$, respectively. 
%For any node that is not the port node, $l \neq m$ the control input follows the same formula as (\ref{eq:laplacian_control_c}); i.e. 
%$$
% u_l^{(i)}= -\mathbf{K} \sum_{\{l,k\} \in \mathcal{E}_1} a_{lk} \left ( z_l^{(i)}-z_k^{(i)} \right ).
%$$
 %use of control law (\ref{eq:laplacian_control_c})  over high order graph $\mathcal{G}_2$ if we replace the matrices with 
 In Fig. \ref{fig:netnet}, we illustrate this composite structure using an example: we have four  modules and inside each of them,  three subsystems are interconnected over  graph $\mathcal{G}_1$, in this case a complete graph. These  subnetworks are then connected  via their ports over graph $\mathcal{G}_2$, which in this case is a path.% in this case. 

\noindent{\it Interpretation of Construction: } %we elaborate on the actual meaning of interconnection between two modules.
 Let say modules $i$ and $j$ are connected, thus  $\{i,j\} \in \mathcal{E}_2$. Then, the ports of these two modules will have access  to the relative difference of their feedback output  $y_m^{(i)}-y_m^{(j)}$ and will reflect this feedback term in their control input. Mathematically speaking, the input to the port node\footnote{~Recall that the port node is arbitrary chosen or labeled   to be  number $m$.}  in  module $i$ is 
\begin{align}
 u_m^{(i)}=& -\mathbf{K}_1 \sum_{\{m,k\} \in \mathcal{E}_1} a_{mk} \left ( y_m^{(i)}-y_k^{(i)} \right ) \\
 & -\mathbf{K}_2  \sum_{\{i,j\} \in \mathcal{E}_2} b_{ij}  \left ( y_m^{(i)}-y_m^{(j)} \right ).  \notag % u^{(i)},  
\end{align} 
The first term is due to initial application of control law (\ref{eq:laplacian_control_c}) over   $\mathcal{G}_1$ with an edge set  $\mathcal{E}_1$, while  the second term is $u^{(i)}$  from (\ref{eq:netnet_u}) based on the  composite network design  over   $\mathcal{G}_2$. 

\subsection{Stability and Performance of Composite Networks} 

%In the following theorem, we characterize stability and performance for these networks of networks.

\begin{theorem} \label{thm:composite} Consider a  dynamical network over graph $\mathcal{G}_1$ with a bounded performance measure $\rho(\mathbf{L}_1,\mathbf{K}_1)$.  Suppose that  in Proposition 1 and Theorem 1 we apply   control law (\ref{eq:laplacian_control_c}) on   systems $S^{(i)}$ defined in (\ref{eq:subsystem_nn}) over $\mathcal{G}_2$ with feedback gain $\mathbf{K}_2$. The resulting composite network  reaches consensus if and only if $\left .\tilde {\mathbf{A}}-\lambda_i(\mathbf{L}_2) \tilde {\mathbf{B}}{\mathbf{K}}_2 \tilde {\mathbf{H}} \right .$ is Hurwitz for nonzero eigenvalues of $\mathbf{L}_2$. 
Moreover,  if $ \phi_{\mathrm{nn}}(\lambda,\mathbf{K})$ is the performance function derived from Theorem 1 for subsystems $S^{(i)}$ defined in (\ref{eq:subsystem_nn}), then% performance measure of the composite network  can be expressed as   
\begin{align}\label{eq:nn_perf}
\rho_{\mathrm{nn}}(\mathbf{L}_2,\mathbf{K}_2)=\rho(\mathbf{L}_1,\mathbf{K}_1)+\sum_{i=2}^{N} { \phi_{\mathrm{nn}} \left (\lambda_i(\mathbf{L}_2),\mathbf{K}_2 \right )}.
\end{align}
\end{theorem}

The significance  of this result is that for a fixed module graph $\mathcal{G}_1$ with Laplacian $\mathbf{L}_1$ and $\mathbf{K}_1$, the value of $\rho(\mathbf{L}_1, \mathbf{K}_1)$ and the form of  composite performance function $\phi_{\mathrm{nn}}$ are fixed. Thus, we can quantify the role of  higher level graph $\mathcal{G}_2$ and feedback gain $\mathbf{K}_2$ in the performance of the composite network by looking at the second term. The extra term compared to Theorem \ref{thm:first_main} appears  because  
\begin{align}
\nu_{\mathrm{nn}} =\left (\mathbf{M}_{Nm} \otimes \mathbf{C}\right )x \neq \left ({\mathbf{M}_N} \otimes \tilde {\mathbf{C}}\right ) x,
\end{align}
where the right-hand side is the  output that would have resulted   in an expression of   form \eqref{eq:spectral_form_gen}.
 
 \begin{remark} If a subnetwork is one subsystem, then  (\ref{eq:nn_perf})   reduces to \eqref{eq:spectral_form_gen}, since each subsystem as a network satisfies % on its own implies that 
 	$\rho(\mathbf{L}_1,\mathbf{K}_1)=0$. %, i.e., the deviation of the only output from the average is  equal to zero. 
 \end{remark}
% 
% \noindent{ \it Multi-Level Networks:} Let us assume we have a network of networks that has been constructed according to the described procedure. We can designate a certain subsystem in the composite network to be its port. Then, we can repeat the procedure on these systems to build  composite networks whose modules themselves are networks of networks.
 
\subsection { Minimum Connectivity Design For Composite Networks}

 %We show that if the feedback gain of the subnetworks $\mathbf{K}_1$ has  an unbounded stability region in terms of the eigenvalues of $\mathbf{L}_1$,
 We show that if $\tilde \lambda(\mathbf K_1)<\infty$, then
   there exists a simple choice for $\mathbf{K}_2$ such that it has also an unbounded stability region in terms of the eigenvalues of  higher level Laplacian $\mathbf{L}_2$; i.e.,  $\tilde \lambda (\mathbf{K}_2)$ exists and  if $\lambda> \tilde \lambda(\mathbf{K}_2)$ then $\left .\tilde {\mathbf{A}}-\lambda \tilde {\mathbf{B}}{\mathbf{K}}_2 \tilde {\mathbf{H}} \right .$ is Hurwitz. This would remedy the concerns about possible complexities in the design of $\mathbf{K}_2$ over  graph $\mathcal{G}_2$ . 

\begin{theorem} \label{thm:min_netnet}  Suppose that the subnetworks are built over any graph $\mathcal{G}_1$ and feedback gain $\mathbf{K}_1$, which has an unbounded stability region. For any $\alpha>0$, let us choose the feedback gain of the composite network to be $\mathbf{K}_2=\alpha \mathbf{K}_1$. Then, $\mathbf{K}_2$ has an unbounded stability region with respect to the eigenvalues of  higher level Laplacian $\mathbf{L}_2$. 
\end{theorem}

This  result is simplified if $\tilde \lambda (\mathbf{K}_1)=0$.

\begin{corollary}  \label{cor:netnet} Suppose that the subnetworks of the network of networks are built with $\mathbf{K}_1$, which induces $\tilde\lambda (\mathbf{K}_1)=0$. Let us choose  $\mathbf{K}_2=\alpha \mathbf{K}_1$ for some $\alpha >0$ in the design of the described composite networks. Then,  higher level feedback gain $\mathbf{K}_2$ satisfies $\tilde \lambda (\mathbf{K}_2)=0$ with respect to the eigenvalues of  higher level Laplacian $\mathbf{L}_2$. 
\end{corollary}

%\noindent{ \it Composite Networks with Decentralized Observer:} We inspect that the current model of construction of a network of networks includes the case of  subsystems which are equipped with the decentralized observer described in Lemmas \ref{lem:obs}  and \ref{lem:kdesign_observable}. 

%\noindent{\it Fused Network of Networks:} If the feedback gain $\mathbf{K}$ has unbounded stability region over the eigenvalues of second graph, then we can analyze the performance of  $N$ identical subnetworks, which  have a mutual port node (instead of connection between the ports). In fact, this corresponds to the case when the ports of all subsystems have \emph{fused} in each other as the weights of the graph $\mathcal{G}_2$ has approached the infinity.  The performance of the network in this case would be 
%\begin{align*}
%& \rho(\mathbf{L}_1,\mathbf{K})+\sum_{i=2}^{N} \lim_{\lambda_i(\mathbf{L}_2) \rightarrow \infty }{\tilde \phi (\lambda_i(\mathbf{L}_2),\mathbf{K})}=\\
%& \rho(\mathbf{L}_1,\mathbf{K})+(N-1) \lim_{\lambda  \rightarrow \infty} \tilde \phi (\lambda,\mathbf{K}).
%\end{align*} 
%Now, we define the last limit as $$ \tilde \rho(\mathbf{L}_1,\mathbf{K}):=\lim_{\lambda  \rightarrow \infty} \tilde \phi (\lambda,\mathbf{K}). $$ Then the performance of the resulting network would be
%$$
%\rho(\mathbf{L}_1,\mathbf{K})+(N-1)  \tilde \rho(\mathbf{L}_1,\mathbf{K}).
%$$
% As expected, it is only a function of the graph of the blocks and $\mathbf{K}$. 

\subsection { Examples of Networks of Networks}  %In the first example, we consider complete modules and compute the performance functions.

\begin{table}
\centering
\resizebox{0.45 \textwidth}{!}{%
\begin{tabular}{|c|c|}
\hline
  &   \\[-1em]
{{Dynamics}}  & Performance Function $ \phi_{\mathrm{nn}} (\lambda)$   \\ \hline \hline  
 &  \\[-1em]
 &   \\[-1em]
single-integrator           & $\dfrac{2(m-1)  \lambda +m^2}{2m k  \lambda }$          \\  
&   \\[-1em]
 &   \\[-1em] \hline
&  \\[-1em]
  &   \\[-1em] 
double-integrator &      $\dfrac{(m-1)(m+2) \lambda^2+2m^2(m-1) \lambda +m^4}{2 m^2 k_1 k_2 \lambda^2}$     \\            
 %& &  \\[-1em] \hline 
%  & &  \\[-1em]
% & & \\[-1em]
  &  \\ \hline
\end{tabular}}
\caption{Performance functions for a composite network with complete graph subnetworks of single and double-integrator agents. Each module has $m$ nodes and the feedback gains are assumed to be identical over both graphs  (see Example \ref{ex:complete_netnet}).} 
\label{table:netnet}
\end{table}

\begin{example} \label{ex:path_netnet} Consider a modular network with subnetworks of single-integrators over an unweighted {path graph} $\mathcal{G}_1$ of $m$ nodes, where the last node of the module is its port. We choose $\mathbf{K}_1=k_1>0, $ so           the open-loop dynamics of the modules before design of the composite network based on (\ref{eq:open_loop_composite}) are  
\begin{align}
\dot x^{(i)}=-k_1 \mathbf{L}_1x^{(i)}+e_m  u^{(i)}+\mathbf{I}_m \xi^{(i)}, 
\end{align}
where $x^{(i)} \in \R^m,\xi^{(i)} \in \R^m$, $u^{(i)} \in \R$, and $\mathbf{L}_1$ is   Laplacian of the unweighted path graph over $m$ nodes. Then, choosing $\mathbf{K}_2=k_2>0$, 
%for $\mathbf{K}_1=k_1$ (over $\mathcal{G}_1)$ and $\mathbf{K}_2=k_2$ (over $\mathcal{G}_2$),
 the performance function of the composite network is 
\begin{align} \label{eq:pathpathperf}
 \phi_{\mathrm{nn}}(\lambda)=\dfrac{ ({m(m-1)}/2)k_2 \lambda +k_1 m}{2k_1 k_2 \lambda }. %\dfrac{\dfrac{m(m-1)}{2}\lambda+m}{2k \lambda }.
\end{align} 
Based on Corollary \ref{cor:netnet} in the higher level $\tilde \lambda (\mathbf{K}_2)=0$. Moreover, we inspect that the resulting family of functions is strictly convex and decreasing for $\lambda>0$ (see the appendix for the details). 
\end{example}

\begin{example} \label{ex:complete_netnet} Consider a modular network, where each subnetwork consists of subsystems with the single or double-integrator dynamics. In this case, we set  $\mathcal{G}_1$  to be the unweighted {complete graph} over $m$ nodes (similar to the example illustrated in Fig. \ref{fig:netnet} for the subnetworks with $m=3$). Therefore, unlike Example \ref{ex:path_netnet}, no matter which node is chosen as  the port, the subnetwork will be the identical. 
For element-wise positive feedback gains $\mathbf{K}_1$ and $\mathbf{K}_2$, Corollary \ref{cor:netnet} again implies  that the minimum connectivity threshold in terms of the eigenvalues of $\mathbf{L}_2$ for both nodal dynamics is zero.  
Moreover, %we choose the feedback gains over $\mathcal{G}_1$ and $\mathcal{G}_2$ to be identical; i.e. $\mathbf{K}_1=\mathbf{K}_2=k$ 
let $\mathbf{K}_1=\mathbf{K}_2=k>0$ for the single-integrators and %$\mathbf{K}_1=\mathbf{K}_2=[k_1,k_2]$ 
$\mathbf{K}_1=\mathbf{K}_2=[k_1,k_2] \succ 0$ for the double-integrators (for simplicity). Using the solution to the Lyapunov equation,  we find performance function $ \phi_{\mathrm{nn}}(\lambda)$ for these subnetworks, which are e given in Table \ref{table:netnet}.  These functions are  strictly convex and decreasing for $\lambda>0$. As a sanity check, for $m=2$  the unweighted path and complete graphs coincide and the first formula in Table \ref{table:netnet} and  \eqref{eq:pathpathperf} produce  identical functions for $k_1=k_2=k$ (see appendix Q for   details). 
\end{example}

%We can find  similar expressions for the performance function of networks with path subnetworks of single-integrators.

%Moreover, considering $\mathbf{K}_\epsilon$ as defined in Theorem \ref{thm:right_mp}, the inequality (\ref{eq:phi_asymptotic}) is asymptotically tight in the sense that 
%\begin{align} \label{eq:limit_1}
%\lim_{\epsilon \rightarrow 0}\phi(\lambda,\mathbf{K}_\epsilon)= \mathrm{Tr}({\mathbf{E}} ^T \mathbf{P}_0 {\mathbf{E}} ),
%\end{align} 

%This limiting behavior is also given by 
%\begin{align} \label{eq:limit_2}
% \inf_{K \text{ has } \tilde \lambda (\mathbf{K}) } \lim_{\lambda \rightarrow \infty} \phi(\lambda,\mathbf{K}) =\mathrm{Tr}({\mathbf{E}} ^T \mathbf{P}_0 {\mathbf{E}} ).
%\end{align}

%Now, we bring an example which lets us understand the process of deriving this lower-bound. 

\section{Performance Bounds and Scaling} \label{sec:bounds}

We look at the cases where combining the information on graph parameters and derived performance functions can  give us macroscopic information on the performance measure.  %of the network. %These are useful for finding the scaling of the performance measure with respect to its parameters (e.g. number of edges), and also as rules of thumb for the design. 
%First, we consider general topologies, while we will specifically look at networks over paths and cycles after that.  These results  give us clues about the asymptotic performance behavior. 

\subsection{ Performance Bounds and Scaling  } 

%In this subsection, the following assumption will be used. 

%\begin{assumption}\label{assumption:convex_decreasing} For the stabilizing feedback gain $\mathbf{K}_0$, the performance $\phi(\lambda)=\phi(\lambda,\mathbf{K}_0)$ is a decreasing continuously differentiable  convex function of $\lambda$.
%\end{assumption} 

%We derive two graph theoretic bounds on the best attainable performance measure of the network. 

 %bound on the best attainable performance of the dynamical networks using the maximum degree of the graph, number of subsystems, number of edges in the graph, and the performance function.%, where we cover both loopless \emph{and} loopy graphs. 
 
 In Sections \ref{sec:example} and \ref{sec:netnet}, a majority of the derived performance functions  are convex. In what follows, we show that this property is useful in derivation of performance bounds. 

\begin{theorem} \label{thm:laplacian_loopless} Consider a network of $N$ subsystems with a performance functions $\phi(\lambda)$ that is convex. The performance measure over an unweighted graph with $M$ edges and maximum nodal degree of  $\Delta $ is lower-bounded according to 
\begin{align} \label{eq:genbound}
%\sum_{i=2}^N \phi(\lambda_i) \geq  \phi(1+\Delta )+(N-2) \phi\left (\frac{2M -1-\Delta}{N-2}\right ),
\rho({\mathbf{L},\mathbf{K}})\, \geq \,  \phi(1+\Delta )+(N-2) \,  \phi\left (\frac{2M -1-\Delta}{N-2}\right ),
\end{align} 
where the equality holds if and only if graph $\mathcal{G}$ is either complete graph or star graph. %This measure is also bounded according to 
%\begin{align}\label{eq:genbound_2}
%\rho({\mathbf{L},\mathbf{K}}) \geq (N-1) \phi\left (\frac{2M}{N-1}\right ),
%\end{align}
%where the equality holds if and only if the graph is complete. 

%Moreover, if the input function $\mu(\lambda)$ is convex, we may replace $(\rho,\phi)$ in this result with $(\rho_u,\mu)$ or $(\rho_\gamma,\phi_\gamma)$ as well.
%On the other hand, for a network over  a loopy Laplacian $\mathbf{L}$ and augmented loopless Laplacian $\hat L$, 
%we may bound the performance measure as
%% \begin{align}
%% &\rho({\mathbf{L},\mathbf{K}})\geq  \\ 
%% & \notag \phi\left (1+\Delta(\hat L) \right)+(N-1) \phi\left (\frac{2M(\hat L) -1-\Delta(\hat L)}{N-1} \right)+  \\ 
%% \notag  & \phi\left (\frac{a_L}{N}\right)-\phi\left (  \frac{2(N+1)}{2+N(N+1)\mathrm{diam}(\hat G)-2M(\hat L)\mathrm{diam}({\hat G}) } \right).
%% \end{align}
%% Moreover, another lower-bound would be 
% \begin{align*}
% &\rho({\mathbf{L},\mathbf{K}})\geq  
% \begin{adjustbox}{max width=200pt}
% $ \phi\left (1+\Delta(\mathcal{ \hat G}) \right)+(N-1) \phi\left (\dfrac{2M -1-\Delta(\mathcal{ \hat G})}{N-1} \right),$
% \end{adjustbox}
% \end{align*}
% where $\Delta(\mathcal{\hat G})$ is the maximum degree of nodes of $\mathcal{\hat G}$. 
 \end{theorem}

 \begin{example} \label{ex:survey}
 Consider a network with nodal dynamics  $\mathfrak{s}_2$,    
 $a_1=0$ and $b_0=b_1=k_1=k_2=a_2=1$. For all connected unweighted graphs with $3$ to $7$ nodes, we do a survey for the ratio of the sides of  inequality \eqref{eq:genbound}, that is %looking at %a ratio %$R$% defined as 
$$
\mathfrak{r}_1:=\dfrac{\rho({\mathbf{L},\mathbf{K}})}{\phi(1+\Delta )+ (N-2) \phi\left (\dfrac{2M -1-\Delta}{N-2}\right )} \geq 1.
$$
%Among these cases, this ratio is 1 for the star and complete graph. 
%that is the ratio of the two sides of the inequality.
 The distribution of $\mathfrak{r}_1$ versus $N$  is illustrated in Fig. \ref{fig:sgencdf}. As $N$ increases, it tends to an almost fixed curve (with a growing tail), where $\sim 90\%$ of the graphs induce a ratio $\mathfrak{r}_1<2$. 

   \begin{figure}
    \centering
\includegraphics[width=7.7cm]{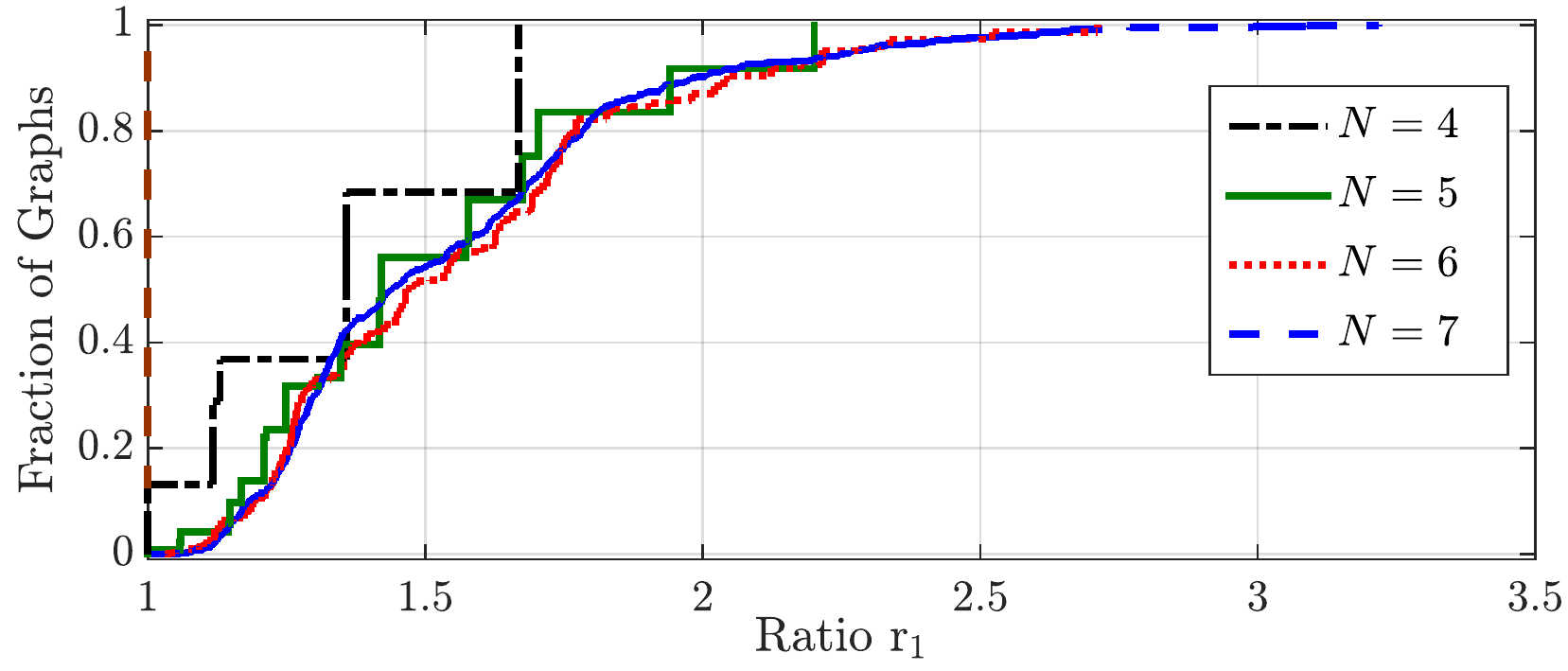}
    \caption{The fraction of connected unweighted graphs for which the ratio $\mathfrak{r}_1$ is less than a threshold (see Example \ref{ex:survey})}
    \label{fig:sgencdf}
\end{figure}

\end{example}

 \begin{theorem} \label{thm:laplacian_loopless_weighted} Consider a network of $N$ subsystems with a performance functions $\phi(\lambda)$ that is convex. The performance measure over any weighted graph with a total weight of $W$ is lower-bounded according to  
 	\begin{align}\label{eq:genbound_2}
 	\rho({\mathbf{L},\mathbf{K}}) \geq (N-1) \phi\left (\frac{2W}{N-1}\right ),
 	\end{align}
 	where the equality holds if and only if the graph is complete and with identical weights. 
 	%Moreover, if the input function $\mu(\lambda)$ is convex, we may replace $(\rho,\phi)$ in this result with $(\rho_u,\mu)$ or $(\rho_\gamma,\phi_\gamma)$ as well.
 	%On the other hand, for a network over  a loopy Laplacian $\mathbf{L}$ and augmented loopless Laplacian $\hat L$, 
 	%we may bound the performance measure as
 	%% \begin{align}
 	%% &\rho({\mathbf{L},\mathbf{K}})\geq  \\ 
 	%% & \notag \phi\left (1+\Delta(\hat L) \right)+(N-1) \phi\left (\frac{2M(\hat L) -1-\Delta(\hat L)}{N-1} \right)+  \\ 
 	%% \notag  & \phi\left (\frac{a_L}{N}\right)-\phi\left (  \frac{2(N+1)}{2+N(N+1)\mathrm{diam}(\hat G)-2M(\hat L)\mathrm{diam}({\hat G}) } \right).
 	%% \end{align}
 	%% Moreover, another lower-bound would be 
 	% \begin{align*}
 	% &\rho({\mathbf{L},\mathbf{K}})\geq  
 	% \begin{adjustbox}{max width=200pt}
 	% $ \phi\left (1+\Delta(\mathcal{ \hat G}) \right)+(N-1) \phi\left (\dfrac{2M -1-\Delta(\mathcal{ \hat G})}{N-1} \right),$
 	% \end{adjustbox}
 	% \end{align*}
 	% where $\Delta(\mathcal{\hat G})$ is the maximum degree of nodes of $\mathcal{\hat G}$. 
 \end{theorem}

 Theorems \ref{thm:laplacian_loopless} and \ref{thm:laplacian_loopless_weighted} give rules of thumb  about the best achievable  performance. To do so, we combine information on the nodal dynamics (through the form of the performance function) and macroscopic    graph information. 
 
%If we are only interested in the scaling implications, we can use a corollary of Theorem \ref{thm:laplacian_loopless}. 
 \begin{corollary} \label{cor:scale}Under the settings of Theorem \ref{thm:laplacian_loopless_weighted}, it holds that 
\begin{align}
\rho({\mathbf{L},\mathbf{K}})=\Omega \left ( N \phi \left ( {W}/{N}\right ) \right ). 
\end{align}
 \end{corollary}

\noindent{\it Performance-Sparsity Tradeoff:} 
Suppose that $\phi(\lambda)$ is also decreasing. For an unweighted graph, $W=M$. Therefore, we can reorganize the result of Theorem \ref{thm:laplacian_loopless_weighted}   and write 
  \begin{align}\label{eq:sineq}
\phi\left ( \dfrac{2M}{N-1}\right ) \, \leq \, \dfrac{\rho({\mathbf{L},\mathbf{K}})}{N-1}.
 \end{align}
 %It is not difficult to see that the similar inequality derived from Theorem \ref{thm:laplacian_loopless} 
% while in case of the loopy graphs it would be 
  % \begin{align}\label{eq:sineq_loopy}
%\phi\left ( \dfrac{2M}{N-1}\right ) <  \frac{\rho({\mathbf{L},\mathbf{K}})}{ N-1}.
% \end{align}
% \begin{remark} Before proceeding with the representation of the tradeoffs, we would like to emphasize that for a stable multi-agent system over a unweighted loopy  topology, as stated earlier, $\lambda_1 \leq 1$, so any design must consider $\tilde \lambda(\mathbf{K})<1$. On the other hand, if the graph (without the self-loops) is connected 
% $$\dfrac{2M}{N-2}\geq \dfrac{2N-2}{N-2}\geq 2+\dfrac{2}{N-2}>1>\tilde \lambda (\mathbf{K}).$$
% This implies computing the performance function $\phi(\lambda)$ at $\lambda=2M/(N-2)$ is indeed relevant. 
% \end{remark}
 This result is useful in quantification of the following tradeoff: as the graph of the network becomes sparser, the best attainable value of the performance measure will increase.  The following example highlights two specific cases.  %and (\ref{eq:sineq_loopy}) as follows. 
 
  \begin{continuance} {ex:three} 
For networks with subsystems that have $\mathfrak{s}_1$ dynamics, Corollary \ref{cor:scale} implies that over any unweighted graph
\begin{align}
\rho({\mathbf{L},\mathbf{K}})=\Omega \left ({N^2}/{M} \right ).
\end{align}
For networks with subsystems of $\mathfrak{s}_2$ dynamics we  deduce 
\begin{align}
\rho({\mathbf{L},\mathbf{K}})=\left\{
    \begin{array}{ll}
        \Omega \left ({N^3}/{M^2} \right ) & \mbox{if } b_0=0 \vspace{1mm} \\ 
        \Omega \left ({N^2}/{M} \right )& \mbox{if } b_0\neq 0
    \end{array}
\right. .
\end{align}
For instance, in the special case of  $a_1=a_2=0$ for $\mathfrak{s}_2$ % in  (\ref{eq:simpletradeofH_2_c})
 we get %over a loopless graph 
%$$
% \dfrac{b_0^2}{2k_2 \dfrac{2M}{N-2}}+\dfrac{b_1^2}{2k_1 k_2 \dfrac{4M^2}{(N-2)^2}} < \dfrac{\rho_c({\mathbf{L},\mathbf{K}})}{N-2},
% $$
% which is equivalent to the inequality 
 \begin{align}
 \dfrac{b_0^2(N-1)^2}{4k_2 {M}}+\dfrac{b_1^2(N-1)^3}{8k_1 k_2 {M^2}} \, \leq \, {\rho({\mathbf{L},\mathbf{K}})}, 
 \end{align}
 that  clearly reflects the sparsity-performance tradeoff for a consensus network of double-integrators (see \cite{siami2016fundamental} for a similar result for single-integrator agents). 
\end{continuance}

\subsection{ Performance Asymptotic over  Path and Cycles} 

%We show that we can grasp the asymptotic behavior of $\rho(\mathbf{L}, \mathbf{K})$  over a path or cycle graph by an appropriate  integration. 

\begin{theorem} \label{thm:integral} For a    network of $N$ subsystems over an unweighted path or cycle graph with $\tilde \lambda (\mathbf{K})=0$, it holds that % a asymptotic expression given by   %we can write %over a path graph scales according to 
\begin{align}\label{eq:rhotheta}
 \rho({\mathbf{L},\mathbf{K}})= \Theta\left (N\,\Gamma_N \right ),
\end{align}
where   $\Gamma_N$ can be computed using a parametric integral 
\begin{align} 
\Gamma_N:=\int_{1/N}^1 \phi\big (2-2\cos(\pi x) \big )~dx. 
\end{align}
Moreover, if $\phi(\lambda)$ is bounded at $\lambda=0$, then it holds that 
\begin{align}\label{eq:limone}
\lim_{N\rightarrow \infty} \dfrac{N\Gamma_N}{\rho({\mathbf{L},\mathbf{K}})}=1.
\end{align}
\end{theorem}

\begin{corollary} \label{cor:path}The performance measure scales similarly with respect to $N$ over unweighted path and cycle graphs. Moreover, if $\phi(\lambda)$ is bounded at  $0$,  the performance measure over the paths and cycles   converge to the same value as $N \rightarrow \infty$.
\end{corollary}

We  should emphasize on few points: {\it (i)} Theorem \ref{thm:integral} {does not} depend on neither convexity nor monotonicity of  $\phi(\lambda)$;  {\it (ii)}  the requirement $\tilde {\lambda} (\mathbf{K})=0$ is natural, since as $N$ increases $\lambda_2(\mathbf{L})=\Theta( 1/N^2)$; i.e., it becomes arbitrary small. Otherwise,  there exist 
%a finite graph size
 $N_1$ such that for $N \geq  N_1$, %we have 
 $\lambda_2< \tilde {\lambda} (\mathbf{K})$. {\it (iii)} This approximation idea  has been previously reported, e.g. in \cite{gutman2007estrada} it is used for estimation of Estrada index. However, we find the reason for which the approximations find the scaling of the sums, even if $\phi(\lambda)$ is singular at  $\lambda=0$.% (that is the case if the subsystems are marginally stable). 

\begin{figure}
	\centering
	\includegraphics[width=7.5cm]{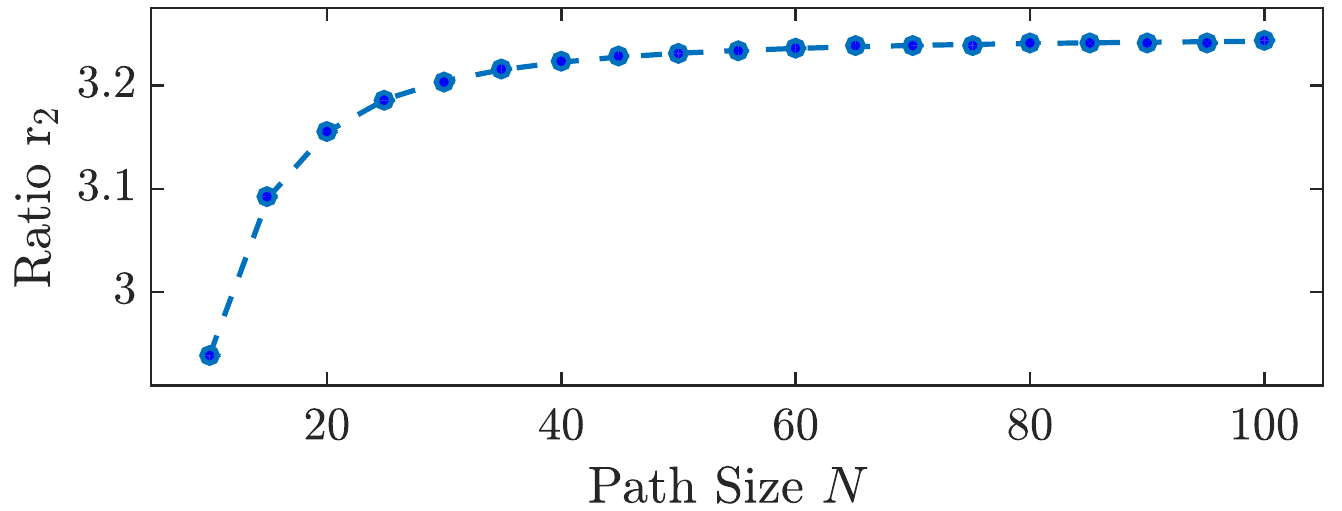}
	\caption{Performance asymptotic over paths in  continuance of Example \ref{ex:three}. }
	\label{fig:pathsc}
\end{figure}

 \begin{continuance} {ex:three}  We apply Theorem \ref{thm:integral} on a network of $\mathfrak{s}_1$ subsystems with $a=0$ (i.e., single-integrators) over an unweighted  path and arrive at the asymptotic expression  
\begin{align}\label{eq:single_int_path}
\rho({\mathbf{L},\mathbf{K}}) =\Theta \left ({ N^2}/k \right ),
\end{align} 
 while if $a>0$, for $\alpha:=a/k$, we have the approximation 
\begin{align}
\rho \sim {N}/(2k\sqrt{\alpha(\alpha+4)}).
\end{align}
For $\mathfrak{s}_2$ agents with $a_0=a_1=0$, the performance  measure satisfies 
\begin{align} \label{eq:h_def}
 \rho=\Theta \left (   \dfrac{ b_0 ^2 N^2}{2 \pi^2 k_2}+\dfrac{b_1^2N^4}{6 \pi^4 k_1 k_2 }   \right ):=\Theta(h(N,k_1,k_2)).
 \end{align} 
Next, for these agents with $b_1=b_0=k_1=k_2=1$ over an unweighted path graph of $N=10,15,\dots,100$ nodes, we investigate the claim of Theorem \ref{thm:integral} by looking at  the ratio
\begin{align}
\mathfrak{r}_2:=\dfrac{\rho({\mathbf{L},\mathbf{K}})}{  h(N,k_1 ,k_2)},
\end{align}
with $h$ given in (\ref{eq:h_def}). 
%for the case of loopless graphs and $S_l$ for this ratio over loopy graphs. 
The result is shown in Fig. \ref{fig:pathsc}, where according to Theorem \ref{thm:integral},  $\mathfrak{r}_2$ indeed goes to a constant.
 \end{continuance}

\begin{figure}
    \centering
\includegraphics[width=7.5cm]{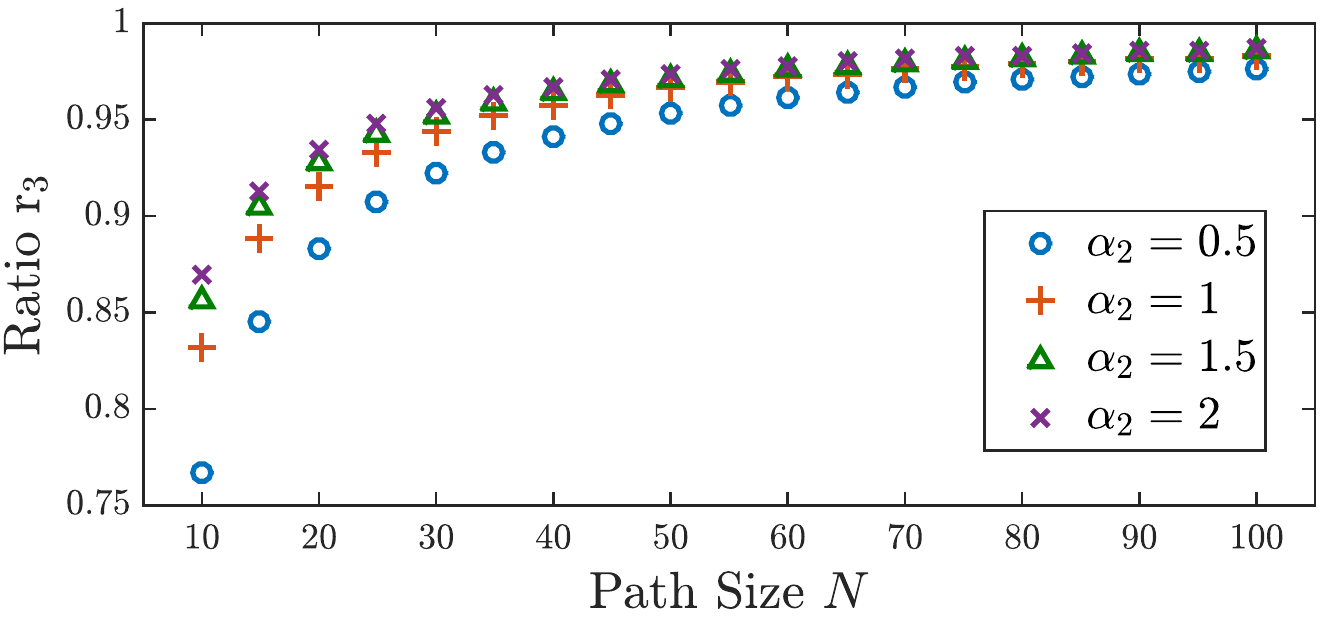}
    \caption{ Ratio $\mathfrak{r}_3$ for a network of harmonic oscillators over a path graph as the network size grows (see the continuance of Example \ref{ex:harmonic})}
    \label{fig:path}
\end{figure}

\begin{continuance}{ex:harmonic} For a network of harmonic oscillators with $\alpha_1 \neq \alpha_2$ over a  path graph, Theorem \ref{thm:integral} implies that %we can write 
\begin{align*}
\begin{adjustbox}{max width=230 pt}
$\rho \sim \dfrac{N}{2k_1k_2(\alpha_1-\alpha_2)}\left (\dfrac{1}{ \sqrt{\alpha_2(\alpha_2+4)}}-\dfrac{1}{ \sqrt{\alpha_1(\alpha_1+4)}}\right )$.
\end{adjustbox}
\end{align*}
We call the right hand side $f(N,\alpha_1,\alpha_2)$. To empirically examine the gap,  we consider \begin{align}
\mathfrak{r}_3:=\dfrac{\rho({\mathbf{L},\mathbf{K}})}{f(N,\alpha_1,\alpha_2)}.
\end{align}
We set $k_1=k_2=1$, $\alpha_1=2\alpha_2$ for $\alpha_2 \in [0.4,4]$, and vary $N$ between $10$ and $200$. Because $\phi(\lambda)$ is bounded, as $N$ increases the approximation becomes tighter as shown in Fig. \ref{fig:path}. 
\end{continuance}

\begin{continuance}{ex:pltn} [Platoon over a Path] For a platoon of vehicles over a path graph,  Theorem \ref{thm:integral} suggests that $\rho$ scales with $N^4$. Thus, the $\mathcal{H}_2$-norm scales with $N^2$. As reported by the authors in \cite{zheng2016platooning}, a similar scaling law in the case of $\mathcal{H}_\infty$ norm of the network over this topology  holds (with an additional leader). %(but with an additional loop)  
\end{continuance} 

\begin{continuance}{ex:fnoise} We can apply Theorem \ref{thm:integral} to find the scaling for the estimation measure as well.  Similar to \eqref{eq:h_def}, we can show that the estimation measure in a network of double-integrators over a path graph satisfies 
	\begin{align}
\mu(\mathbf{L},\mathbf{F}) =\Theta\left (\dfrac{N^4}{f_1 f_2}+N\sigma^2\left (f_1+\dfrac{f_2}{f_1} \right ) \right ). 
	\end{align} 
\end{continuance}

\begin{continuance}{ex:path_netnet} We consider a network of $N$ subnetworks over a path graph, where the subsystems are a network of single integrators, also over a path graph with $m$ subsystems as analyzed in Example \ref{ex:path_netnet}. This network is illustrated in Fig. \ref{fig:pathofpaths}. From  (\ref{eq:single_int_path}), we already know that the performance of isolated subnetworks satisfies 
\begin{align}\label{1524}
 \rho(\mathbf{L}_1,\mathbf{K}_1)= \Theta \left ({ m^2}/k_1 \right ).
\end{align}
Combining \eqref{1524} and Theorem 7, we can show that 
\begin{align}
\rho_{\mathrm{nn}}(\mathbf{L}_2,\mathbf{K}_2)= \Theta \left (\dfrac{ m^2}{k_1}+ \dfrac{m^2N}{k_1}+\dfrac{mN^2}{k_2  } \right ).
\end{align}
  \begin{figure}[t]
    \centering
\includegraphics[width=6.7cm]{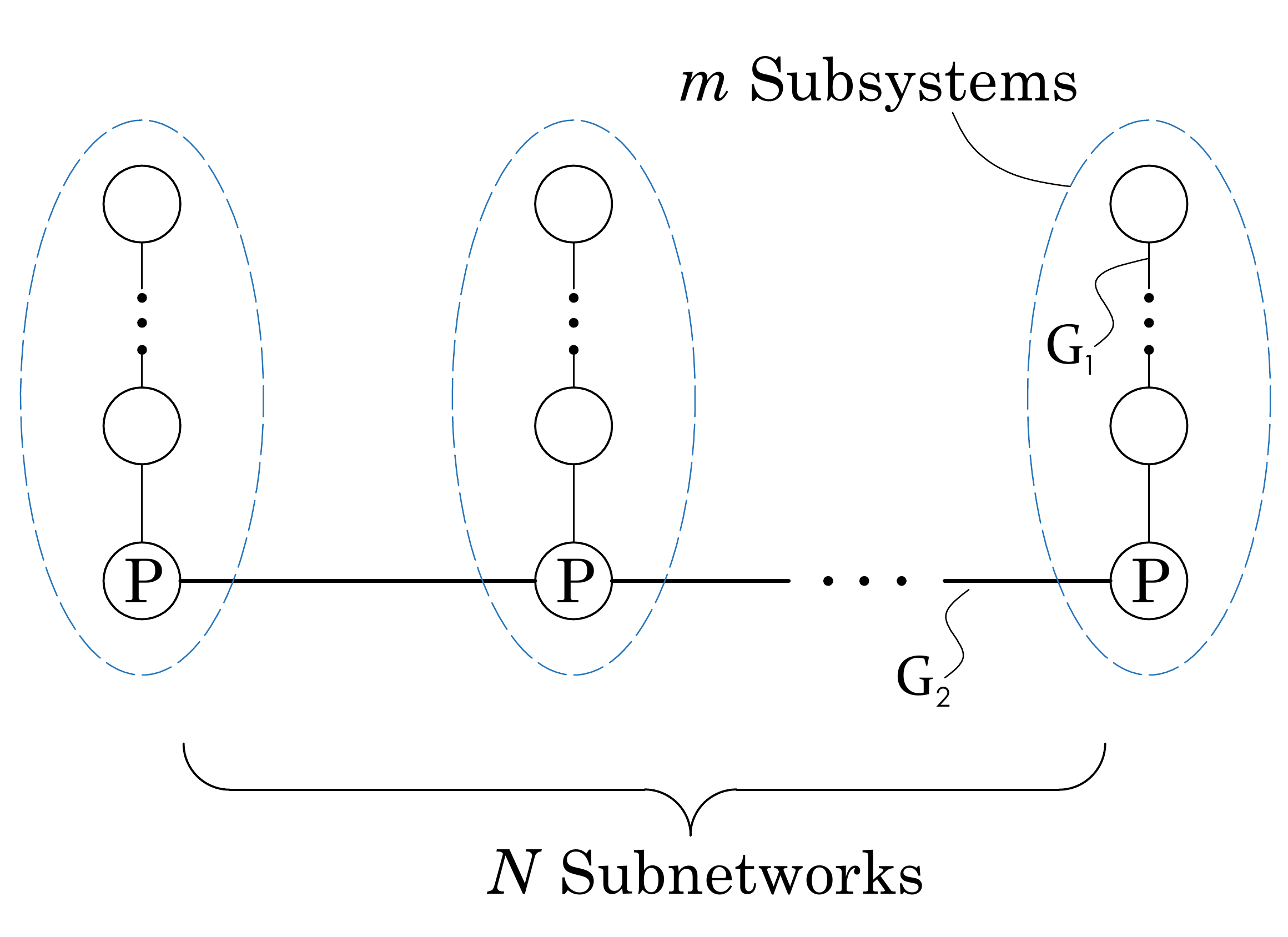}
    \caption{ Schematic of the composite network whose performance is analyzed in the continuance of Example \ref{ex:path_netnet}}
    \label{fig:pathofpaths}
\end{figure}
%$$
%\Gamma_N= \dfrac{1}{2 \pi } {\int_{\pi^2/N^2}^{4} \phi_{nn} (\lambda)\dfrac{1}{\sqrt{\lambda-\lambda^2/4}}~d\lambda}
%$$
\end{continuance}

\section{Application to Formation of Aircraft}

%In the next example, while a certain choice of $\mathbf{K}$ results in an unbounded stability region,  it gives a performance function that is neither   convex nor monotone.%  beyond its connectivity threshold. 

\begin{example}[Formation of Aircraft] \label{ex:aircraft} We consider a linearized model for the dynamics of an aircraft \cite{boyd2008lecture} expressed as 
	\begin{align*}
	\dot X& =
	\mathbf{A} X
	+
	\mathbf{B}  \begin{bmatrix}
	u_1 & u_2
	\end{bmatrix}^T+{\mathbf{E}} \begin{bmatrix}
	\xi_1 &
	\xi_2 
	\end{bmatrix}^T, 
	\end{align*}
	where $X=\begin{bmatrix}
	\mathrm{u} ~
	\mathrm{v} ~
	\dot \theta  ~
	\theta ~
	\mathrm{x}  ~
	\mathrm z 
	\end{bmatrix} ^T$ (see Appendix Q for the numerics). The variable $\mathrm{u}$ is the horizontal velocity component  from its set point and $\mathrm{v}$ is the component normal to that. The pitch angle is denoted by $\theta$. The control inputs $u_1$ and $u_2$ are the elevator angle and thrust force, respectively. The scalars $\xi_1$ and $\xi_2$ denote the wind velocity in the longitudinal and lateral directions, respectively.  We consider the formation shown in Fig. \ref{fig:aircraft}. Once each vehicle takes into account the relative distances from their neighbors (in computation of the position feedbacks), we can use these dynamics to analyze the performance of this  network with the performance output 
	$
	z=\begin{bmatrix}
	\alpha \mathrm x \,\,\,
	\beta \mathrm z
	\end{bmatrix}^T.$

	\noindent{\it Relative State-Feedback: }  We use the convex optimization toolbox CVX \cite{grant2010cvx} to find $\mathbf{K}$ for $c=0.25$ using Theorem \ref{lem:kdesign}. If $\xi_1$ and $\xi_2$ have  intensity of unity, we get  % the performance function can be evaluated as 
	\begin{align}
	\phi(\lambda)=\alpha^2 \phi_1(\lambda)+\beta^2 \phi_2(\lambda),
	\end{align}
	where $\phi_1(\lambda)$ and $\phi_2(\lambda)$  describe the magnitude of the fluctuations in the formation in  $\mathrm x$ and $ \mathrm z$ directions, respectively.   The performance functions are rational functions with the numerator and denominator of order 9. While it is guaranteed to get $\tilde \lambda(\mathbf{K}) \in [0,0.25]$, we have $\tilde \lambda (\mathbf{K})=0$. In Fig. \ref{fig:aircraftphi}, we plot $\phi_1$ and $\phi_2$, where for larger values of $\lambda$, they are different by more than an order of magnitude. The function $\phi_2(\lambda)$ is convex and decreasing, while $\phi_1 (\lambda)$ is neither strictly convex nor monotone for $\lambda>0$. This suggests that properties of these functions in general could be beyond a simple classification. 
	
	\noindent{\it Observer-Based Relative Output-Feedback: } next, we consider the observer-based output-feedback on the last two states of each subsystem (i.e., horizontal and vertical relative positions). For the value of $\mathbf{K}$ we reuse its value from the previous design. We choose  observer gain $\mathbf{F}$ for $c=0.25$ using Theorem \ref{lem:kdesign_observable} and find that 
	\begin{align}
	\phi(\lambda)=\alpha^2 \hat \phi_1(\lambda)+\beta^2 \hat \phi_2(\lambda),
	\end{align}
	where these two functions are also depicted in Fig. \ref{fig:aircraftphi}. In this case,  $\tilde \lambda (\mathbf{F})=0$ as well. In Fig. \ref{fig:aircraft_traj}, we demonstrate two sample longitudinal output plots based on these two designs, where we have $5$ planes that are supposed to travel with $\Delta \mathrm{x}=0.6$.  The graph is a path with weights of $4$ and identical disturbance samples are fed into the subsystems in two cases. The different level of fluctuations is justifiable upon   comparison of the values of $\phi_1$ and $\hat \phi_1$ in Fig. \ref{fig:aircraftphi}. 
	
	\begin{figure}
		\centering
		\includegraphics[width=6.7cm]{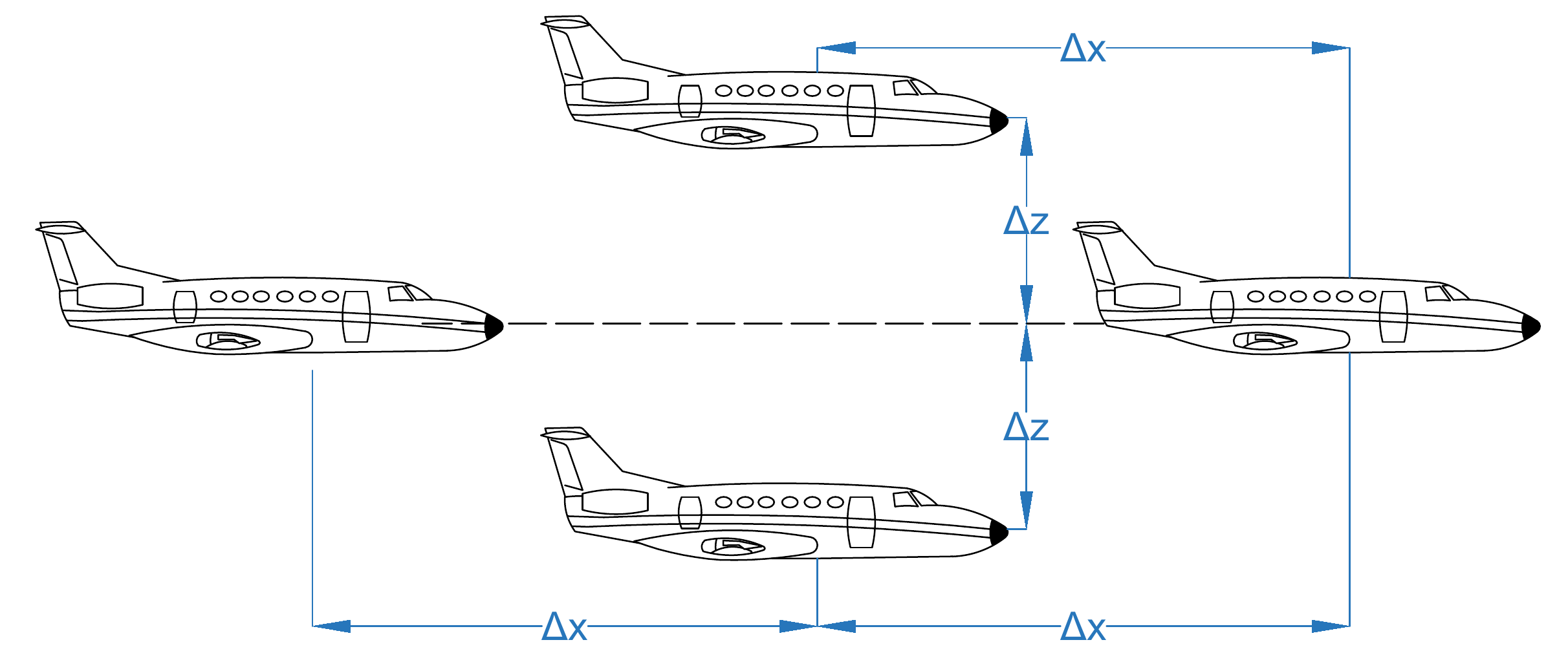}
		\caption{The formation of interest in Example \ref{ex:aircraft}} 
		\label{fig:aircraft}
	\end{figure}

	\begin{figure}[t]
		\centering
		\includegraphics[width=7.6cm]{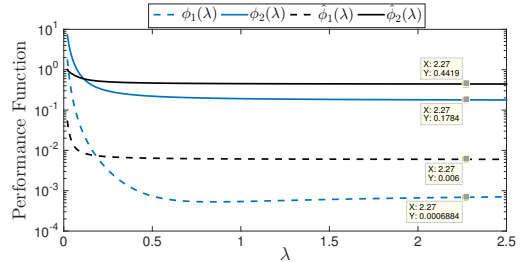}
		\caption{The performance functions for Example \ref{ex:aircraft}} 
		\label{fig:aircraftphi}
	\end{figure}
	
	\begin{figure}[t]
		\centering
		\includegraphics[width=7cm]{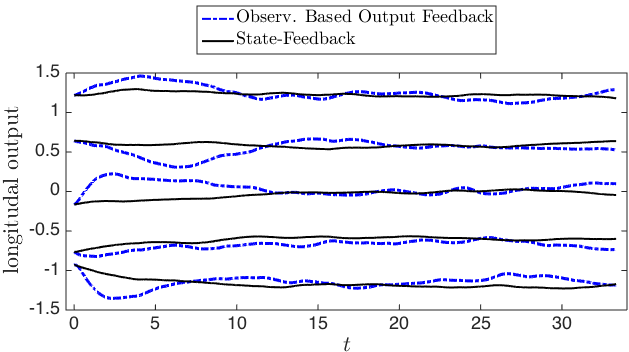}
		\caption{The sample outputs based on the designs in Example \ref{ex:aircraft}} 
		\label{fig:aircraft_traj}
	\end{figure}
	
	%  \begin{figure}
	%    \centering
	%\includegraphics[width=2.7cm]{images/interconnectiongraph.pdf}
	%    \caption{The interconnection graph in Example \ref{ex:aircraft}, where the consecutive longitudinal  distances of the vehicles from left to right is $1$} 
	%    \label{fig:graph}
	%\end{figure}
	
	%  \begin{figure}
	%    \centering
	%\includegraphics[width=6cm]{images/phi1.png}
	%    \caption{A plot of $\phi_1(\lambda)$ showing it is neither convex nor decreasing.} 
	%    \label{fig:phi1}
	%\end{figure}
	
\end{example}

\section{Conclusion and Discussion} 

We brought a unifying framework for performance analysis of a class of networked control systems. The resulting spectral expressions let us derive bounds and scaling laws for the performance of the system. We would like to include a a number final remarks: 

\noindent {\it (i)} The spectral expressions for the performance measure can be used to find the optimal values of  feedback gain $\mathbf{K}$.  In fact, for large networks, solving   the Lyapunov equation for the $\mathcal{H}_2$ performance measure once the value of feedback gain is updated could be computationally expensive. Instead, suppose that we find the spectral expressions for the performance measure for a fixed graph. Then, our objective function will be a scalar function of the feedback gain. The resulting problem can be effectively approached using general nonlinear problem methods. This approach is also useful when solving for optimal observer gains $\mathbf{F}$ or feedback gains for composite networks $\mathbf{K}_1$ and $\mathbf{K}_2$ when the graph is fixed. The gains derived from the linear matrix inequalities  given in Section \ref{sec:K} can be used as a starting point of the optimization procedure.

\noindent {\it (ii)} %Our spectral analysis can be extended to characterize the magnitude of the input signals as well. In fact,
We can derive similar spectral expressions for the variance of the  control input that is consumed throughout the network in the steady-state, which is given by 
\begin{align}
\rho_u:=\lim_{t\rightarrow \infty}\mathbb{E}\left \{ \|u(t) \|_2^2 \right \}
\end{align}
Then, we can show that 
\begin{align}\label{eq:u_function}
\rho_u({\mathbf{L},\mathbf{K}})=\sum_{i=2}^N \phi_u(\lambda,\mathbf{K}).
\end{align}
for rational input function $\phi_u(\lambda,\mathbf{K})$ given by 
$$
\mu(\lambda,\mathbf{K}):=\mathrm{Tr} \left (\lambda^2 \mathbf K \mathbf H {\mathbf{P}}(\lambda,\mathbf{K})\mathbf H ^T \mathbf K^T \right ).
$$ 
The map ${\mathbf{P}}(\lambda,\mathbf{K}) \succ 0$ is the   solution to    (\ref{eq:sub_lyap}).   %R's entries. %Furthermore, for a network over loopless graph that reaches consensus the input measure can be evaluated as
%\begin{align}\label{eq:\mathbf{U}_function_c}
%\rho_u({\mathbf{L},\mathbf{K}})=\sum_{i=2}^N \mu(\lambda_i,\mathbf{K}).
%\end{align}
%\end{theorem}
For instance, we can show  
the input functions   for networks of single-integrators and double-integrators are given by 
	 $$
	 \phi_u(\lambda,\mathbf{K})=\dfrac{k\lambda }{2},~~~~ \phi_u(\lambda,\mathbf{K})=\dfrac{k_1}{2 k_2 }+\dfrac{k_2  \lambda}{2},
	 $$
respectively. The developments in this paper which has to do with the performance functions can be applied to the input functions as well (e.g. asymptotic   control input over a path).

{\emph{(iii)}  In   cases that symbolic evaluation of the performance functions is computationally prohibitive, an alternative option is to conduct regression to estimate   the coefficients of these rational  performance functions numerically.   }

\bibliography{mybib}

\newpage

\section*{Appendix A}

\noindent{\it Proof of Theorem \ref{thm:first_main}:}
%\begin{proof}
	Let us define $m=m_1+m_3$
	The transfer matrix from disturbance and noise $[\xi^T,\eta^T]^T$ to consensus output $\nu$ can be expressed as  
	\begin{align*} 
	& {\mathbf{G}}(s)=%=({\mathbf{M}_N} \otimes \mathbf{I}_{m_2}){\mathbf{G}}(s)= \\
	%({\mathbf{M}_N} \otimes \mathbf{I}_{m_2})(\mathbf{U} \otimes \mathbf{I}_{m_2})\mathrm{diag}(\hat H_1,\dots ,\hat H_N)(\mathbf{U}^T\otimes \mathbf{I}_{m_1})=\\
	({\mathbf{M}_N} \mathbf{U} \otimes \mathbf{C})\mathrm{diag}\left (\tilde {\mathbf{G}}_1,\dots ,\tilde {\mathbf{G}}_N \right )(\mathbf{U}^T\otimes \mathbf{I}_{m}),
	\end{align*}
	where $\tilde {\mathbf{G}}_i(s)$ is the transfer matrix from $\left [\chi_i^T,\gamma_i^T \right ]^T$ to $r_i$.  
	This lets us compute the following quantity
	\begin{align*}
	&\mathbf{G}^*(j\omega){\mathbf{G}}(j\omega)=(\mathbf{U} \otimes \mathbf{I}_{m})\mathrm{diag}\left (\tilde {\mathbf{G}}_1^*,\dots ,\tilde {\mathbf{G}}_N^* \right )\\
	& \left ( \mathbf{U}^T{\mathbf{M}_N}{\mathbf{M}_N}\mathbf{U} \otimes \mathbf{C}^T \mathbf{C} \right)\mathrm{diag}\left (\tilde {\mathbf{G}}_1,\dots ,\tilde {\mathbf{G}}_N \right )\left (\mathbf{U}^T\otimes \mathbf{I}_{m} \right ).
	\end{align*}
	The matrix $\mathbf{U}^T{\mathbf{M}_N}{\mathbf{M}_N}\mathbf{U}$ is simply given by 
	$$
	\mathbf{U}^T{\mathbf{M}_N}{\mathbf{M}_N}\mathbf{U}=\mathrm{diag}(0,1,\dots,1) \in \R^{N\times N}.
	$$
	Taking  ${1}/(2\pi)\int_{-\infty}^{\infty}\mathrm{Tr}(.)~d\omega$ from the both sides results in %of that inequality gives
	\begin{align*}
&	\rho({\mathbf{L},\mathbf{K}})= 
	\dfrac{1}{2\pi}\int_{-\infty}^{\infty}\mathrm{Tr}\big ((\mathbf{U} \otimes \mathbf{I}_{m})\\ & \mathrm{diag}\big(0,\tilde {\mathbf{G}}_2^* \mathbf{C}^T\mathbf{C} \tilde {\mathbf{G}}_2,\dots ,\tilde {\mathbf{G}}_N^*\mathbf{C}^T\mathbf{C}\tilde {\mathbf{G}}_N\big)  
	(\mathbf{U}^T\otimes \mathbf{I}_{m})\big )~d\omega.
	\end{align*}
	Due to cyclic property of the trace, the first and last matrix in the trace argument cancel out and we get 
	$$
	\dfrac{1}{2\pi}\int_{-\infty}^{\infty}\mathrm{Tr}\left (\tilde {\mathbf{G}}_i^* \mathbf{C}^T\mathbf{C} \tilde {\mathbf{G}}_i \right )~d\omega=\left \|\mathbf{C}\tilde {\mathbf{G}}_i(s) \right \|_{\mathcal{H}_2}^2:=\phi(\lambda_i,\mathbf{K}).
	$$
	The last  $\mathcal{H}_2$ norm term can be computed using the state-space formulation of systems $\Sigma_i$. This will be  the Lyapunov equation \eqref{eq:sub_lyap} \cite{doyle1989state}. Therefore, we have managed to prove 
	$$ 
	\rho({\mathbf{L},\mathbf{K}})=\sum_{i=2}^N \left \|\mathbf{C}\tilde {\mathbf{G}}_i(s) \right \|_{\mathcal{H}_2}^2=\sum_{i=2}^N \phi(\lambda_i,\mathbf{K}).  
	$$
	%Therefo, boundedness of $\rho({\mathbf{L},\mathbf{K}})$ 
	%dictates asymptotic stability of  $\Sigma_2$ to $\Sigma_N$
	
	Next, we prove that $\phi(\lambda,\mathbf{K})$ is a rational function. The Lyapunov equation (\ref{eq:sub_lyap}) upon vectorization becomes
	\begin{align*}\label{eq:vec_lyp}
	\begin{adjustbox}{max width=250 pt}
	$
	\left({\mathbf{A}}_\lambda \otimes \mathbf{I}_n+\mathbf{I}_n\otimes {\mathbf{A}}_\lambda \right)\mathrm{vec}(\mathbf{P})=
	-\mathrm{vec}\big ({\mathbf{E}} {\mathbf{E}} ^T+\lambda^2 \sigma^2 {\mathbf{B}}{\mathbf{K}}  ({\mathbf{B}}  {\mathbf{K}})^T \big ).  
	$
	\end{adjustbox}
	\end{align*}
	Using the Cramer's rule, for $j =1,2,\dots,n^2$, we may compute the $j$'th element of $\mathrm{vec}(\mathbf{P})$ as
	\begin{align}
	\mathrm{vec}(\mathbf{P})_j=\dfrac{\mathrm{det}(({\mathbf{A}}_\lambda \otimes \mathbf{I}_n+\mathbf{I}_n\otimes {\mathbf{A}}_\lambda)_{-j})}{\mathrm{det}({\mathbf{A}}_\lambda \otimes \mathbf{I}_n+\mathbf{I}_n\otimes {\mathbf{A}}_\lambda )},
	\end{align}
	where $(\mathbf{D})_{-j}$ is the matrix derived by replacing column $j$ of $\mathbf{D}$ with $-\mathrm{vec}({\mathbf{E}} {\mathbf{E}} ^T+\lambda^2 \sigma^2 {\mathbf{B}}{\mathbf{K}}  ({\mathbf{B}}  {\mathbf{K}})^T)$. Both numerator 
	and denominator are polynomials of $\lambda$ with coefficients that are polynomials of the elements of $\mathbf{K}$. Therefore, the same conclusions holds about  $\phi(\lambda,\mathbf K)=\mathrm{Tr}(\mathbf{C}{\mathbf{P}}(\lambda,\mathbf{K})\mathbf{C}^T)$. %The form of spectral function that is shown in (\ref{eq:spectral_form}) is followed by the definition of $\phi(\lambda)$. 
	%Now, we look at the performance measure over loopless graphs..
%\end{proof}

\section*{Appendix B}

\noindent{\it Proof of Theorem \ref{lem:obs}:}
%
%\begin{proof} 
	 If we apply control law (\ref{eq:laplacian_control_c}) for the new subsystem $\hat S_i$ with observer gain $\hat {\mathbf{F}}=[-\mathbf{F},\mathbf{F}]$ (partitioned based on $\hat y_i$), we have the following formula
	$$
	\hat u_i=\mathbf{F}\mathbf{H} \Big  (\sum_{j\in \mathcal{N}_i} a_{ij} ({x_i}-x_j)-\sum_{j\in \mathcal{N}_i} a_{ij} ({\hat x_i}-\hat x_j)\Big  ).
	$$
	This means that  $\Sigma_i$ corresponding to the dynamics  are 
	$$
	\begin{adjustbox}{max width=240pt}$
	\begin{bmatrix}\dot r_i \\ 
	\dot {\hat r}_i \end{bmatrix}  =
	\begin{bmatrix}{\mathbf{A}}  & -{\mathbf{BK}}  \\ \lambda \mathbf{F} \mathbf{H} & {\mathbf{A}}-{\mathbf{BK}}-\lambda \mathbf{F} \mathbf{H} \end{bmatrix}
	\begin{bmatrix} r_i \\  {\hat r}_i \end{bmatrix} +\begin{bmatrix} {\mathbf{E}} \\ \mathbf{0} \end{bmatrix}  \chi_i+\begin{bmatrix} {\mathbf{0}} \\ -\sigma  {\lambda_i} \mathbf{F}  \end{bmatrix}   \gamma_i
	$
	\end{adjustbox}
	$$
	Defining the error as $e_i:=r_i -\hat r_i$, we get that
	\begin{align} \label{eq:decoupled_r_e}
	\begin{bmatrix}\dot r_i \\ \dot e_i \end{bmatrix}  =
	\begin{bmatrix}{\mathbf{A}}-\mathbf{BK}  & {\mathbf{BK}}  \\ \mathbf{0} & {\mathbf{A}}-\lambda_i{\mathbf{F}} \mathbf{H} \end{bmatrix}
	\begin{bmatrix} r_i \\  e_i \end{bmatrix} +
	\begin{bmatrix} {\mathbf{E}} & \mathbf{0} \\ {\mathbf{E}} &    \sigma \lambda_i  \mathbf F \end{bmatrix} \begin{bmatrix} \chi_i \\  \gamma_i \end{bmatrix}.
	\end{align}
	The subsystems in the consensus problem have reduced to the familiar decoupled Leunberger  observer/regulator form (e.g. see \cite{hespanha2005lecture}). Therefore,  we need to simultaneously have: ${\mathbf{A}}-\lambda_i \mathbf{F}\mathbf{H}$ for $i=2,\dots,N$ and ${\mathbf{A}}-{\mathbf{BK}}$ to be Hurwitz. Then, the  corresponding $\Sigma_i$ is asymptotically stable for $i=2,\dots,N$ and the network reaches the consensus. 
%\end{proof}

\section*{Appendix C}

\noindent{\it Proof of Theorem \ref{thm:estimation}:}
%
%\begin{proof}
	If we consider the dynamics of $e_i$ in \eqref{eq:decoupled_r_e}, they are identical to dynamics of $\Upsilon_i$ in \eqref{eq:decoupled_obs}. The rest of the proof is similar to Theorem \ref{thm:first_main} once we replace $\Sigma_i$ with $\Upsilon_i$. 
%\end{proof}

\section*{Appendix D}

\noindent{\it Proof of Theorem \ref{eq:lip_const}:}
%
%\begin{proof} 
	  The definition of $\tilde \lambda(\mathbf{K})$ implies that for all $\lambda>\tilde \lambda(\mathbf{K})$, subsystems $\Sigma_2$ to $\Sigma_N$ are asymptotically stable. Therefore, the performance function is bounded. Because  $\phi(\lambda)$ is rational, it is analytic  everywhere in its domain, including this interval.   %The dynamics  of $\Sigma_i$ without the noise are 
	%$
	%\dot {r}_i=({\mathbf{A}}-\lambda_i{\mathbf{B}}{\mathbf{K}})r_i.
	%$
	%  Its root locus is continuous  in terms of $\lambda_i$. Thus,  definition of $\tilde \lambda(\mathbf{K})$ and its positivity imply that the eigenvalues of the closed-loop for $\lambda_i \in [0,\tilde \lambda(\mathbf{K})]$ must have crossed the imaginary axis first time 
	%at $\tilde \lambda(\mathbf{K})$. Otherwise, it is not the infimum based on its definition. 
	%%The same argument works for the input function. 
%\end{proof} 

\section*{ Appendix E}
 	\noindent{\it Proof of Theorem \ref{lem:kdesign}:}
% 	]
 	 First, for a linear time invariant control system the feasibility of the linear matrix inequality and the stabilizability are equivalent \cite{boyd1994linear}. The second part of the claim is a special case of Theorem 11 in \cite{li2011h} with only accounting for the stabilizablity of the subsystems, so we do not repeat the proof in this manuscript. The converse argument holds because if $\tilde \lambda(\mathbf K)<\infty$, then for $\mathbf K^*=2\tilde\lambda (\mathbf K) \mathbf{K}$, $\mathbf A-\mathbf{BK}^*$ is Hurwitz; i.e., $(\mathbf{A}, \mathbf{B})$ is stabilizable.  

 \section*{Appendix F}
 
%\begin{proof}[
\noindent{\it Proof of Theorem \ref{lem:kdesign_observable}:}
%	]
 Due to duality of between the stabilizability and detectability, if the pair $({\mathbf{A}},{\mathbf{H}})$ is detectable, then $({\mathbf{A}}^T,{\mathbf{H}}^T)$ is stabilizable. Now, we can use the same argument as Theorem \ref{lem:kdesign} to complete the proof. 
% \end{proof}

%\begin{proof}[Proof of Lemma \ref{lem:one}]
%Based on the range of eigenvalues of $\mathbf{L}$, we can see that choosing $\delta_{\max}=\lambda_{\mathrm{max}}/\lambda_{\mathrm{min}}$ gives a continuum of  feedback gains, for which every $\Sigma_i$ for $i=2,\dots,N$ is asymptotically stable. 
%\end{proof} 

 \section*{Appendix G}

%  \begin{proof}[
  \noindent{\it Proof of Theorem \ref{thm:right_mp}:} %]
  	 Consider the control system 
\begin{align} \label{eq:systemlike} \left \{
\begin{array}{l}
  \dot x= {\mathbf{A}}x+ {\mathbf{B}} u+{\mathbf{E}}\xi, \\
   z=\begin{bmatrix}{\mathbf{C}} x \\ \epsilon u \end{bmatrix}.  
 \end{array}  \right .
\end{align}
 Saberi et. al. \cite{saberi1995h} have shown that the minimum value of the $\mathcal{H}_2$ norm for this system is
$
\gamma^*(\epsilon)=\sqrt{\mathrm{Tr} \left ( {\mathbf{E}} {\mathbf{P}}_\epsilon {\mathbf{E}}^T \right )},
$
where ${\mathbf{P}}_{\epsilon}$ can be computed as the solution to an algebraic Riccati  equation
\begin{align} \label{eq:ric}
 {\mathbf{A}}^T {\mathbf{P}}_\epsilon +{\mathbf{P}}_\epsilon {\mathbf{A}}+\mathbf{C}^T \mathbf{C}-\dfrac{1}{\epsilon^2}{\mathbf{P}}_\epsilon \mathbf{B} \mathbf{B}^T {\mathbf{P}}_ \epsilon=0.
\end{align}
 Moreover, it has been shown that  if $({\mathbf{A}},\mathbf{B})$ is stabilizable and $({\mathbf{A}},\mathbf{C})$ is detectable,  then  ${\mathbf{P}}_\epsilon$ converges to zero if and only if the mentioned transfer matrix is right-invertible and minimum-phase. Now, we should note that the performance function is the $\mathcal{H}_2$ norm squared of a system similar to (\ref{eq:systemlike}), expect that we have
$ \mathbf{B} \rightarrow \lambda \mathbf{B} $. 
Under this modification, the limiting case for ${\mathbf{P}}_\epsilon$ in  Riccati equation (\ref{eq:ric}) does not change. %, hence we can see that it is still a tight bound.

\section*{Appendix H}

\noindent{\it Proof of Theorem \ref{thm:composite}:}
%
%\begin{proof} 
	 Let us denote  orthonormal eigendecompostion of   Laplacian $\mathbf{L}_2$ by $\mathbf{L}_2 =\mathbf{U}_2 \mathbf{\Lambda}_2 \mathbf{U}_2^T$. The decoupled system $\Sigma_i$ in this case is
	$$
	\dot {r}^{(i)}=\left (\tilde {\mathbf{A}} -\lambda _i(\mathbf{L}_2) \tilde {\mathbf{B}}{\mathbf{K}}_2 \tilde {\mathbf{H}} \right ) r^{(i)}+ \tilde {\mathbf{E}}\, \chi^{(i)}.
	$$
	Let us call the transfer matrix from $\chi^{(i)}$ to  $ {r}^{(i)}$ by $\tilde {\mathbf{G}}^{(i)}$. Then, the transfer matrix from  disturbance  $\xi$ to  performance output $\nu_{\mathrm{nn}}$  in the case of network of networks can be written as 
	\begin{align*} 
	&	{\mathbf{G}}_\mathrm{nn}(s)= \\ &\begin{adjustbox}{max width=250 pt}$
	%=({\mathbf{M}_N} \otimes \mathbf{I}_{m_2}){\mathbf{G}}(s)= \\
	%({\mathbf{M}_N} \otimes \mathbf{I}_{m_2})(\mathbf{U} \otimes \mathbf{I}_{m_2})\mathrm{diag}(\hat H_1,\dots ,\hat H_N)(\mathbf{U}^T\otimes \mathbf{I}_{m_1})=\\
	(\mathbf{M}_{Nm}\otimes \mathbf{C})( \mathbf{U}_2\otimes \mathbf{I}_{mn})\mathrm{diag}(\tilde {\mathbf{G}}^{(1)},\dots ,\tilde {\mathbf{G}}^{(N)})(\mathbf{U}_2^T\otimes \mathbf{I}_{mm_1}).$
	\end{adjustbox}
	\end{align*}
	This lets us compute the following quantity
	\begin{align*}
&	({\mathbf{G}}_\mathrm{nn})^*(j\omega){\mathbf{G}}_\mathrm{nn}(j\omega)=\\ & (\mathbf{U} \otimes \mathbf{I}_{m_1})\mathrm{diag}((\tilde {\mathbf{G}}^{(1)})^*,\dots ,(\tilde {\mathbf{G}}^{(N)})^*)\\
	& (\mathbf{U}_2^T \otimes \mathbf{I}_{nm}) (\mathbf{M}_{Nm} \otimes \mathbf{C}^T \mathbf{C}) (\mathbf{U}_2\otimes \mathbf{I}_{nm})\\ 
	& \mathrm{diag}(\tilde {\mathbf{G}}_1,\dots ,\tilde {\mathbf{G}}_N)(\mathbf{U}^T\otimes \mathbf{I}_{m_1}).
	\end{align*}
	We take the trace and move the first two terms  of the trace argument to the right to get
	\begin{align*}
	\mathrm{Tr} \left ({\mathbf{G}}_\mathrm{nn}^*{\mathbf{G}}_\mathrm{nn} \right)= &  \mathrm{Tr}  \Big (  \mathrm{diag}((\tilde {\mathbf{G}}^{(1)})^*,\dots , (\tilde {\mathbf{G}}^{(N)})^*) \\
	& (\mathbf{U}_2^T \otimes \mathbf{I}_{nm}) (\mathbf{M}_{Nm} \otimes \mathbf{C}^T \mathbf{C}) (\mathbf{U}_2\otimes \mathbf{I}_{nm}) 
	\\ & \mathrm{diag}(\tilde {\mathbf{G}}^{(1)} ,\dots ,\tilde {\mathbf{G}}^{(N)}) \Big ) .
	\end{align*}
	%, we can show that 
	% \begin{align*}
	%& \rho_{nn}(\mathbf{L}_2,\mathbf{K}_2)= \\
	%&  \mathrm{Tr} \left \{  (\mathbf{M}_{Nm} \otimes \mathbf{C}) (\mathbf{U}_2\otimes \mathbf{I}_{nm})  \mathrm{diag} \left (\tilde {\mathbf{P}}(\lambda_i(\mathbf{L}_2),\mathbf{K}_2) \right ) \right . \\
	%&~~~~~~~~~~~~~~~~~ ~~~~~~~~~~~~~~ \left .(\mathbf{U}_2^T\otimes \mathbf{I}_{nm}) (\mathbf{M}_{Nm} \otimes \mathbf{C}^T) \right \},
	%\end{align*}
	%where $\tilde {\mathbf{P}}(\lambda_i,\mathbf{K})$ is the solution to the Lyapunov equation X once we do the replacement the matrices defined in X. 
	%If we move the two matrices on the right side of the trace argument to the right, we need to evaluate 
	The intermediate term can be simplified according to 
	\begin{align*}
	& \left (\mathbf{U}_2^T \otimes \mathbf{I}_{nm} \right ) \left (\mathbf{M}_{Nm} \otimes \mathbf{C}^T \mathbf{C} \right) \left (\mathbf{U}_2\otimes \mathbf{I}_{nm}\right) \\ 
	& =\mathrm{diag}\Big ( \mathbf{M}_m, \underbrace{\mathbf{I}_m,\dots \mathbf{I}_m}_{N-1 \text{ times }}  \Big )  \otimes \mathbf{C}^T\mathbf{C}.
	\end{align*}
	Hence, we can further write 
	\begin{align*}
	\mathrm{Tr} \left ({\mathbf{G}}_\mathrm{nn}^*{\mathbf{G}}_\mathrm{nn} \right)=\, & \mathrm{Tr} \Big ( (\tilde {\mathbf{G}}^{(1)})^*(\mathbf{M}_m \otimes \mathbf{C}^T)(\mathbf{M}_m \otimes \mathbf{C}) \tilde {\mathbf{G}}^{(1)} \Big ) \\
	& +\sum_{i=2}^{N} \mathrm{Tr} \Big (  (\tilde {\mathbf{G}}^{(i)})^*(\mathbf{I}_m \otimes \mathbf{C}^T)(\mathbf{I}_m \otimes \mathbf{C}) \tilde {\mathbf{G}}^{(i)} \Big ).
	\end{align*}
	If we take the map ${1}/(2\pi)\int_{-\infty}^{\infty}\mathrm{Tr}(.)~d\omega$ from the sides% of this equality, we get
	$$
	\rho_{\mathrm{nn}}(\mathbf{L}_2,\mathbf{K}_2)=\left \|(\mathbf{M}_m \otimes \mathbf{C}) \tilde {\mathbf{G}}^{(1)}\right \|_{\mathcal{H}_2}^2+\sum_{i=2}^N \left \|(\mathbf{I}_m \otimes \mathbf{C}) \tilde {\mathbf{G}}^{(i)} \right \|_{\mathcal{H}_2}^2.
	$$
	For $i=1$, $\lambda_1(\mathbf{L}_2)=0$   and the system $\Sigma_1$ will have a transfer matrix from the disturbance to output  $(\mathbf{M}_m \otimes \mathbf{C}) \tilde {\mathbf{G}}^{(1)}$. Moreover, in this case $\Sigma_1$ has the closed-loop dynamics of the subnetworks. Therefore, we inspect that 
	$$
	\left \|(\mathbf{M}_m \otimes \mathbf{C}) \tilde {\mathbf{G}}^{(1)} \right \|_{\mathcal{H}_2}^2=\rho(\mathbf{L}_1,\mathbf{K}_1).
	$$
	Additionally, we observe that for $i=2,\dots,N$, we have 
	$$
	\left \|(\mathbf{I}_m \otimes \mathbf{C}) \tilde {\mathbf{G}}^{(i)} \right \|_{\mathcal{H}_2}^2:=\phi_{\mathrm{nn}}(\lambda_i(\mathbf{L}_2),\mathbf{K}_2),
	$$
	provided that $\phi_{\mathrm{nn}}$ is the performance function computed using    matrices  in (\ref{eq:replacements}). 
%\end{proof} 

\section*{Appendix I}

\noindent{\it Proof of Theorem \ref{thm:min_netnet}:}
%
%\begin{proof} 
	 If $\mathbf{K}_2=\alpha \mathbf{K}_1$, then we can consider the network of network to be a single network with feedback gain $\mathbf K_1$, over a graph $\mathcal{G}_3=\mathcal{G}_1 \cup \mathcal{G}_2$. The weights of links in $\mathcal{G}_1$ are preserved, while the weights of the links in $\mathcal{G}_2$ are scaled by $\alpha$. 
	% 	based on the procedure for construction of these composite networks, we can see that  the whole network of networks can be described with a single composite weighted graph, where the parameter $\alpha$ should be multiplied by the weights between the two subnetworks. In fact, if $\mathbf{K}_2=\alpha \mathbf{K}_1$, we can scale the weight of the links in the higher level graph $\mathcal{G}_2$ by $1/\alpha$. Then the equivalent network would have a simple network that is built using the feedback gain $\mathbf{K}_1$.
	Hence, if we increase the weights in the higher level network (equivalently, the eigenvalues of $\mathbf{L}_2$), at some point the second smallest eigenvalue of  equivalent Laplacian $\mathbf L_3$ will pass $\tilde \lambda(\mathbf{K}_1)<\infty$. 
%\end{proof}

\section*{Appendix J}

\noindent{\it Proof of Corollary \ref{cor:netnet}:}
%\begin{proof} 
	 Following the same lines as in the proof of Theorem \ref{thm:min_netnet}, if the minimum connectivity threshold is zero, for any choice of $\mathbf{K}_2=\alpha \mathbf{K}_1$, the network with a single equivalent graph and feedback gain $\mathbf{K}_1$ has zero minimum connectivity threshold, while the equivalent Laplacian would always have a nonzero $\lambda_2$. Hence, the connectivity threshold in terms of the eigenvalues of $\mathbf{L}_2$ is zero as well.  
%\end{proof} 

\section*{Appendix K}

\noindent{\it Proof of Theorem \ref{thm:laplacian_loopless}:}
%
%\begin{proof}
	First we proof an inequality that is an extension of one   in \cite{zhou2008sum} in the case of $f(\lambda)=\lambda^{\alpha}$ for $\alpha \notin [0,1]$. For a  continuously differentiable convex function $f(x)$ and a Laplacian $\mathbf{L}$ with $M$ edges and maximum degree $\Delta $,  we show that 
	\begin{align}  \label{eq:ineqs}
	%\sum_{i=2}^N \phi(\lambda_i) \geq  \phi(1+\Delta )+(N-2) \phi\left (\frac{2M -1-\Delta}{N-2}\right ),
	\sum_{i=2}^N f(\lambda_i)  \geq  f(1+\Delta )+(N-2) f\left (\frac{2M -1-\Delta}{N-2}\right ),
	\end{align} 
	and the equality holds if and only if $\mathcal{G}$ is complete or star. The steps provided in the proof of this lemma are essentially the same steps reported for Theorem 3 in \cite{zhou2008sum} (only for power functions). Since $f$ is convex and continuous, we use Jensen's inequality to write  %we may write
	$$
	f\left ( \frac{1}{N-2} \sum_{i=2}^{N-1}\lambda_i \right ) \leq \frac{1}{N-2} \sum_{i=2}^{N-1} f\left (  \lambda_i \right ),
	$$
	where if the function $f(\lambda)$ is not affine, then the equality holds if and only if $\lambda_1=\dots=\lambda_{N-1}$. This implies we can write
	\begin{align*}
	\sum_{i=2}^N f(\lambda_i )& \geq f\left (  \lambda_N\right )+(N-2) f\left ( \frac{1}{N-2} \sum_{i=2}^{N-1}\lambda_i \right ) \\
	&=f\left (  \lambda_N\right )+(N-2) f\left (\frac{2M -\lambda_N}{N-2}\right ):=s(\lambda_N),
	\end{align*}
	where the auxiliary function $s(x)$ is defined as 
	\begin{align}\label{eq:sx}
	s(x):=f\left ( x\right )+(N-2) f\left (\frac{2M -x}{N-2}\right ).
	\end{align}
	Because $f(x)$ is continuously differentiable, so is $s(x)$ and
	$$
	s'(x)=f'\left ( x\right )-f'\left ((2M -x)/(N-2)\right ).
	$$
	The function $f(x)$ is convex, thus $f'(x)$ is nondecreasing. Then, $s(x)$ is strictly increasing, since for any $x\geq 2M/(N-1)$
	\begin{align*}
	s'(x)  & \geq f'\left ( \frac{2M}{N-1}\right )- f'\left ( \frac{2M-{2M}/(N-1)}{N-2}\right )  \\
	&\geq f'\left ( \frac{2M}{N-1}\right )- f'\left ( \frac{2M}{N-2}\right ) >0,
	\end{align*}
	In an unweighted graph, $\lambda_N \geq 1+\Delta \geq {2M}/(N-1)$ 
	(see \cite{zhou2008sum} and also \cite{brouwer2008lower}). Therefore, 
	\begin{align}
	\sum_{i=2}^N f(\lambda_i ) \geq s(\lambda_N) \geq s(1+\Delta), 
	\end{align}
	which proves \ref{eq:ineqs}. Applying this on a convex $\phi(\lambda)$,   \eqref{eq:genbound} is followed.  The equality holds if and only if $\lambda_2=\dots=\lambda_{N-1}$ and $\lambda_N=1+\Delta$, which happens if and only if $\mathcal{G}$ is either complete or star (again, see both \cite{zhou2008sum} and \cite{brouwer2008lower}). 
	% For Class \ref{cs:1}, since $\mu$ and $\phi_\gamma$ are convex, the result of lemma  \ref{prop:laplacian_loopless} is also applicable  (for loopless graphs) by replacing $(\rho,\phi)$ with $(\rho_u,\mu)$ or $(\rho_\gamma,\phi_\gamma)$. 
	%For loopy graphs, we start from (\ref{eq:interlace}), where the right hand side belongs to a  loopless $\hat L$ with $N+1$ nodes. $f(x)$ is convex, thus  Lemma \ref{prop:laplacian_loopless} gives 
	%\begin{align*} 
	%&\sum_{i=1}^N f(\lambda_i) \geq \sum_{i=2}^{N+1} f(\lambda_i(\hat L)) \geq \\ & f\left (1+\Delta( \mathcal{ \hat G} )\right )+(N+1-2) f\left (\frac{2M -1-\Delta( \mathcal{ \hat G})}{N+1-2}\right ),
	%\end{align*} 
	%where we have used the facts that $f(x)$ is decreasing and number of edges $M$ in $\mathcal{G}$ and $\hat G$ is identical  (while $\Delta$ could be different). 
	%%that is the claimed inequality. 
	%We apply this inequality to  $\phi(\lambda)$'s that are not only convex but also decreasing, and the claim is followed. 
%\end{proof}

\section*{Appendix L}

\noindent{\it Proof of Theorem \ref{thm:laplacian_loopless_weighted}:}
%
%\begin{proof} 
	Consider any convex function $f$. We start from Jensen's inequality in the form of 
	$$
	f\left ( \frac{1}{N-1} \sum_{i=2}^{N}\lambda_i \right ) \leq \frac{1}{N} \sum_{i=2}^{N} f\left (  \lambda_i \right ).
	$$ 
	Because the eigenvalues sum to $2W$, replacing $f$ with $\phi$ gives us the   inequality \eqref{eq:genbound_2}. The equality holds if and only if $\lambda_2=\dots=\lambda_{N}$ that happens if and only if $\mathcal{G}$ is complete graph with identical weights. 
%\end{proof}

\section*{Appendix M}

\noindent{\it Proof of Theorem \ref{thm:integral}:}
%
%\begin{proof}  
	 For an unweighted path graph % has the Laplacian eigenvalues  
	$$\lambda_{i}=2-2\cos(\pi (i-1) /N), \text{ for } i=1,\dots,N$$  
	We define the equidistant partition of interval $[1/N,1]$ as $$\mathcal{P}=\bigcup\limits_{i=1}^{N-1} \left [{i}/{N},(i+1)/{N}\right]:=\bigcup\limits_{i=1}^{N-1} \mathcal{P}_i(N).$$
	We define the following quantities for each interval:
	\begin{align*}
	& \overline \phi_{i,N}:=\max_{x\in \mathcal{P}_i(N)} \phi(2-2\cos(\pi x)), \\
	&  \phi_{i,N}:= \phi(2-2\cos(\pi i/N)), \\
	& \underline \phi_{i,N}:=\min_{x\in \mathcal{P}_i(N)} \phi(2-2\cos(\pi x)). 
	\end{align*}
	They induce the following summations
	$$
	\overline S_N=\sum_{i=1}^{N-1} \overline \phi_{i,N},~S_N=\sum_{i=1}^{N-1}  \phi_{i,N},~\underline S_N=\sum_{i=1}^{N-1} \underline \phi_{i,N},
	$$
	where $\rho({\mathbf{L},\mathbf{K}})=S_N$. These sums imply the natural ordering
	$$
	\overline S_N \leq S_N \leq \underline S_N.
	$$
	Moreover, compared to $\Gamma_N$, we observe that 
	$$
	\underline S_N \cdot 1/N \leq \Gamma_N \leq \overline S_N\cdot 1/N. 
	$$
	We bring a lemma whose proof is given in the next appendix. %which we will prove after this proof. 
	\begin{lemma}\label{eq:maxminratio}
		For a rational  function $\phi(\lambda)$ that is bounded for any $\lambda \in (0, \infty)$, 
		$
		{\overline \phi_{i,N}}/{\underline \phi_{i,N}} \leq \delta_{\phi},
		$
		uniformly over $i$ and $N$ for  some $\delta_\phi>0$ depending on $\phi$. 
	\end{lemma}
	We can apply Lemma \ref{eq:maxminratio}, since $\tilde \lambda (\mathbf K)=0$. As a results, we   find that 
	$$
	1\leq \min_{i=1,\dots,N-1} \dfrac{\phi_{i,N}}{\underline \phi_{i,N}} \leq \dfrac{\overline S_N}{\underline S_N} \leq \max_{i=1,\dots,N-1} \dfrac{\phi_{i,N}}{\underline \phi_{i,N}} \leq \delta_{\phi}.
	$$
	This means that $\Gamma_N$ and $S_N$ are both bounded according to 
	\begin{align*}
	& 1  \leq \dfrac{N \Gamma_N}{ \underline S_N} \leq \delta_\phi, ~ 1  \leq \dfrac{S_N}{ \underline S_N} \leq \delta_\phi.
	\end{align*} 
	If we combine these two inequalities, we find that 
	$$
	\dfrac{1}{\delta_\phi}  \leq \dfrac{N \Gamma_N}{S_N} \leq \delta_\phi \Rightarrow S_N=\Theta(N\Gamma_N). 
	$$
	If $\phi(\lambda)$ is bounded, then one deduces that 
	$$\lim_{N \rightarrow \infty} {\overline \phi_{i,N}}/{\underline \phi_{i,N}}=1\Rightarrow \lim_{N \rightarrow \infty} {\overline S_N}/{\underline S_N}=1.$$
	Therefore, we can write the  following two inequalities 
	\begin{align*}
	\lim_{N \rightarrow \infty} \dfrac{N \Gamma_N}{S_N}=\lim_{N \rightarrow \infty} \dfrac{N \Gamma_N/\underline S_N}{S_N/\overline S_N} \leq \lim_{N \rightarrow \infty} \dfrac{ \overline S_N/\underline S_N}{\underline S_N/\overline S_N} =1
	\end{align*} 
	\begin{align*}
	\lim_{N \rightarrow \infty} \dfrac{N \Gamma_N}{S_N}=\lim_{N \rightarrow \infty} \dfrac{N \Gamma_N/\overline S_N}{S_N/\underline S_N} \geq \lim_{N \rightarrow \infty} \dfrac{ \underline S_N/\overline S_N}{\overline S_N/\underline S_N} =1.
	\end{align*}
	Thus, we can write 
	$$
	\displaystyle \lim_{N \rightarrow \infty} {N \Gamma_N}/{S_N}=1.
	$$ For  unweighted cycle graphs, the Laplacian eigenvalues are 
	$$\lambda_{i}=2-2\cos(2 \pi (i-1) /N), \text{ for } i=1,\dots,N.$$ 
	Without loss of generality, for deriving the scaling purposes, we may assume that $N$ is $odd$.
	Then, we will have $(N-1)/2$ distinct values for the eigenvalues of $\mathbf{L}$, where each value is repeated exactly twice. Moreover, we can write
	$$
	\rho({\mathbf{L},\mathbf{K}})=2\sum_{i=1}^{(N-1)/2} \phi \left (  2-2\cos \left ( \dfrac{2\pi i}{N}\right  )\right ).
	$$ 
	This time, we need to partition $[1/N,1/2]$ and proceed with identical steps to find out that 
	$
	2 \int_{1/N}^{1/2} \phi  \left (  2-2\cos \left ( 2 \pi x \right  )\right ) ~dx,
	$
	does the same job in the case of cycle graphs. 	
	If we replace $2x\rightarrow x$, the factor $2$ is canceled. This means that   \eqref{eq:rhotheta} and  \eqref{eq:limone} upon replacement of $\Gamma_N$ with $\Psi_N$ are achieved, where 
	\begin{align}
	\Psi_N:=\int_{2/N}^1 \phi\big (2-2\cos(\pi x) \big )~dx;
	\end{align}
	Since $\Gamma_N=\Theta(\Psi_N)$, the same conclusion apply for the networks over cycle graphs as well. 
	%The \emph{density} of the eigenvalues that are located at $\lambda \in (0,4)$ can be observed that is proportional to 
	%$$f(x)=\dfrac{1}{d (2-2\cos \pi x)/dx}=\dfrac{1}{2\pi \sin \pi x},$$
	%where $x$ is the parameter $x\in (0,1)$ such that $2-2\cos (\pi x)=\lambda$. Solving for $\sin (\pi x)$ from this equation,  we find that this density at $\lambda$ is 
	%$$f(\lambda)=\dfrac{1}{2\pi \sqrt{1-(1-\lambda/2)^2}}=\dfrac{1}{2 \pi \sqrt{\lambda-\lambda^2/4}},$$
	%We can see that the asymptotic of the $\rho({\mathbf{L},\mathbf{K}})$ would be 
	%\begin{align*}
	%& \rho({\mathbf{L},\mathbf{K}})= \sum_{i=1}^{N-1} \phi(2-2\cos(\pi i /N))\approx \\
	%& \dfrac{\int_{2-2\cos (\pi /N)}^{2-2\cos (\pi (N-1) /N)} \phi (\lambda) (N-1) f(\lambda)~d\lambda}{\int_{2-2\cos (\pi /N)}^{2-2\cos (\pi (N-1) /N)} f(\lambda)~d\lambda},
	%\end{align*}
	%where $(N-1) f(\lambda)$ accounts for "number" of the eigenvalues that are located at location $f(\lambda)$, and the denominator is required for normalizing the approximation. As $N$ increases, the lower and upper bounds of the integrals tend to $\pi^2 /N^2$ and $4$, respectively, and one can find the the denominator becomes $1$. This means we can write
	%$$
	%\rho_c({\mathbf{L},\mathbf{K}}) \propto N {\int_{\pi^2 /N^2}^{4} \phi (\lambda)f(\lambda)~d\lambda}
	%$$
%\end{proof}

 \section*{Appendix N}

\noindent{\it Proof of Lemma \ref{eq:maxminratio}:}  Let us define  
	$\chi:=\log (2-2\cos(\pi x)),$ and% the function 
	$
	\varphi(\chi):=\log(\phi(2-2\cos(\pi x))). 
	$
	%that can be shown that is equal to
	%$$
	%\varphi(z)=\log\left (\phi \left (\dfrac{1}{\pi} \cos^{-1}(\exp(z)/2)\right)\right ).
	%$$
	Denote the order of the pole of $\phi(x)$ at $x=0$ by $\alpha \in \mathbb{Z}_+$. This means that we can decompose $\varphi$ according to 
	\begin{align*}
	\varphi(\chi)&= \log\left (\dfrac{1}{(2-2\cos(\pi x))^\alpha} \hat \varphi(2-2\cos(\pi x))\right )\\
	&=\log\left ( \hat \varphi(2-2\cos(\pi x))\right )-\alpha \log(2-2\cos(\pi x)),
	\end{align*}
	for some strictly positive function $\hat \varphi(.)$ that is bounded and rational. We can write this in terms of $\chi$ as follows
	$$
	\varphi(\chi)=\log\left ( \hat \varphi(\exp(\chi))\right )-\alpha \chi.
	$$
	In the interval of interest, $\log(.)$, $\hat \varphi(.)$ (a bounded and positive rational function) and $\exp(.)$ are all Lipschitz continuous, so is their composition $\log\left ( \hat \varphi(\exp(\chi))\right )$. 
	This implies that $\varphi(\chi)$ is Lipschitz-continuous. Hence,  for $\chi_1=\log(2-2\cos(\pi x_1))$ and $\chi_2=\log(2-2\cos(\pi x_2))$, the corresponding  Lipschitz continuity inequality will be 
	$$
	|\varphi(\chi_2)-\varphi(\chi_1)| \leq \delta_{\varphi}|\chi_2-\chi_1|,
	$$
	for some Lipschitz constant $\delta_{\varphi} \geq 0$. This is equivalent to
	\begin{align*}
	\left |\log \left(\dfrac{\phi(x_2)}{\phi(x_1)}\right)\right |&  \leq \delta_{\varphi}\left |\log \left(\dfrac{2-2\cos(\pi x_2)}{2-2\cos(\pi x_1)}\right)\right |\\
	& =\left |\log \left(\dfrac{2-2\cos(\pi x_2)}{2-2\cos(\pi x_1)}\right)^{\delta_{\varphi}}\right |.
	\end{align*}
	This means  that alternatively we may write 
	\begin{align}\label{eq:starrrr}
	& \max \left ( \dfrac{\phi(x_2)}{\phi(x_1)} , \dfrac{\phi(x_1)}{\phi(x_2)} \right ) \leq  \\
	& \max  \left ( \left(\dfrac{2-2\cos(\pi x_2)}{2-2\cos(\pi x_1)}\right)^{\delta_{\varphi}},\left(\dfrac{2-2\cos(\pi x_1)}{2-2\cos(\pi x_2)}\right)^{\delta_{\varphi}}\right ). \notag
	\end{align}
	%Now, we can show that the ratios on the right hand side for $x_1,x_2 \in \mathcal{P}_i(N)$ are bounded from above by $4$. 
	Let us define 
	$$
	h_{N}(x):=\dfrac{2-2\cos\left (\pi \left (x+{1}/{N}\right )\right)}{2-2\cos(\pi x)}.
	$$
	Since $2-2\cos(\pi x)$ is increasing in $ \mathcal{P}_i(N)$, for $x_1,x_2 \in \mathcal{P}_i(N)$
	$$
	\dfrac{2-2\cos(\pi x_2)}{2-2\cos(\pi x_1)} \leq h_{N}(i/N).
	$$
	Moreover, one can find that $h_{N}(x)$ is decreasing for any $x \in [1/N,(N-1)/N]$. Thus, we can write
	\begin{align}\label{eq:star}
	\dfrac{2-2\cos(\pi x_2)}{2-2\cos(\pi x_1)} \leq h_{N}(i/N) \leq h_{N}(1/N). 
	\end{align}
	We can see note that the right hand side of \eqref{eq:star} is 
	$$
	h_N(1/N)=\dfrac{2-2\cos\left ({2\pi}/{N}\right )}{2-2\cos({\pi}/{N})}.
	$$
	Computing its \emph{formal} derivative with respect to $N$, we get
	$$
	\dfrac{d h_N(1/N)}{d N}=\dfrac{2\pi \sin \left ({\pi}/{N}\right ) }{N^2}>0.
	$$
	Thus, its supremum should be evaluated based on the limit 
	\begin{align}\label{eq:four}
	\dfrac{2-2\cos(\pi x_2)}{2-2\cos(\pi x_1)} \leq \lim_{N \rightarrow \infty} h_N(1/N)=4.
	\end{align}
	Let us take the maximum of the left hand side of \eqref{eq:starrrr}   over $x_1,x_2 \in \mathcal{P}_i(N)$ and combine it with \eqref{eq:four}. We conclude that   
	$$
	{\overline \phi_{i,N}}/{\underline \phi_{i,N}} \leq 4 ^{ \delta_{\varphi}}:=\delta_\phi,
	$$
	which completes the proof. 
%	For networks over the cycle graphs, the eigenvalues of the Laplacian matrix can be enumerated as  
%	$$\lambda_{i}=2-2\cos(2 \pi (i-1) /N), \text{ for } i=1,\dots,N.$$ 
%	Without loss of generality, for deriving the scaling purposes, we may assume that $N$ is $odd$. Then, we will have $(N-1)/2$ distinct values for the eigenvalues of $\mathbf{L}$, where each value is repeated exactly twice. Moreover, we can write
%	$$
%	\rho({\mathbf{L},\mathbf{K}})=2\sum_{i=1}^{(N-1)/2} \phi \left (  2-2\cos \left ( \dfrac{2\pi i}{N}\right  )\right ).
%	$$ 
%	This time, we need to partition $[1/N,1/2]$ and proceed with identical steps to find out that 
%	$
%	2 \int_{1/N}^{1/2} \phi  \left (  2-2\cos \left ( 2 \pi x \right  )\right ) ~dx,
%	$
%	does the same job in the case of cycle graphs. If we replace $2x\rightarrow x$, the factor $2$ is canceled and the claim is followed. 
	%This completes the proof of the lemma. 
%\end{proof} 

 \section*{Appendix O}

\noindent{\it Proof of Corollary \ref{cor:path}:} Because $\Gamma_N$ and $\Psi_N$ scale similarly with respect to $N$, the first part of the claim follows. If the performance function is bounded, then both of them become the same quantity, because we can replace the lower bound of these two integrals with a common limit of $0$. 
%\end{proof}

 \section*{Appendix P} 
 
\begin{proof}[Proof of Result on Input Measures] 
	We can see that 
	\begin{align*}
	\|u\|_2^2=\mathrm{Tr}\left (uu^T \right)= (\mathbf L \otimes \mathbf K \mathbf H) xx^T (\mathbf L \otimes \mathbf H^T  \mathbf K^T).
	\end{align*}
	We can further write 
	\begin{align*}
	uu^T=(\mathbf L \otimes \mathbf K \mathbf H) xx^T (\mathbf L \otimes \mathbf H^T  \mathbf K^T). 
	\end{align*}
	Because $\mathbf M_N \mathbf L=  \mathbf L \mathbf M_N=\mathbf L$, we can write 
	\begin{align*}
	uu^T =(\mathbf L \otimes \mathbf K \mathbf H) (\mathbf M_Nx)(\mathbf M_Nx)^T (\mathbf L \otimes \mathbf H^T  \mathbf K^T).
	\end{align*}
	If we take the expected value and tend the time to infinity, using the same lines as the proof of Theorem 1, we find that 
	\begin{align*}
	\lim_{t \rightarrow \infty} 	\mathbb{E} \{u(t)u(t)^T \}=& (\mathbf L \otimes \mathbf K \mathbf H)  (\mathbf U \otimes \mathbf I_n) \\ & \mathrm{diag} (\mathbf 0,\mathbf P(\lambda_2, \mathbf K), \dots,\mathbf P(\lambda_N, \mathbf K)) \\ & (\mathbf U^T \otimes \mathbf I_n) (\mathbf L \otimes \mathbf H^T  \mathbf K^T). 
	\end{align*}
	Let us take the trace from the both sides. We find that 
	\begin{align*}
	\rho_u(\mathbf L,\mathbf K)=&\mathrm{Tr} \left ( \mathrm{diag} (\mathbf 0,\mathbf P(\lambda_2, \mathbf K), \dots,\mathbf P(\lambda_N, \mathbf K)) \right . \\
	& \left .   (\mathbf U^T \mathbf L^2 \mathbf U \otimes \mathbf H^T  \mathbf K^T \mathbf K \mathbf H)  \right ). \end{align*} 
	Because $\mathbf U^T \mathbf L^2 \mathbf U=\mathbf \Lambda^2$, we conclude that 
	\begin{align*}
	\rho_u(\mathbf L,\mathbf K)=&\mathrm{Tr} \left ( \mathrm{diag} (\mathbf 0,\mathbf \lambda_2^2 \mathbf P(\lambda_2, \mathbf K) \right . \mathbf H^T  \mathbf K^T \mathbf K \mathbf H, \\ & \left .\dots,\lambda_N^2 \mathbf P(\lambda_N, \mathbf K)\mathbf H^T  \mathbf K^T \mathbf K \mathbf H) \right . \\ &= \sum_{i=2}^N \mathrm{Tr} \left ( \lambda_i^2 \mathbf{P}(\lambda_i,\mathbf K) \mathbf H^T  \mathbf K^T \mathbf K \mathbf H \right ). 
	\end{align*}
	Let us move   $\mathbf{KH}$ to the left hand side of the trace arguments. The claim is followed. 
\end{proof}

\section*{Appendix Q: \\ Additional Details of Examples}

\noindent{\it Details of Example \ref{ex:three}:} For networks with nodal dynamics  $\mathfrak{s}_1$, the noiseless dynamics of $\Sigma_i$ are
 $$
 \dot {r}=(-a-\lambda_ik)r,
 $$
that are asymptotically stable if 
 $a-\lambda_i k<0$. 
This confirms $\tilde \lambda (\mathbf{K})=\max\left (-{a}/{k},0\right )=0$. The solution to (\ref{eq:sub_lyap}) is $ {\mathbf{P}}={1}/{2(k\lambda+a)},$
 which result in the claimed form for $\phi$. We  can see that for $\lambda >\tilde \lambda(\mathbf{K})=0$,  it is strictly convex and strictly decreasing. % Moreover, we can show that
%$$
%{\mathrm{d}^2\mu}/{\mathrm{d}\lambda^2} ={k^2 \lambda^2}/(2(k \lambda+a))>0,
%$$
%Thus, the input function is also strictly convex. 
The  dynamics of the subsystem $\Sigma_i$ for nodal dynamics $\mathfrak{s}_2$ without disturbance and noise  are 
 \begin{align*}
 \dot {r}=\begin{bmatrix}
0 & 1 \\
-a_2-\lambda k_1 & -a_1 \lambda k_2 
\end{bmatrix} r.
 \end{align*}
The corresponding characteristic polynomial is 
 $$
p_\lambda(s)= s^2 + (a_1 + k_2\lambda)s + a_2 + k_1\lambda,
 $$
 which is a stable polynomial if and only if 
 $
 a_1 + k_2\lambda>0,\text{ and } a_2 + k_1\lambda>0,
 $
that imply $\tilde \lambda (\mathbf{K})=\max \left (-{a_1}/{k_2},-{a_2}/{k_1},0\right )=0$.  Using (\ref{eq:sub_lyap}), we get that 
%$$({\mathbf{A}}-\lambda {\mathbf{B}}{\mathbf{K}}) +{\mathbf{P}}(\lambda,\mathbf{K})({\mathbf{A}}-\lambda {\mathbf{B}}{\mathbf{K}})^T+E E ^T=0,$$
\begin{align*} 
\begin{adjustbox} {max width=250 pt}
${\mathbf{P}}(\lambda,\mathbf{K})=\mathrm{diag} \left (
 \dfrac{1}{2( k_2 \lambda+a_1)(k_1\lambda+a_2 )},\dfrac{1}{2(k_2 \lambda +a_1 )} \right ).$
 \end{adjustbox}
 \end{align*}
Substitution of this matrix into  \eqref{eq:fidef} gives us the  results of the table. Now, noting that 
$$
\phi(\lambda)=\dfrac{b_0^2}{2( k_2 \lambda+a_1)}+\dfrac{b_1^2}{2( k_2 \lambda+a_1)(k_1 \lambda+a_2)},
$$
that is sum of two  strictly convex and strictly decreasing functions after their negative poles. Those poles are less than or equal to $\tilde \lambda (\mathbf{K})$. Hence, the $\phi(\lambda)$ is strictly decreasing and strictly convex in the claimed domain. 

 For the double integrator with observer, note that $$
\mathbf{A}_\lambda=\mathbf{A}-\lambda \mathbf{F} {\mathbf{H}}=\begin{bmatrix}-f_1\lambda & 1 \\  -f_2\lambda & 0 \end{bmatrix}.
$$
Observe that its characteristic polynomial is 
$
p_\lambda(s)=s^2+f_1\lambda s+f_2\lambda,
$
which is stable if and only if $f_1,f_2>0$. 

\begin{table*}[t]%\large
	\begin{center}
		\resizebox{0.5 \textwidth}{!}{
			\begin{tabular}{ p{3cm}p{3cm}p{3cm}p{3cm}  }
				
				\multicolumn{4}{c}{$
					\dfrac{1}{2}\mathlarger{\begin{bmatrix}
						\dfrac{k_3 \lambda}{ ( k_1 \lambda) (k_2k_3 \lambda^2 \lambda)-   k_1 \lambda)} &                                            0 & * \\
						0 & \dfrac{1}{ k_2 k_3\lambda^2  -  k_1 \lambda } &                                             0 \\
						\dfrac{-1}{ k_2 k_3\lambda^2  -  k_1 \lambda }&                                            0 &   \dfrac{k_2 \lambda}{ k_2 k_3\lambda^2  -  k_1 \lambda }
						\end{bmatrix}} 
					$} \\
				
		\end{tabular}}
	\end{center}
	\caption{The value of ${\mathbf{P}}(\lambda,\mathbf{K})$ as the solution of Lyapunov equation for triple integrators.} \label{tableP_S_3}
\end{table*}

\noindent{\it  Details of Example \ref{ex:triple}:} For the network of triple-integrators, the characteristic polynomial $\Sigma_i$ is 
$$
p_\lambda(s)=s^3 + k_3 \lambda s^2 +   k_2 \lambda s +  k_1 \lambda.
$$
The stability requires $k_i>0$ and 
$$
(k_2 \lambda)(k_3 \lambda)- ( k_1 \lambda)>0 \Rightarrow  \lambda >\dfrac{k_1}{k_2 k_3} \Rightarrow \tilde{\lambda}(\mathbf{K})=\dfrac{k_1}{k_2 k_3}. 
$$
The solution to for ${\mathbf{P}}(\lambda,\mathbf{K})$ from the Lyapunov equation  is shown in Table \ref{tableP_S_3}. The formula for the performance function then is followed by computing $\mathrm{Tr}(\mathbf{C}^T\mathbf{P}\mathbf{C})$.% If the condition $b_1^2-2b_2b_0 \geq 0$ holds, then 
The performance function in this case is  strictly decreasing and convex. 

%For the network with nodal dynamics $\mathfrak{s}_3$, the characteristic polynomial $\Sigma_i$ is 
%$$
%p_\lambda(s)=s^3 + (a_1 + k_3 \lambda) s^2 + (a_2 + k_2 \lambda) s + a_3 + k_1 \lambda.
%$$
%If $k_i$'s are positive, then for large enough values of $\lambda$, this polynomial is stable. In fact, we need
%$$
%(k_2 \lambda+a_2)(k_3 \lambda+a_1)- ( k_1 \lambda+a_3)>0. 
%$$
%If the quadratic equation on the left hand side has positive roots, $\tilde \lambda (\mathbf{K})$ would be equal to the largest root. Otherwise, this inequality would hold for any value of $\lambda$, and we have $\tilde \lambda(\mathbf{K})=0$. The solution to for ${\mathbf{P}}(\lambda,\mathbf{K})$ from the Lyapunov equation  is shown in Table \ref{tableP_S_3}. The formula for the performance function then is followed by computing $\mathrm{Tr}(\mathbf{C}^T\mathbf{P}\mathbf{C})$. If the condition $b_1^2-2b_2b_0 \geq 0$ holds, then the performance function will be the sum of three strictly convex and strictly decreasing functions.

\noindent{\it  Details of Example \ref{ex:pltn}:} The  realization  of the agents is 
\begin{align*}
\begin{adjustbox}{max width=220 pt}
${\mathbf{A}}=\begin{bmatrix}0 & 1 & 0 \\ 0 & 0 & 1 \\ 0 & 0 & -{1}/{\tau} \end{bmatrix},~{\mathbf{B}}={\mathbf{E}} =\begin{bmatrix}0 \\ 0 \\ {1}/{\tau}\end{bmatrix},~{\mathbf{C}}=\begin{bmatrix} 1 & 0 & 0 \end{bmatrix}.$
\end{adjustbox}
\end{align*}
The characteristic polynomial of the systems in this case becomes
$$
p_{\lambda}(s)={\tau s^3 + (1+\lambda k_3)s^2 + \lambda k_2 s +\lambda k_1},
$$
 (see also \cite{zheng2016platooning}). Based on Routh-Hurwitz criteria, for an unbounded stability region,  we should impose the following restrictions on $\mathbf{K}$.
%(i.e. for stability at large enough values of $\lambda_i$)
$
k_1,k_2>0,\text{ and } k_3\geq 0.
$ 
Moreover, we need the inequality 
\begin{align}\label{eq:non_const}
(1+\lambda k_3)\lambda k_2>\tau \lambda k_1\Rightarrow k_3\lambda>{\tau k_1/k_2-1},
\end{align}
from which we may infer  the bicriteria definition for $\tilde \lambda(\mathbf{K})$ in (\ref{eq:bi}). To find the performance functions, we find ${\mathbf{P}}(\lambda,\mathbf{K})$ which is %shown in Table \ref{tableP_platoons}, 
$$
\begin{adjustbox}{max width=250 pt}
$\mathsmaller{\begin{bmatrix}
\frac{1}{2k_1}\frac{k_3\lambda + 1}{k_2k_3\lambda^3 + (k_2 - k_1\tau)\lambda^2} &                                            0 & \frac{-1}{2(k_2k_3\lambda^2 +(k_2 - k_1\tau)\lambda)} \\
0 &  \frac{1}{2(k_2k_3\lambda^2 +(k_2 - k_1\tau)\lambda)} &                                             0 \\
                              \frac{-1}{2(k_2k_3\lambda^2 +(k_2 - k_1\tau)\lambda)}&                                            0 &         \frac{k_2}{2\tau( k_2k_3\lambda+k_2 - k_1\tau)} 
\end{bmatrix}}$ \end{adjustbox}
$$
which lets us compute the performance function. If $k_3>0$, the function $\phi(\lambda)$  is a positive combination of 
\begin{align*}
\begin{adjustbox}{max width=250 pt}
$f_1(\lambda)=\left (k_3\lambda^2 + \dfrac{k_2 - k_1\tau}{k_2}\lambda\right )^{-1},~~f_2(\lambda)=\left (k_3\lambda^3 + \dfrac{k_2 - k_1\tau}{k_2}\lambda^2\right )^{-1}. 
$
\end{adjustbox} 
\end{align*}
Moreover, $f_1$ and $f_2$ are products of functions that are strictly convex and decreasing  for $\lambda >\tilde \lambda (\mathbf{K})$ according to 
\begin{align*}
\begin{adjustbox}{max width=240 pt}
$f_1(\lambda)=\dfrac{1}{k_3 \lambda} \cdot \dfrac{1}{\lambda + \dfrac{k_2 - k_1\tau}{k_3k_2}}, ~f_2(\lambda)=\dfrac{1}{k_3 \lambda^2} \cdot \dfrac{1}{\lambda + \dfrac{k_2 - k_1\tau}{k_3k_2}}.$
\end{adjustbox} 
\end{align*}
Thus, the performance function is this case is also strictly convex and strictly decreasing for $\lambda >\tilde \lambda(\mathbf{K})$.

\noindent{\it Details of Example \ref{ex:exampleone}:} 
 The input-output transfer function is $H(s)=(s-\zeta)/{s^2}$ with a right-hand plane zero at $\zeta>0$.  To evaluate $\mathbf {P}_0$, we use a method suggested by \cite{qiu1993performance}. First, we decompose the transfer function according to $H=H_1H_2$ where the components are 
$$
H_1(s)=\dfrac{s-\zeta}{s+\zeta},~H_2(s)=\dfrac{s+\zeta}{s^2}.
$$
The transfer function $H_1$ has the balanced realization \footnote{A minimal realization of a stable transfer matrix is called balanced if its controllability and observability Gramians are diagonal and equal (such a realization exists for a stable transfer matrix)}. 
$$
\hat {\mathbf{A}}_1=-\zeta, ~\hat {\mathbf{B}}=\sqrt{2\zeta},~\hat {\mathbf{C}}_1=-\sqrt{2\zeta},~\hat {\mathbf{D}}_1=1,
$$
while $H_2$ has a stabilizable and detectable realization
$$
\hat {\mathbf{A}}_2=\begin{bmatrix}
0 & 1 \\
0 & 0 
\end{bmatrix},~\hat {\mathbf{E}} =\begin{bmatrix}
0  \\
1 
\end{bmatrix},~\hat {\mathbf{C}}_2=\begin{bmatrix}
\zeta  & 1
\end{bmatrix},~\hat {\mathbf{D}}_2=0. 
$$ 
Suppose that the factorized realizations have the state vectors $X$ and $x$, respectively. Then, we can show that they are related based on
$$
X=\begin{bmatrix} \sqrt{2\zeta } & 0 \\ 1 & 0 \\ 0 & 1 \end{bmatrix} x:=\mathbf{T}x.
$$
Then, the reference shows that
\begin{align} \label{eqP_0_def}
\mathbf{P}_0=\mathbf{T}^T \begin{bmatrix} \mathbf{I}_b & \mathbf{0}  \\ \mathbf{0} & \mathbf{0} \end{bmatrix} \mathbf{T}. 
\end{align}
Therefore, we get 
$
\mathbf{P}_0=\mathrm{diag} \left ( 2\zeta ,0 \right ).
$

\noindent{\it Details of Example \ref{ex:path_netnet}:} We can see that $\tilde {\mathbf{E}} =\tilde {\mathbf{C}}=\mathbf{I}_m$ and 
$$
\begin{adjustbox} {max width=240 pt}$
\tilde {\mathbf{A}}_\lambda =\begin{bmatrix}   - k_1 &    k_1 & 0 & 0&  \dots      &   0 \\
k_1 & -2k_1 & k_1 & 0&   \dots & 0 \\
\vdots  & \ddots & & \ddots & \ddots &  \vdots \\
    0 & \dots &   0  & 0 &  k_1 & -k_1- k_2 \lambda \end{bmatrix}.$
    \end{adjustbox}
$$
The solution to the Lyapunov equation in this case is 
$$
\tilde {\mathbf{P}}=\begin{adjustbox} {max width=220 pt}$
\begin{bmatrix}  \dfrac{k_1 + (m-1) k_2 \lambda}{2 k_1 k_2 \lambda} & \dfrac{k_1 + (m-2) k_2 \lambda}{2 k_1 k_2 \lambda} & \dots & \dfrac{k_1 + k_2 \lambda}{2 k_1 k_2  \lambda} & \dfrac{1}{2 k_2  \lambda} \\ \dfrac{k_1 + (m-2) k_2 \lambda}{2 k_1 k_2 \lambda} & \dfrac{k_1 + (m-2) k_2 \lambda}{2 k_1 k_2 \lambda} & \dots & \dfrac{k_1 + k_2 \lambda}{2 k_1 k_2  \lambda} & \dfrac{1}{2 k_2  \lambda}  \\
\vdots & \vdots &\ddots & \dfrac{k_1 + k_2 \lambda}{2 k_1 k_2  \lambda} & \dfrac{1}{2 k_2  \lambda} \\
\vdots & \vdots &\ddots & \vdots & \vdots \\
 \dfrac{k_1 + k_2 \lambda}{2 k_1 k_2  \lambda} &  \dfrac{k_1 + k_2 \lambda}{2 k_1 k_2  \lambda} &\dots & \dfrac{k_1 + k_2 \lambda}{2 k_1 k_2  \lambda} & \dfrac{1}{2 k_2  \lambda} \\
\dfrac{1}{2 k_2  \lambda} & \dfrac{1}{2 k_2  \lambda} &\dots & \dfrac{1}{2 k_2  \lambda} & \dfrac{1}{2 k_2  \lambda} \\
\end{bmatrix}. $
\end{adjustbox}
$$
Now, similar to the previous example, we can see that 
\begin{align*}
 \phi_\mathrm{nn}& = \mathrm{Tr}(\tilde {\mathbf{C}}\tilde{\mathbf{P}}\tilde {\mathbf{C}}^T)=\mathrm{Tr}(\tilde{\mathbf{P}})=\sum_{i=1}^m \dfrac{k_1 + (i-1) k_2 \lambda}{2 k_1 k_2 \lambda}\\
& =\dfrac{k_1 m+ k_2 \lambda \displaystyle \sum_{i=1}^m (i-1)}{2k_1 k_2}=\dfrac{ \dfrac{m(m-1)}{2}k_2 \lambda +k_1 m}{2k_1 k_2}.
\end{align*}

\begin{table*}[t]%\large
	\begin{center}
		\resizebox{0.9 \textwidth}{!}{%
			\begin{tabular}{ p{3cm}p{3cm}p{3cm}p{3cm}  }
				
				\multicolumn{4}{c}{$
					\dfrac{1}{2}\begin{bmatrix}  \mathbf{J}_{m-1} \otimes \begin{bmatrix} \dfrac{(m+1) \lambda^2 + 2m^2 \lambda + m^4}{m^2 k_1 k_2 \lambda^2} & 0 \\0 &  \dfrac{\lambda +m}{mk_2 \lambda }  \end{bmatrix}+   \mathbf{I}_{m-1} \otimes \begin{bmatrix} \dfrac{1}{m^2 k_1 k_2} & 0 \\0 &  \dfrac{1}{mk_2 }  \end{bmatrix}& * \vspace{1mm}  \\
					1_{m-1}^T \otimes \begin{bmatrix} \dfrac{\lambda+m}{ k_1 k_2 \lambda^2} & 0 \\0 &  \dfrac{1}{k_2\lambda }  \end{bmatrix} & \begin{bmatrix} \dfrac{m}{ k_1 k_2 \lambda^2} & 0 \\0 &  \dfrac{1}{k_2\lambda } \end{bmatrix}  \end{bmatrix}
					$} \\
				
			\end{tabular}
		}
	\end{center}
	\caption{The solution to Lyapunov equation $\tilde{\mathbf{P}}$ for complete subnetworks with $m$ double-integrator agents ($*$ implies  symmetric element).} \label{table:Q_d}
\end{table*} 

\noindent{\it Details of Example \ref{ex:complete_netnet}:} For subnetworks of single-integrators over $\mathcal{G}_1$ that is complete,  $\tilde {\mathbf{E}} =\tilde {\mathbf{C}}=\mathbf{I}_m$ and 
$$
\begin{adjustbox} {max width=240 pt}$
\tilde {\mathbf{A}}_\lambda =\begin{bmatrix}   -(m-1) k &    k & \dots      &         k \\
\vdots  & \ddots & & \vdots \\
    k & \dots &   k  & - (m-1) k - k \lambda \end{bmatrix}.
$
\end{adjustbox}
$$
We can verify that the solution to the Lyapunov equation is 
$$
\begin{adjustbox} {max width=240 pt}$
\tilde {\mathbf{P}}=\dfrac{1}{2k}\begin{bmatrix} \dfrac{\lambda +m}{m \lambda } \mathbf{J}_{m-1}+ \dfrac{1}{m } \mathbf{I}_{m-1}& \dfrac{1}{ \lambda}1_{m-1}  \vspace{1mm}  \\
\dfrac{1}{ \lambda}1_{m-1}^T & \dfrac{1}{ \lambda}   \end{bmatrix}.
$
\end{adjustbox}
$$
Then, we can write 
\begin{align*}
 \phi_\mathrm{nn}&= \mathrm{Tr}(\tilde {\mathbf{C}} \tilde{\mathbf{P}} \tilde {\mathbf{C}}^T)=\mathrm{Tr}( \tilde{\mathbf{P}} ) \\ &=(m-1) \dfrac{1}{2k}   \left (   \dfrac{\lambda +m}{m \lambda } + \dfrac{1}{m } \right )+\dfrac{1}{2k\lambda}\\&  =\dfrac{2(m-1)  \lambda +m^2}{2m k  \lambda }.
\end{align*}
For double-integrators over complete graph modules,  similar expressions for the matrices $\tilde {\mathbf{A}},$ $\tilde {\mathbf{B}}$, and $\tilde {\mathbf{C}}$ holds, while the solution to the Lyapunov equation in this case is shown in Table \ref{table:Q_d}.  Because the output of the double-integrator is on the first state, we can write
\begin{align*}
& \phi_\mathrm{nn}= \mathrm{Tr}(\tilde {\mathbf{C}}\tilde{\mathbf{P}}\tilde {\mathbf{C}}^T)=%\mathrm{Tr}(\tilde{\mathbf{P}})
 \\
  &  =\dfrac{m-1}{2}   \big (   \dfrac{(m+1) \lambda^2 + 2m^2 \lambda + m^4}{m^2 k_1 k_2 \lambda^2}+\dfrac{1}{m^2 k_1 k_2} \big )+\dfrac{m}{2k_1k_2\lambda^2} \\
  & =\dfrac{(m-1)(m+2) \lambda^2+2m^2(m-1) \lambda +m^4}{2 m^2 k_1 k_2 \lambda^2}.
\end{align*}
This proves the claims in the example.

\noindent{\it Details of Continuance of Example \ref{ex:three}:} 
%For the first class we can write 
%$$
%\phi_\gamma= \dfrac{1}{2k(\lambda+z)} +\gamma \dfrac{k^2\lambda^2}{2k(\lambda+z)}.
%$$
First, we prove that we can replace $\Gamma_N$ 
%and $\Psi_N$ 
 in Theorem \ref{thm:integral} with
% By the change of variable , the function $g(N)$ in Theorem \ref{thm:integral} can be alternatively computed as
\begin{align*}
& \Gamma_N= \dfrac{1}{2 \pi } {\int_{\pi^2/N^2}^{4} \phi (\lambda)\dfrac{1}{\sqrt{\lambda-\lambda^2/4}}~d\lambda}.
%; \\
%& \Psi_N= \dfrac{1}{2 \pi } {\int_{4 \pi^2/N^2}^{4} \phi (\lambda)\dfrac{1}{\sqrt{\lambda-\lambda^2/4}}~d\lambda}.
\end{align*}
Considering $\lambda=2-2\cos(\pi x)$, we get that 
 $$
 d \lambda =2\pi \sin (\pi x)\, dx=2\pi \sqrt{1-\cos^2(\pi x)}\, dx .
 $$
Given $\lambda =2-2\cos(\pi x)$, 
$
d\lambda ={2\pi}\sqrt{\lambda-\lambda^2/4}~d x,
$ and 
\begin{align*}
\left \{\begin{array}{l}
 x=1/N \Rightarrow \lambda=2-2\cos(\pi/N) \sim \pi^2/N^2 \\
 x=1 \Rightarrow \lambda=2+2=4
\end{array}\right . .
\end{align*}
that are integral limits of interest. %in the first case. For the case over the cycle graphs, the only difference is in the lower limit of the integral as shown in the result.

Now, if $a=0$, for the single integrators %first class  we need%  the integral
$$
\int \dfrac{1}{\lambda}\dfrac{1}{ \sqrt{\lambda-\lambda^2/4}}~d\lambda=-\dfrac{\sqrt{4-\lambda}}{\sqrt{\lambda}},
$$
we can compute $\Gamma_N$ for $\phi(\lambda)$ as
$$\Gamma_N=\dfrac{1}{2k}\dfrac{1}{2\pi}\left (  \dfrac{\sqrt{4-\pi^2/N^2}}{\sqrt{\pi^2/N^2}}   \right )\sim \dfrac{ N}{2 \pi^2 k}.$$
%Moreover, because we can compute 
%\begin{align*}
%\begin{adjustbox}{max width=230 pt}
%$ \int  \dfrac{\lambda}{ \sqrt{\lambda-\lambda^2/4}}~d\lambda=-4 \left(\sqrt{\lambda-{\lambda^2}/{4}}+\sin^{-1} \left (1-{\lambda}/{2} \right ) \right ).$
%\end{adjustbox} 
%\end{align*}
Now, if $a>0$, then we need the integral 
\begin{align*}
& \int_{0}^4 \dfrac{1}{\lambda+\alpha}\dfrac{1}{ \sqrt{\lambda-\lambda^2/4}}~d\lambda =% \\ 
%& \left .\dfrac{4}{ \sqrt{\alpha(\alpha+4)}} \tan^{-1}\left (\sqrt{\dfrac{\alpha+4}{\alpha}\dfrac{\lambda}{4-\lambda}}\right ) \right |_0^4=
\dfrac{2\pi }{ \sqrt{\alpha(\alpha+4)}}.
\end{align*}
This implies that $\Gamma_N$ for $\phi(\lambda)$ in this case satisfies 
$$
\Gamma_N \sim \dfrac{1}{2k}\dfrac{1}{2\pi}\dfrac{2\pi }{ \sqrt{\alpha(\alpha+4)}}=\dfrac{1}{2k\sqrt{\alpha(\alpha+4)}}.
$$

%Also, we compute the integral  
%$$
% \int_0^4 \dfrac{1}{ \sqrt{\lambda-\lambda^2/4}}~d\lambda =2\pi  \int_0^1 1-\cos(\pi x)~d x=2\pi. 
%$$
%One can note that we can write the input function as  
%$$
%\mu(\lambda)=\dfrac{k}{2} \left ( \lambda-z+\dfrac{z^2}{\lambda+z} \right ).
%$$
%This lets us deduce that $\Gamma_N$ for $\mu(\lambda)$ satisfies 
%\begin{align*}
%& \Gamma_N\sim \dfrac{k}{2} \dfrac{N}{2\pi}  \left (  4\pi /2 -2\pi z+  \dfrac{2z^2}{ \sqrt{z(z+4)}} \dfrac{\pi}{2}   \right )=\\
%& \dfrac{kN}{4}  \left (  2(1-z)+ \dfrac{z^2}{\sqrt{z(z+4)}} \right ) 
%\end{align*}
%For the second order agents, we have 
%$$
%\phi(\lambda)=\dfrac{b_0^2}{2k_2 (\lambda+z_1)}+\dfrac{b_1^2}{2k_1 k_2 (\lambda+z_1)(\lambda+z_2)},
%$$
%$$
%\mu(\lambda)=\dfrac{k_1^2 \lambda^2}{2k_2 (\lambda+z_1)}+\dfrac{k_2^2\lambda ^2}{2k_1 k_2 (\lambda+z_1)(\lambda+z_2)}.
%$$
%For dynamics $\mathfrak{s}_2$ with 
For $\mathfrak{s}_2$ agents with $a_0=a_1=0$ we need %$z_1=z_2=0$ we need 
$$
\int \dfrac{1}{\lambda^2}\dfrac{1}{ \sqrt{\lambda-\lambda^2/4}}~d\lambda=-\dfrac{\sqrt{4-\lambda}(\lambda+2)}{6\lambda\sqrt {\lambda}}.
$$
Now, we compute $\Gamma_N$ for $\phi(\lambda)$ as follows 
\begin{align*}
& \Gamma_N \sim 
\begin{adjustbox}{max width=220 pt } 
$\dfrac{1}{2\pi}\dfrac{b_0^2}{2k_2} \left( \dfrac{\sqrt{4-\pi^2/N^2}}{\sqrt{\pi^2/N^2}} \right )+\dfrac{1}{2\pi}\dfrac{b_1^2}{2k_1 k_2} \left( \dfrac{2 \sqrt{4-\pi^2/N^2}}{6 \pi^2/N^2 \sqrt{\pi^2/N} } \right )$
\end{adjustbox}
\\ &~~~~  \sim  \dfrac{ b_0 ^2 N}{2\pi^2 k_2}+\dfrac{b_1^2N^3}{6 \pi^4 k_1 k_2 }. 
\end{align*}

\noindent{\it Details of Continuance of Example \ref{ex:harmonic}:}
In this case  similar computations reveals that for $\phi(\lambda)$
\begin{align*}
& \Gamma_N \sim %\dfrac{b_0^2}{ 2k_2\sqrt{\alpha_1(\alpha_1+4)}}+ \\
  \dfrac{1 }{2k_1k_2(\alpha_1-\alpha_2)}\left (\dfrac{1}{ \sqrt{\alpha_2(\alpha_2+4)}}-\dfrac{1}{ \sqrt{\alpha_1(\alpha_1+4)}}\right ).
\end{align*}

\noindent{\it Details of the Continuance of Example \ref{ex:path_netnet}:}
In this case, 
\begin{align*}
\Gamma_N& =\dfrac{1}{2\pi} \int_{\pi^2/N^2}^{4}\dfrac{ \dfrac{m(m-1)}{2}k_2 \lambda +k_1 m}{2k_1 k_2 \lambda }\dfrac{1}{\sqrt{\lambda-\lambda^2/4}}~d\lambda\\
& \sim \dfrac{m(m-1)}{8 \pi k_1 } \times 2\pi +\dfrac{m}{4k_2 \pi} \times \dfrac{2N}{ \pi }\sim \dfrac{m^2}{4 k_1}+\dfrac{mN}{2k_2 \pi^2}.
\end{align*} 
Thus, based on the formula for the performance of the network of networks in (\ref{eq:nn_perf}), the claim is followed.

\noindent{\it  Details of Example \ref{ex:aircraft}:}  The state space matrices of the aircraft model are borrowed from \cite{boyd2008lecture} are given below.
\begin{align*} 
& \begin{adjustbox} {max width=220 pt}${\mathbf{A}}=\begin{bmatrix}
-0.003 & 0.039 & 0 & -0.322 & 0 & 0\\
-0.065 & -0.319 & 7.74 & 0 & 0 & 0\\
0.02 & -0.101& -0.429 & 0 & 0 & 0\\
0 & 0 & 0 & 1 & 0 & 0\\
1 & 0 & 0 & 0 & 0& 0 \\
0 & -1 & 0 & 7.74 & 0  & 0
\end{bmatrix} $\end{adjustbox}
\\
& \begin{adjustbox} {max width=220 pt}$ {\mathbf{B}}=\begin{bmatrix}
0.01 & 1\\
-0.18 & -0.04 \\
-1.16 & 0.598 \\
0 & 0\\
0 & 0 \\
0 & 0
\end{bmatrix},~~
{\mathbf{E}} =\begin{bmatrix}
0.003 & -0.039 \\
0.065 & 0.319 \\
-0.02 & 0.101 \\
0 & 0 \\
0 & 0 \\
0 & 0
\end{bmatrix}.$\end{adjustbox}
\end{align*}
The result of feedback gain design is
$$
\mathbf{K}= 
\begin{adjustbox}{max width=230 pt}$\begin{bmatrix} 
1.1894  &  0.7756  &  -2.0834  & -7.5558  &  0.3675  & -0.2017 \\
2.8779 &   -0.0193   &  0.1032 &    0.1276  &  0.7532  &   0.0872  \end{bmatrix}.$
% $\begin{bmatrix} 
%0.2395  &  0.2473 &  -1.0974  &  -2.5547 &   0.0313 &   -0.0259 \\
%    0.7958  & -0.0802  &  0.3871 &   0.8986   & 0.0776   & 0.0602
%    \end{bmatrix}.$
\end{adjustbox}
$$ 
%The functions $\phi_1$ and $\phi_2$ in this case have been reported in Table \ref{table:aircraft}. The derivation of these functions follows from symbolic solution to the corresponding Lyapunov equation using MATLAB. 
For the case of observer-based relative output feedback, the following value of $\mathbf{F}$  gives us depicted performance functions.
$$
\mathbf{F}=
\begin{adjustbox}{max width=200 pt}$
\begin{bmatrix} 9.6772 &  -0.3789 \\
1.0285 &  12.6584 \\ 
0.4233  & -1.9982 \\
0.1418   & 3.3839 \\
9.4718  & -0.0616 \\
-0.0616  &  9.0089 \end{bmatrix} $
\end{adjustbox}.
$$

\end{document}